# THE DYNAMICS OF NEURAL CODES

## GUILLERMO BARRIOS MORALES

### Advisor: MIGUEL A. MUÑOZ

From criticality to machine learning
and drifting representations

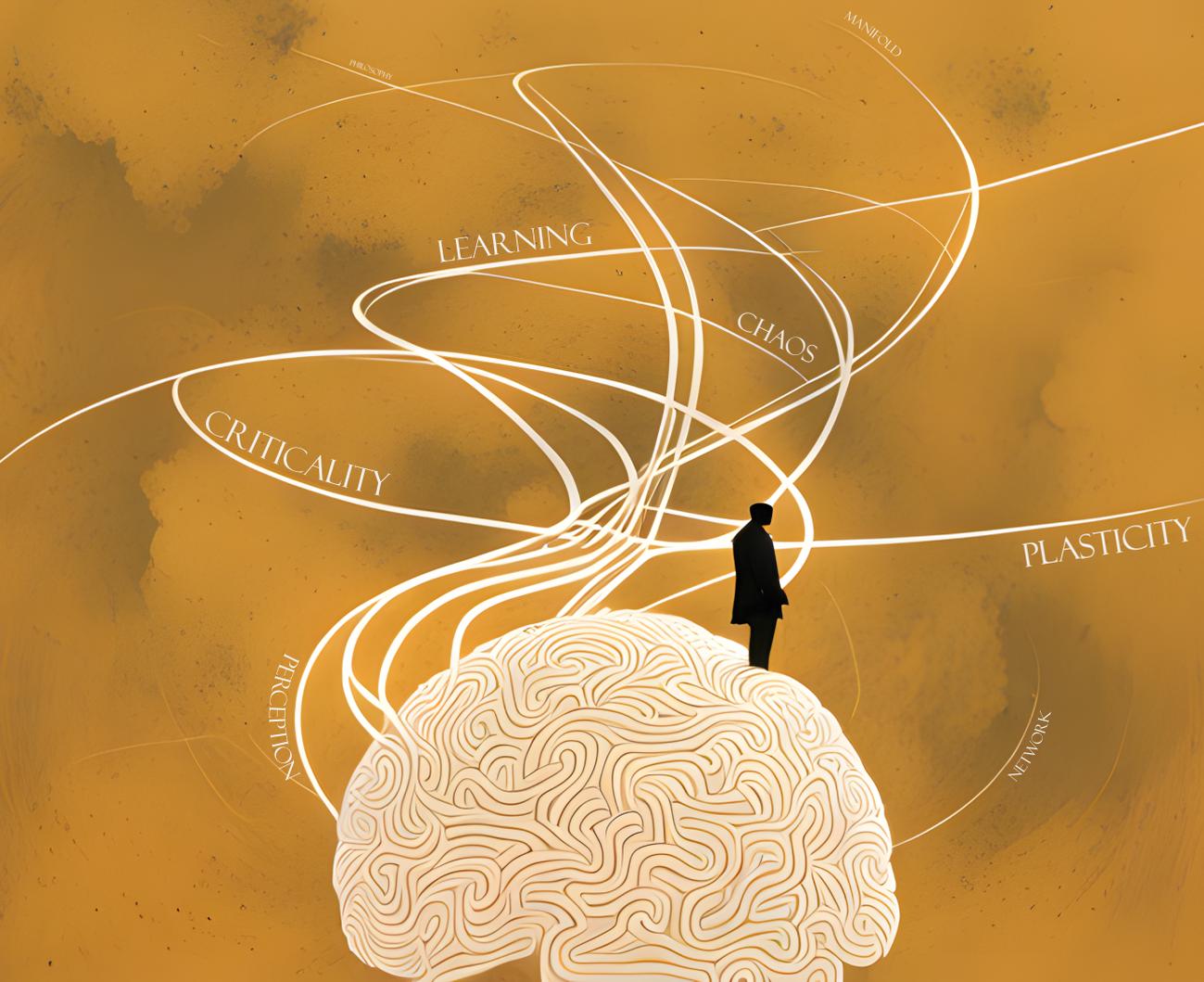

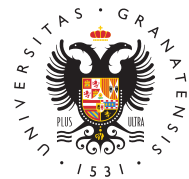

**UNIVERSIDAD
DE GRANADA**

GUILLERMO BARRIOS MORALES

# THE DYNAMICS OF NEURAL CODES
## IN BIOLOGICAL AND ARTIFICIAL
## NEURAL NETWORKS

Advisor: Miguel Ángel Muñoz
Programa de doctorado en Física y Matemáticas

A dissertation submitted to the University of Granada
in partial fulfillment of the requirements for the degree of
DOCTOR OF PHILOSOPHY.

A mi madre,
por enseñarme el don de la curiosidad.

# Abstract

Advancing our knowledge of how the brain processes information remains a key challenge in neuroscience. This thesis combines three different approaches to the study of the dynamics of neural networks and their encoding representations: a computational approach, that builds upon basic biological features of neurons and their networks to construct effective models that can simulate their structure and dynamics; a machine-learning approach, which draws a parallel with the functional capabilities of brain networks, allowing us to infer the dynamical and encoding properties required to solve certain input-processing tasks; and a final, theoretical treatment, which will take us into the fascinating hypothesis of the "critical" brain as the mathematical foundation that can explain the emergent collective properties arising from the interactions of millions of neurons.

Hand in hand with physics, we venture into the realm of neuroscience to explain the existence of quasi-universal scaling properties across brain regions, setting out to quantify the distance of their dynamics from a critical point. Next, we move into the grounds of artificial intelligence, where the very same theory of critical phenomena will prove very useful for explaining the effects of biologically-inspired plasticity rules in the forecasting ability of Reservoir Computers. Halfway into our journey, we explore the concept of neural representations of external stimuli, unveiling a surprising link between the dynamical regime of neural networks and the optimal topological properties of such representation manifolds. The thesis ends with the singular problem of representational drift in the process of odor encoding carried out by the olfactory cortex, uncovering the potential synaptic plasticity mechanisms that could explain this recently observed phenomenon.

Throughout this journey, amid fractal networks of neurons and chaotic strange attractors, the reader will encounter historical accounts of scientific breakthroughs, brain-inspired literary quotes and philosophical arguments on the existence of reality.

# Resumen

Avanzar en nuestro conocimiento sobre cómo el cerebro procesa información sigue siendo un desafío clave en la neurociencia. Esta tesis aúna tres enfoques diferentes para el estudio de la dinámica de las redes y códigos neuronales: un enfoque computacional, basado en características biológicas básicas de las neuronas y sus redes para construir modelos efectivos que puedan simular su estructura y dinámica; un enfoque desde la inteligencia artificial, en paralelo con las capacidades funcionales del cerebro, permitiéndonos inferir las propiedades dinámicas requeridas para resolver ciertas tareas involucradas en el procesamiento de *inputs* externos; y un tratamiento teórico, bajo la hipótesis del cerebro "crítico" como base matemática para explicar las propiedades colectivas emergentes que surgen de las interacciones de millones de neuronas.

De la mano de la física, nos aventuraremos en el reino de la neurociencia para explicar la existencia de propiedades invariantes de escala que son cuasi-universales entre diferentes regiones cerebrales, buscando cuantificar cuan lejos se encuentra la dinámica de estas últimas de un punto crítico. Luego pasaremos al terreno de la inteligencia artificial, donde la misma teoría de fenómenos críticos resultará muy útil para explicar los efectos de reglas de plasticidad neuronal en la capacidad de predicción de algoritmos tipo Reservoir Computing. A mitad de camino en nuestro viaje, introducimos el concepto de representaciones neuronales de estímulos externos, revelando un sorprendente vínculo entre el régimen dinámico de las redes neuronales y las propiedades topológicas que deben presentar estas representaciones para ser óptimas. La tesis culmina con el singular problema de la deriva representacional, observado recientemente en el proceso de codificación de olores por parte de la corteza olfativa, analizando los posibles mecanismos de plasticidad sináptica que podrían explicar dicho fenómeno.

A lo largo de este viaje, entre redes neuronales fractales y atractores caóticos extraños, el lector debe estar preparado para encontrar relatos históricos de hitos científicos, citas literarias inspiradas en el cerebro y algún que otro argumento filosófico sobre la existencia de la realidad.

# Preface

The aim of this book is to present the reader with some of the most fascinating problems that I have encountered during four years of PhD at the boundary between physics, neuroscience and machine learning, with special emphasis in the theory of criticality.

The first chapter serves as an introduction to many of the concepts that will be later used across the book. Starting from basic aspects of neuroscience, we will build the necessary mathematical scaffold to describe and model neural networks and their dynamics, as well as the plasticity mechanisms that allow such networks to learn from external inputs. Moving into the realm of artificial intelligence, the chapter also provides a historical view into the field of machine learning, up to the advent of Reservoir Computing techniques. It concludes with a section that serves as an overview on the theory of critical phenomena, laying down some of the fundamental concepts (phase transitions, scale-invariance, renormalization group...) that will be crucial for the results presented in Chapter 2, but also relevant to Chapters 3 and 4.

In Chapter 2 we roll up our sleeves and take a journey into the fascinating hypothesis of the "critical brain". We begin by showing that, under a phenomenological renormalization group, the activity in different regions of the brain shows quasi-universal scaling properties. We then set out to quantify how close these networks actually are from a critical dynamics, closing with a minimal model that allows us to link the empirically observed scale invariance with the dynamical regime of the networks. The core ideas of this chapter have been presented in Morales et al. 2023, but we also included two new sections that will be published independently elsewhere: Section 2.6 makes use of topographically-designed neural cultures to unveil the role of network structure in the emergence of scale invariance and close-to-critical dynamics; Section 2.7, on the other hand, presents a test for frequency-dependent criticality in MEG human data, spotlighting the differences observed between healthy individuals and patients with Parkinson's syndrome.

In Chapter 3 we jump from biological to artificial neural networks, delving into the effects of biologically-inspired plasticity rules on a type of machine learning (ML) framework known as Reservoir Computing (RC). While the main ideas behind this chapter have been published in Morales et al. 2021a, we present the results therein under a new light. In particular, we move further away from the ML drive for "best performance", searching instead for a deeper understanding of the effects of plasticity rules over the dynamics of the reservoir units. For this reason, instead of analyzing several plasticity rules and combinations of them to develop a best best-performing training protocol, we focus our study on just two rules that serve as paradigmatic examples of synaptic and non-synaptic plasticity. Moreover, the maximum Lyapunov exponent (MLE) is now introduced as a control parameter to characterize changes in the dynamical regime of reservoirs across the plasticity-induced learning phase, showing that transitions to a low-performance regime in networks "over-trained" with plasticity are actually of different nature for the synaptic and non-synaptic rules.

Some justice is done to the historical meaning of the word "PhD" in Chapter 4, which takes us into the philosophical problem of perception. This will be understood as a two-fold process, beginning with the translation of incoming sensory stimuli into *neural representations* (i.e., patterns of activity elicited by such inputs), then followed by a perceptual phase in which the subjective conscious experience of reality emerges. Using the RC approach laid out in the previous chapter, we study how optimal representations of external inputs are constrained by requirements of continuity and differentiability of the manifold in the neural space, finding an astounding link between the topological properties of these manifolds and the dynamical regime of the networks.

The thesis concludes with a chapter on the recently discovered phenomenon of representational drift. In particular, we focus our attention in the problem of odor encoding, constructing a biologically-realistic spiking neural model of the olfactory cortex. A mechanism based on synaptic plasticity is proposed, that is able to explain with surprising accuracy a series of recent empirical results related to the existence of a population encoding drift in this region, including the dependence of the drift rate on the frequency of stimulus presentation.

# Acknowledgements

As a physicist working in the field of complex systems, I cannot help but see this book as an emergent phenomenon, not just born as a result of four years of scientific work, but stemming from hundreds of meaningful interactions with truly wonderful people that gave shape to the person I am today.

To my advisor, Miguel Ángel, I can only say how fortunate I feel for having had the opportunity to learn from you all these years. I could thank you for all your passionate new ideas, for your constant scientific guidance, for making always time for a discussion, or just for the unbreakable optimism that, even in the lowest moments of any research project, seems to tell: "you are doing great, just one more step and you will get there!". And although this on itself makes for a great advisor, it still barely scratches the surface of all you have done for me. Today, what I really owe you, is my gratitude for the freedom you gave me. For always letting me work in things that I could be truly passionate about; for encouraging me to pursue my own ideas; for believing in them even when I was myself skeptical about their value. For all this and much more, thank you.

Following the natural path of science-related acknowledgements, this thesis would not be what it is today without the support of all my colleagues at the group of Statistical Physics in Granada. To my collaborator and friend, Serena: no PNAS is as valuable as knowing that I will always have a roof in whatever country I may find you; to Matteo and his indomitable energy, for those unforgettable dinners at his house in the Albaycin; to the indivisible Carles and Rubén, for whom correlations hold no secrets, thank you for your clarity of mind; to Rubén (the ever-present Rubén), for working full-time as the oracle-of-all-doubts in the department; and to Anna, the newcomer, whose sense of humor and willingness for coffee at any moment of the day earned her an acknowledgement in record time.

From my year as a visiting researcher in New York, I am deeply grateful to Yuhai, my advisor at IBM and mastermind behind the results presented in Chapter 5. Thank you for giving me the opportunity to have a glimpse into such a brilliant mind; I promise that I will be practicing my tennis skills for our next match. To my roommates with whom I shared a house in Harlem, to Anneysa, who opened me a door into the Brooklyn poetry and jazz scene, and to all my Fulbright friends (Paula, Patricia, Ale, Malgosia, Jordi, Sofía, Gonzalo, Nacho, Marta, Alex, Diana and a very long etcetera...); thank you for becoming the true meaning of what was likely the most stimulating year of my life.

One of the greatest things of doing a PhD is the opportunity to constantly meet fellow scientists in conferences, seminars and summer schools. Today, I am lucky to be able to call some of these scientists friends. Special thanks go to Tessa and her wonderful sense of humor, to whom I owe my newly acquired aerial-hoop skills (I can't wait for the next Biennale!). To Karolina I am not only grateful for some of the best conversations about human nature and ethical dilemmas I have ever had, but also for her invaluable comments on the introduction of Chapter 4, which helped me concealing some of my ignorance in the philosophical matters discussed there.

To all my friends with whom I share a passion for dancing, and very specially to Clara, Giselle, Nathalie and Raquel: thank you for becoming a new irreplaceable part of my life. To my group of friends from San Diego, thank you for showing that there is no time or space that can break what we started —I will be always looking forward for the next New Year's Eve.

A special place is reserved for those who have been constantly present in my life since my years of college and before. To Paula and Roberto, my confidants and support in every step (be it up or down) along this journey. Because you, among all people, know me for who I truly am, no words can describe how much I value having you in my life. And to my most idealistic friend, Beth, who also had the infinite patience to proofread important parts of this thesis: thank you for staying at my side, layer upon layer of abstraction, always pushing me to become a better person.

Umberto Eco once said that "translation is the art of failure". Es por ello que este párrafo, dedicado a mi familia, no podría sino ser escrito en mi lengua materna. A mi padre, quiero agradecerle su fe ciega en mí, así como su apoyo en cada una de las decisiones que he tomado a lo largo de mi vida, las cuales me han llevado, en última instancia, hasta donde hoy me encuentro. A mi hermana le debo el mostrarme, vez tras vez, contra viento y marea, que la familia de verdad es aquello que perdura incluso cuando la tempestad más fuerte arrastra todo lo demás. Os quiero muchísimo más de lo que jamás seré capaz de haceros ver.

Finally, I want to thank the person who probably had the greatest impact in this thesis without ever having written a word of it. For being by my side throughout these years: in my first class as teaching assistant, in my first paper rejection, but also in the day my first work was published. For listening to the endless rehearsals for my first conference talk, despite how little the term "renormalization group" meant to her. For reminding me the beauty of poetry, but even more specially, for showing me the pleasure of writing. To María, I can only say:

> Aprender el nombre del viento que seca tu cara
> fue el precio de llevarnos el mar en los bolsillos.

Y a todos una vez más, de corazón, muchísimas gracias.

"Follow me, reader! Who told you that there is no true, faithful, eternal love in this world! May the liar's vile tongue be cut out! Follow me, my reader, and me alone, and I will show you such a love!"

Mikhail Bulgakov,
*The Master and Margarita*

# Contents

# A physicist's view into the brain

[...]

The Brain is just the weight of God—
For —Heft them— Pound for Pound—
And they will differ—if they do—
As Syllable from Sound—

EMILY DICKINSON, c. 1862

"Rabbit's clever," said Pooh
thoughtfully.
"Yes," said Piglet, "Rabbit's clever."
"And he has Brain."
"Yes," said Piglet, "Rabbit has Brain."
There was a long silence.
"I suppose," said Pooh, "that that's why
he never understands anything."

A. A. MILNE, *Winnie the Pooh*

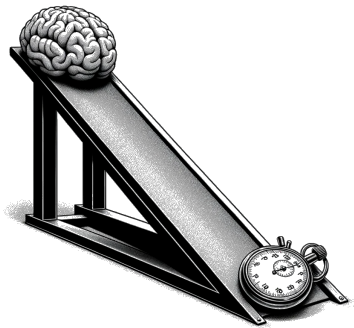

## 1.1 From hearts to neurons

The history of neuroscience is a tale of long-lasting misconceptions, serendipitous discoveries and confronted nobel prizes. Although nowadays the brain is widely accepted to be the source and vessel of our own identity, fears and emotions, likings and dislikings, and, in essence, our window to experience reality; this was not always the case.

Back in the ancient Egypt, about 3,000 years ago, the brain was ruthlessly thrown away during embalming (part drawn out through the nostrils with a piece of iron, part rinsed with drugs; Barnes 2012; X. Fan et al. 2019). According to the *Book of the Dead* (Faulkner 1985), the heart was considered to be the headquarters of emotions and thoughts, and so it was the only thing that the gods cared about when deciding the destiny of the deceased.

This *cardiocentric* hypothesis (the belief that the heart controls sensation, thoughts, body movement, etc.) would remain unchallenged for another five centuries, until the works of Alcmaeon of Croton (460 BC), in the ancient Greece. Alcmaeon, a physician and philosopher (and possibly a pupil of Pythagoras; Panegyres et al. 2016), was the first one to describe the existence of "channels" —$\pi\acute{o}\rho o\iota$ (poroi)— connecting the brain to the sensory organs (Panegyres et al. 2016; Zemełka 2017)[1]. Not only he postulated that the brain was the center of all thoughts, emotions and intelligence, but also considered it as a possible source of disorders in sensory organs (Zemełka 2017).

Despite the efforts of Alcmaeon, the *encephalocentric* hypothesis (sustaining that cognition occurs exclusively inside the brain) would still struggle to settle down for over 2,000 years, with philosophers of the stature of Aristotle or Avicenna advocating for a cardiocentric view of the body.

Even after the scientific revolution and the birth of modern science, in the early 19th century the nervous system was still an enigma, to the point of resisting inclusion in the cell theory proposed by Theodor Schwann in 1839, which posited that all tissues were made up of cells. A significant breakthrough would arrive from Camillo Golgi's invention of a silver staining technique in 1873, allowing nerve cells and their complex structures to be clearly visualized (Golgi 1879; Zhong et al. 2019). While Golgi supported the reticular theory, which viewed the nervous system as a continuous network, Ramón y Cajal, making use of the Golgi stain in the neural tissue of birds, provided evidence for the discontinuity of the nervous system and the presence of individual nerve cells (Ramón y Cajal 1904) —oddly enough, both scientists would be awarded the Nobel Prize in 1906 despite their diametrically opposing views. Ultimately, the neuron doctrine as a fundamental principle of neuroscience was conclusively validated in the 1950s with the advent of the electron microscopy, which demonstrated that nerve cells were individual units connected through synapses.

---

[1]It is likely that, while studying dissections of animals, Alcmaeon came across the optic the nerve, although he never named directly (Doty 2007).

At the present time, modern cell-counting techniques estimate that human brains are composed of around 80-100 billion neurons, and as many "support" cells known as *glia* (including, but not limited to, astrocytes, oligodendrocytes, microglia, Schwann cells...; see Bartheld et al. 2016 for a recent review). Neurons are considered the fundamental functional unit of the nervous system, responsible for transmitting and processing information in the form of electrical and chemical signals. Therefore, understanding their dynamics is crucial to comprehend the emergence of complex cognitive processes, such as perception, thought, and behavior, from the collective activity of billions of neurons in the human brain.

Broadly speaking, nearly all neurons can be divided into three functionally distinct parts (see Fig. 1.1a).

In the first place, one finds the *dendrites*, branching extensions of the cell body that receive the incoming signals from other neurons or sensory receptors. Dendrites are covered with small protrusions called dendritic spines, which collect inputs coming from other neurons and transmit them to the soma, playing a fundamental role in signal processing and learning.

The *soma* is the main body of the neuron, containing the nucleus of the cell, and it can be thought as the "central processing unit". In most cases, the soma is responsible of the non-linear integration of incoming signals from other neurons; only if the total input arriving to the soma exceeds a certain threshold, and output signal will be generated.

Finally, a long, slender projection, termed the *axon*, extends from the soma of most neurons. Axons are responsible of carrying the electrical impulses, known as action potentials, away from the soma towards other neurons, muscles, or glands, and so they can be thought as the "output devices" of neurons. In several cell types, the transmission of action potentials is sped up by mean of a fatty substance called myelin, which insulates the axon (see Fig. 1.1a).

Communication between neurons take place in regions called synapses, where the axon of a presynaptic neuron makes contact with the dendrites (or in some cases soma) of the postsynaptic neurons. Chemical synapses involve the release of neurotransmitters from the presynaptic neuron, which traverse the *synaptic cleft* (a small gap of $\sim 20-30$nm in width) and bind to receptor proteins on the postsynaptic neuron, leading to a change in the postsynaptic neuron membrane potential (see Fig 1.1a, inset). Depending on the type of neurotransmitter and receptors involved, this change can result in either excitation or inhibition of the postsynaptic neuron. Electrical synapses, on the other hand, are less common and feature gap junctions that allow ions to pass directly from one neuron to the next, allowing for a much faster and direct communication.

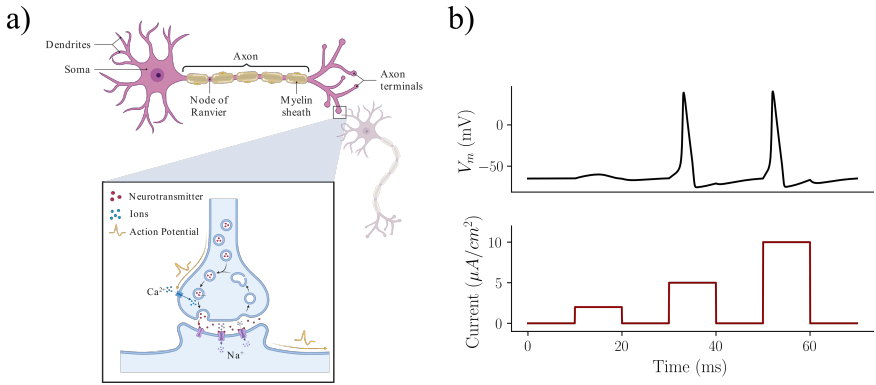

Figure 1.1: **Structure and dynamics of an archetypal neuron. (a)** Diagram highlighting the three distinct parts of a neuron: dentrites, soma and axon, as well as an schematic depiction of the basic processes involved in the transmission of an action potential from a presynaptic to a postsynaptic neuron in a chemical synapse. **(b)** Action potentials in the membrane voltage (black) of a Hodgkin-Huxley (H&H) model neuron, under a square-pulse input current (red). Notice how the first pulse is insufficient to rise the voltage over the threshold value, and therefore does not elicit an action potential. Once the firing threshold is reached, the shape and amplitude of the peak is shown to be fairly independent of the intensity of the input.

The dynamics of a neuron is primarily governed by the flow of ions (charged particles) across its cell membrane. At rest, neurons maintain a negative charge inside their membrane as compared to the extracellular fluid, referred to as the resting membrane potential. This resting potential, which is typically $\sim -70$mV, is primarily due to the differential distribution and transport of ions such as sodium ($Na^+$), potassium ($K^+$), and chloride ($Cl^-$) across the membrane, and maintained by the selective permeability of the cell membrane and the activity of ion pumps.

When a neuron receives a signal, typically in the form of neurotransmitters released by neighboring presynaptic neurons, ion channels open, allowing ions to flow across the membrane. This can cause a small, localized change in the membrane voltage known as post-synaptic potential (PSP), which can be positive (EPSP) in excitatory synapses inducing *depolarization*, or negative (IPSP) in inhibitory synapses, in a process of *hyperpolarization*. For an excitatory input, if depolarization reaches a critical threshold (typically $\sim -55$mV), it triggers a rapid and transient change in the membrane voltage, known as *action potential* or "spike". Sodium channels open during this transient, allowing a massive influx of $Na^+$ that reverses the membrane potential (see Fig 1.1a, inset). Once a peak value is reached, potassium channels open, allowing $K^+$ ions to flow out of the neuron and restoring the negative potential of the membrane, a process known as *repolarization*. Typically, an excess $K^+$ outflow after the action potential leads to a brief hyperpolarization of the membrane, giving rise to a refractory period in which the neuron cannot fire again.

Action potentials integrated in the soma continue their journey along the axon. In the majority of neurons, a fatty substance called *myelin* envelops segments of the axon, interspersed with regions with no insulation known as *Nodes of Ranvier*. This myelination allows the action potential to "jump" from one Node of Ranvier to the next in a process known as saltatory conduction. This mechanism ensures faster transmission than would occur in non-myelinated axons, where the action potential must be regenerated at every point along the axon, consuming more time and metabolic energy. At the axon terminals, the depolarization of the axon triggers the release of neurotransmitters into the synaptic cleft, which will bind to receptors on postsynaptic neuron dendrites, possibly leading to the initiation of new spiking events.

Armed with this biological knowledge, in the next section we introduce the basic aspects regarding the mathematical and computational modeling of networks of neurons, paving the way towards a statistical-physics view of the brain.

## 1.2 Modeling the brain

If neuroscience originated in a nerve, then computational neuroscience began with an axon. In the 1950s, Alan Lloyd Hodgkin and Andrew Fielding Huxley, at Cambridge University, set out to explain the ionic mechanisms underlying the initiation and propagation of action potentials in neurons. Upon observations in voltage-clamp experiments on the giant axon of the squid —large enough to be measured with the technology available at the time—, they constructed a mathematical model that described how the flow of ions across the neuronal membrane contributes to the electrical conductivity and the capacitance of the neuron, leading to the generation of action potentials (Hodgkin et al. 1939; Schwiening 2012). At its core, the model comprised a set of nonlinear differential equations describing the kinetics of voltage-gated sodium and potassium channels, each critical for the rapid rise and fall of the action potential, respectively (Hodgkin et al. 1939).

The Hodgkin-Huxley (H&H) model would later become one the most beautiful and paradigmatic examples of theory-before-experiment breakthroughs, accurately predicting the existence of these ion channels even before they were directly observed. Moreover, this new approach illustrated for the first time how detailed biophysical processes could be captured quantita-tively and simulated with computers, giving birth to a new field of computational neuroscience. At the time of writing, the most complex single-neuron models are able to combine electrophysiological, morphological, and even transcriptomical features, predicting gene expression and electrical activity in major cortical neurons, while highlighting unique variability across and within cell-types (Nandi et al. 2022; Brette 2015b).

Now, the journey from individual neurons to neural networks represents a fundamental shift in the complexity and computational power of the nervous system. It is a this transition from the micro to the macro scale in neuroscience that higher-order cognitive functions and behaviors can emerge through the collective interactions of neurons. Given the formidable advances on modeling the dynamics of individual neurons, one could naively wonder whether a similar approach could not be taken with the whole brain, just by defining some rules that could account for the synapses and then connecting the previously modeled neurons up to a full-brain simulation

To throw some numbers into the problem, a high-resolution, real-time molecular simulation of the entire human brain would need to account for 90 billion neurons, 1,000 trillion synapses, 90 billion glial cells, 450 billion vascular end feet (supporting 450 trillion synapses), 1 trillion molecules per cell undergoing 1,000 reactions per molecule per second, at 10,000 time steps per second. All together gives us the spectacular value of $4 \times 10^{29}$ FLOPS (floating-point operations per second) for a fully realistic simulation of the brain (X. Fan et al. 2019).[2]

The above estimation, far from discouraging, should help us to realize the importance of developing mathematical formulations and simple models that, in the spirit of physics (and in particular of statistical mechanics), seek to understand the macroscopical phenomenology arising from the interaction of millions of neurons, without the need for a detailed, microscopic description of the later. Thus, the models used nowadays to simulate the activity of networks of neurons can be typically categorized into two primary classes, which differ in their level of biological realism and the types of phenomena they can capture.

On the one hand, *rate-based* models, also known as firing rate or mean-field models, focus on the average firing rate of a population of neurons. In these models, neurons (or sometimes larger, coarse-grain regions in the brain) are typically represented as single units and their activity can be described by a continuous or discrete variable, $x_i(t)$, representing the firing rate or probability of firing per unit of time in neuron $i$. Continuous rate models are described using differential equations, which take the general

---

[2]To give an intuition on the magnitude of the problem, to this day, the *Folding@home* network holds the largest combined computing power with over $2.3 \times 10^{1}8$ FLOPS, still 11 orders of magnitude below the required computational capacity.

form:

$$\frac{d\mathbf{x}(t)}{dt} = f(\mathbf{x}(t), \mathbf{u}(t)) \ , \qquad (1.1)$$

where $\mathbf{x}(t)$ is a vector containing the firing rate of all neurons in the network at time $t$, $\mathbf{u}(t)$ is the total input (possibly multivariate) current to the network, and $f$ is a transfer or activation function (typically linear, sigmoidal or ReLU). Similarly, discrete-time rate models can be described in terms of difference equations:

$$\mathbf{x}(t+1) = f(\mathbf{x}(t), \mathbf{u}(t)) \ , \qquad (1.2)$$

where now $t$ is an integer value ranging from 0 to the maximum time-step, $T$, considered in the simulation.

Notably, these types of models are very efficient from a computational perspective, so they are often used in large-scale network simulations. In particular, they have proved to be very useful for investigating the overall behavior of neural populations, as they can capture population dynamics phenomena such as average activity, oscillations, and the emergence of stable states. Moreover, rate models often yield analytical solutions, allowing for mathematical analysis and insights into network behavior (see, for instance, Cook et al. 2022 for a recent review). Nevertheless, because rate models disregard the precise timing and amplitude of spikes among individual neurons, they are not suitable for understanding phenomena such as spike-timing-dependent plasticity (STDP) or temporal encoding.

*Spike-based* models, on the other hand, aim to capture the detailed temporal dynamics of individual cells, describing how neurons generate action potentials (spikes) in response to input currents, and how these spikes influence the network. The H&H model is an example of this type. However, while groundbreaking, the H&H model has several limitations due to its complexity, involving multiple differential equations and parameters for each neuron, which makes it computationally too costly for simulations of large networks.

In that sense, integrate-and-fire (I&F) models offer a useful alternative, significantly reducing the computational load by simplifying the representation of neuronal activity. They abstract the individual neuron behavior to a single equation, integrating incoming currents until a threshold is reached, point in which an action potential is fired, and the voltage is reset. More concretely, the simple leaky integrate-and-fire (LIF) model can be described by the following equation:

$$C\frac{dV(t)}{dt} = -\frac{V(t) - V_{\text{rest}}}{R} + I(t) \ , \qquad (1.3)$$

where $C$ is the membrane capacitance, $V(t)$ is the membrane potential, $V_{\text{rest}}$ is the resting membrane potential, $R$ is the membrane resistance, and $I(t)$ is the input current. When the membrane potential $V$ reaches a threshold value, $V_{th}$, a spike is generated, and the membrane potential is reset to a value, $V_{reset}$, that is typically lower than the resting potential (hyperpolarization).

The biological plausibility of these types of models as compared to rate networks comes as a cost, not only from the computational perspective, but also in terms of complexity. They typically require a larger number of parameters and are hard to treat analytically. On the other hand, their high-temporal precision allows the study of phenomena such as temporal encoding or synchrony detection, as well as forms of neural plasticity that depend on the exact timing of the spikes in pre and postsynaptic neurons.

Summarizing, rate models are a valuable tool for exploring population-level dynamics while being amenable to analytical treatment, whereas spiking neuron models are essential for investigating precise timing, spike-based learning rules, and other mechanisms at the cellular level (see Brette 2015a for a beautiful article on the epistemological underpinnings of rate-based vs spike-based codes). I hope that, by the end of this thesis, I will have convinced the reader that a combination of these two approaches can provide us with a more comprehensive understanding of the dynamics of neural networks in the brain.

In particular, we will make use of a simple, continuous linear rate model of recurrently connected neurons in Chapter 2, where the analytical tractability of the model will prove very useful linking the observed neural dynamics in actual experimental data to the theory of critical phenomena (introduced in the last section of this chapter). Chapters 3 and 4, on the other hand, make use of a machine learning (ML) framework, known as Reservoir Computing (RC), which can be understood as a time-discrete rate model of non-linear interacting units.

The spiking model paradigm will be the protagonist of Chapter 5, where we set out to build a biologically-realistic model of the olfactory cortex using LIF neurons. This will allow us to study the role of input-induced changes in the neural encoding of odors, using plasticity rules that depend on the exact timing of the spikes.

From the most biologically-realistic model of a single neuron, to a coarse rate model describing the activity of macroscopic patches in the cerebral cortex, all approaches considered so far have something in common: they reproduce the dynamics and structure of neurons or neural networks within a synthetic model, trying to close the epistemological gap comparing the emergent phenomenology with an underlying biological reality.

We now introduce a fundamentally different paradigm that can also be used in the theoretical study of the brain. Instead of looking to reproduce some observables of biological networks of neurons, one can try to generate models that can mimic some of the expected *functions* or capabilities of the brain from a task-solving perspective (including, e.g., memory, prediction ability or classification accuracy). With this purpose in mind, let us take a step together into the realm of machine learning.

## 1.3 The advent of learning machines

From the first bird-inspired "flying machines" of Leonardo da Vinci to the latest advances in artificial photosynthesis, humankind has constantly sought to mimic nature in order to solve complex problems. It is therefore not surprising that the dawn of artificial intelligence (AI) and machine learning (ML) was also characterized by the idea of emulating the functionalities and characteristics of the human brain.

Within his book *The Organization of Behavior*, Donald Hebb proposed in 1949 a neurophysiological model of neural interactions that attempted to explain the way associative learning takes place (Hebb 1949). Hebb suggested that the simultaneous activation of cells would lead to the reinforcement of the involved synapses, a hypothesis often summarized in the today's famous statement: "neurons that fire together, wire together". Thus, Hebbian theory would be swiftly taken by neurophysiologists and early brain modelers as the foundation upon which to build the first working artificial neural networks. In 1950, Nat Rochester at the IBM research lab embarked in the project of modeling an artificial cell assembly following Hebb's rules (Milner 2003). However, he would soon be discouraged by a critical issue in Hebb's theory: as learning progresses and connections strengthen, neural activity could overwhelm the entire assembly, leading to network saturation.

A solution to this problem would not arrive until 1957, when Frank Rosenblatt, seeking to find a more "model-friendly" version of Hebb's assemblies, came up with the Perceptron, the first example of a feed-forward Neural Network (FFNN) (Rosenblatt et al. 1958). Rosenblatt would be the first one to introduce different types of units within the network, which today would correspond to what we know as input, hidden and output layers in FFNNs. Mathematically, for a given input, $\mathbf{x}$, the output of the perceptron is a single binary value $f(\mathbf{x})$ that can be computed as:

$$f(\mathbf{x}) = \begin{cases} 1 & \text{if } \mathbf{w} \cdot \mathbf{x} + \mathbf{b} > 0 \\ 0 & \text{otherwise} \end{cases} \tag{1.4}$$

where $\mathbf{w} \cdot \mathbf{x}$ is the dot product of the input, $\mathbf{x}$, with the weight vector, $\mathbf{w}$, and the bias term, $\mathbf{b}$, acts like a moving threshold. In a modern FFNN, the step function is usually substituted by a non-linearity $\varphi(\mathbf{x})$ termed the activation function. Being computationally more applicable than the original ideas of Hebb, Rosenblatt paved the way that would progressively detach ML from its biological inspiration.

Despite the initial excitement about perceptrons, in 1969 Marvin Minsky and Seymour Papert proved that these type of networks could only be trained to recognize linearly separable patterns (Minsky et al. 1969). The authors already recognized the potential of multilayer perceptrons (MLPs) to tackle non-linear classification problems, but the lack of suitable learning algorithms at the time lead to the stagnation of the field, in what is known as the first of the AI winters (Kurenkov 2015).

The interest in ML would not thaw out until 1974, with the advent of today's widely known backpropagation algorithms. Backpropagation was derived by multiple researchers in the early 60's and implemented to run on computers as early as 1970 by Seppo Linnainmaa (Linnainmaa 1970). Neverthelss, it was Paul Werbos —back then a PhD student at Harvard University— the first one to propose it as a way of effectively training MLPs (Paul Werbos 1974), although its modern formulation, as we know it today, would not arrive until a decade later in the work of Rumelhart et al. 1986. Understood as a supervised learning method in multilayer networks, backpropagation aims to adjust the internal weights in each layer to minimize the error or loss function at the output, using a gradient-descent approach based on the chain rule to propagate the error from the outer to the inner layers. Interestingly enough, although backpropagation is usually criticized for being biologically unrealistic, Werbos himself originally found inspiration in the psychological theories of Freud, as he would later recount (Paul Werbos 2006):

> In 1968, I proposed that we somehow imitate Freud's concept of a backwards flow of credit assignment, flowing back from neuron to neuron . . . I explained the reverse calculations using a combination of intuition and examples and the ordinary chain rule, though it was also exactly a translation into mathematics of things that Freud had previously proposed in his theory of psychodynamics!

It is known, however, that signals in the human brain do not cross from one layer of neurons to the next following a feed-forward architecture, nor learning takes place "backpropagating" errors from deeper to outer layers. Instead, biological neural networks are recurrently connected, which allows neurons to send feedback signals among each other. This is idea that motivated the appearance of recurrent neural networks (RNNs). From a theoretical point of view, RNNs are not only more biologically plausible, but also computationally more powerful than FFNNs. While FFNNs can approximate mathematical functions, RNNs can approximate entire dynam-ical systems —i.e., functions with an added time component (Grezes 2014).

It was John Hopfield, in 1992, the first one to come up with a successful implementation of an RNN: the Hopfield network (Hopfield 1982). Developed as a content-addressable (i.e., "associative") memory, the model consisted of a fully connected network of binary units implementing a Hebbian learning rule during training. Nevertheless, the real milestone in the field of ML would arrive with the discovery of the backpropagation-through-time (BPTT) algorithm for RNNs, derived by numerous researchers, but popularized in 1990 by Paul Werbos (P.J. Werbos 1990). In short, the idea behind the BPTT is to "unfold" the RNN in time, treating the state of the network at time $t$, $\mathbf{x}(t)$, as the input to another copy of the same network at time $t + 1$, $\mathbf{x}(t + 1)$ (Kurenkov 2015). Although the appearance of this new method seemed promising at first, RNNs were still performing worse than simpler FFNNs and at a higher computational cost, due to what is known today as the vanishing gradient problem of the BPTT algorithm (Hochreiter et al. 2001).

It wouldn't be until the beginning of the 21st century when a new paradigm in RNNs design and training appeared. In 2000, a learning algorithm aiming at overcoming the problems of the BPTT was proposed under the name of Atiya-Parlos Recurrent Learning (APRL) (Atiya et al. 2000). Among its results, it showed that the dominant changes always appeared in the weights of the output layer, while training in the weights of deeper layers converged slowly. This idea motivated two fundamentally new approaches to RNNs that would appear independently on the following years: the Echo State Network (ESN) of Herbert Jaeger (Jaeger 2001c) and the Liquid State Machine (LSM) of Wolfgang Maass (Maass et al. 2002), both constituting trailblazing models of what is known today as the Reservoir Computing (RC) paradigm (Grezes 2014). The main idea behind the new approach was simple: if only the changes in the output layer weights are significant, then the treatment of the weights of the inner network can be completely separated from the treatment of the output layer weights. We will cover all the intricacies of this RC framework in Chapter 3, when studying the effect of plasticity mechanisms in their dynamics.

We have seen so far two different "simulation-based" approaches — computational neuroscience and machine learning (ML)— that can be used to tackle the complexity of the brain. In this last section of the first chapter we provide a third approach, leaning into the theoretical side, that will allow us to investigate the dynamics of biological networks using formal mathematical descriptions based on the theory of statistical mechanics and the behavior of systems near critical points.

## 1.4 The theory of critical phenomena

The origins of the theory of critical phenomena can be traced back to the early observations of critical behavior in fluids. One of the earliest recorded studies was by Thomas Andrews in 1869, who described the critical point of carbon dioxide, beyond which gas and liquid phases become indistinguishable. This was followed by the work of Johannes Diderik van der Waals, who introduced an equation of state in 1873 that could predict the behavior of fluids near the critical point (Honig et al. 2016).

In the early 20th century, the development of quantum mechanics provided a deeper understanding of the microscopic interactions that give rise to critical behavior. Paul Ehrenfest introduced the concept of phase transitions and classified them according to the discontinuity in thermodynamic derivatives, such as volume or entropy. Historically, phase transitions were divided by Ehrenfest into first-, second-, and higher-order transitions. Thus, a transition is said to be of first order if the first derivative of the free energy with respect to some thermodynamics variable is discontinuous at the critical point. Likewise, in a second order transition the first derivative is continuous, but the second and successive derivatives show a singularity at the critical point (Honig et al. 2016; Binney et al. 1992).

Nevertheless, this division of phase transitions according to their order is not universal. For instance, in the case of spin-glasses, the second derivative of the free energy with respect to an external magnetic field is singular, but it is continuous with respect to the temperature. Moreover, there are also examples of topological phase transitions, like the Berezinskii-Kosterlitz-Thouless (BKT), which are effectively of infinite order (Honig et al. 2016). Therefore, in the reminder of this thesis we will favor Landau's convention, which categorizes phase transitions into *continuous* and *discontinuous* ones, where the later can be considered of first order if all first derivatives of the function of state are discontinuous at the same critical point (Landau et al. 2013).

The post-World War II era saw significant advancements with the introduction of the renormalization group theory by Leo P. Kadanoff (Kadanoff 1966) and Kenneth Wilson (K. Wilson et al. 1974; Kenneth G. Wilson 1975). As we will see in the next section, this theory showed that systems with vastly different microscopic details could exhibit similar macroscopic behavior near the critical point, setting the mathematical scaffold to understand the emergence of scale-invariance and universality near critical points.

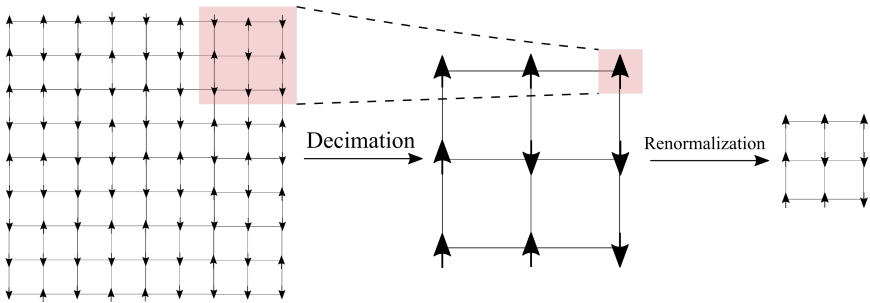

Figure 1.2: **Kadanoff's renormalization group approach.** At each step of the RG, neighboring spins are clustered together and replaced by an effective spin in a process of coarse-graining or decimation. Then all lengths are re-scaled with the new lattice spacing, such that the new lattice is identical to the original with a reduced number of spins by coarse-graining.

### 1.4.1 From *micro* to *macro*: the Renormalization Group

The fundamental concept of the Renormalization Group (RG) is rooted in the works of physicists such as Ernst Stueckelberg and André Petermann in the late 1940s (Stueckelberg et al. 1953), and Murray Gell-Mann and Francis Low in the early 1950s (Gell-Mann et al. 1954), who sought to understand the behavior of quantum fields at different energy scales. The term "renormalization" originally referred to the process of removing infinities arising in quantum electrodynamics (QED) calculations, but it was Kenneth G. Wilson who, in the early 1970s, realized the broader implications of these ideas in the context of critical phenomena and phase transitions in statistical physics (K. Wilson et al. 1974; Kenneth G. Wilson 1975).

In this section we will provide a brief, pedagogical introduction into the ideas of the real space or "block-spin" Renormalization Group, as originally introduced by Leo P. Kadanoff in 1966 (Kadanoff 1966).

Let us consider a two-dimensional microscopic system of spins, described by state variables $\{s_i\}$ taking values $\pm 1$ at a certain temperature $T$, and assume that each spin can only interact with its nearest neighbours. The physics of this system will be then characterized by a certain Hamiltonian, $H(T, J)$, where $J$ is known as the coupling strength, describing the interactions between neighboring spins. Because we saw in the previous chapter how macroscopic collective phases can emerge from the interactions of individual elements at the microscopic level, we will be now interested in understanding how the properties of our original system, as described by the original Hamiltonian $H(T, J)$, change when we move the *observation scale* towards more coarse-grained descriptions of the system.

The idea of Kadanoff was to block the original variables into "clusters" of neighboring spins, replacing them by a single effective spin, in a process termed coarse-graining or *decimation*. This transformation scales the length of the system by a factor $b$, and the new effective Hamiltonian, $\tilde{H}$, for the system can be written in terms of a renormalized coupling $\tilde{J}$. The changes in the coupling constant under renormalization are implemented by a certain *beta function* $\tilde{J} = \beta(J)$, which is said to induce a *renormalization group flow* on the space of couplings.[3] Thus, the above procedure can be iterated a number of times, alternating steps of "coarse-graining" and renormalization, effectively evolving the system towards a more macroscopical description at each step.

In more general terms, the beta function can be understood as a mathematical tool that describes how the interactions of the system change with the observation scale. Mathematically, if $g$ is a coupling constant and $\mu$ represents the scale (which is usually energy or length), then the beta function is given by:

$$\beta(g) = \mu \frac{dg}{d\mu} \ . \tag{1.5}$$

If the beta function is positive, the coupling gets stronger when moving towards larger scales. In the other hand, if the beta function is negative the coupling weakens as we move towards more coarse-grained descriptions of the system. The fixed points, $g^*$, of the RG transformation are the *critical points* of the dynamics, characterized by:

$$\beta(g^*) = 0 \ . \tag{1.6}$$

At these points, the system is said to be *scale-invariant*, and the coupling constants no longer change as we coarse-grain our system. The critical exponents, which describe how different macroscopic observables diverge near the critical point, can be derived from the behavior of the beta function near its fixed points. For instance, it can be shown that the correlation length, $\xi$, that is the scale over which spins are correlated, diverges with a critical exponent $\nu$ as the system approaches the critical temperature $T_c$:

$$\xi \sim |T - T_c|^{-\nu} \ . \tag{1.7}$$

Although we will not go into further details for the purpose of this short introduction, one of the main strengths of the RG theory lies in its ability to show that systems with different microscopic details (like different lattice structures or spin configurations) can show the same critical behavior, characterized by *universal* critical exponents. For an in-depth treatment of the RG theory and the concepts of scaling, relevant and irrelevant operators and universality classes, we refer the reader to the classic books of Binney et al. 1992 and Chaikin et al. 2000, as well as Honig et al. 2016 for a more recent (and gentle) introduction to the theory of critical phenomena and RG.

---

[3]More generally, given a certain coarse-grain transformation $\{s_i\} \longleftrightarrow \{\tilde{s}_i\}$, a theory is said to be *renormalizable* if we can rewrite the partition function $Z$ for the new coarse-grained system only in terms of the new variables $\{\tilde{s}_i\}$ for a certain transformation in the coupling parameters $\{J_k\} \longleftrightarrow \{\tilde{J}_k\}$.

## 1.4.2 The critical hypothesis in neuroscience

By now, I hope I could convince the reader that. when systems are composed of numerous microscopic elements, they can exhibit a variety of macroscopic collective behaviors or phases, whose properties are intrinsically different from those of their individual components. By the end of the 20th century, this phenomenon, famously encapsulated in the title of the quintessential article by P. W. Anderson: *More is different* (Anderson 1972), soon led to the speculation that different forms of biological states may similarly be considered as collective phases, with structural and dynamical reconfigurations between these states akin to phase transitions (Anderson 1972; Hopfield 1994; Pollack et al. 2008; Solé 2011).

Indeed, such transitions have been shown to be ubiquitous across many different biological systems (Muñoz 2018), from the synchronization phase transitions characteristic of biochemical rhythms (Garcia-Ojalvo et al. 2004)) or fireflies flashing (J. Buck et al. 1968), to melting phase transition in DNA strands (Y. C. Li et al. 2006), or percolation transitions in collagen networks (Alvarado et al. 2013).

Now the question we will be naturally interested in —given the topic of this thesis— is the following: can we find some similar phenomenology, resembling the concepts of phases and phase transitions in statistical physics, which takes place inside the brain?

When neuroscientists poke into the electrophysiological processes of the brain — from microscopic individual neurons to large-scale whole-brain measurements— they systematically detect, even under conditions of quiet rest, an ongoing background of noisy and variable neural activity (Softky et al. 1993; Arieli et al. 1996; Raichle 2011; Deco et al. 2008; Deco et al. 2012). The fact that this ceaseless activity is energetically so costly had researchers long wondering what crucial functionalities it may entail, giving rise to a diversity of theoretical explanations.

One prominent hypothesis, which will be at the core of most of the results presented in this thesis, proposes that neuronal networks operate close to a *critical point*, i.e., at the edge between two different types of collective behavior (Plenz et al. 2014; Chialvo 2010; Tagliazucchi et al. 2012; Haimovici et al. 2013; Cocchi et al. 2017; Wilting et al. 2019; Hidalgo et al. 2014; Z. Ma et al. 2019; Ponce-Alvarez et al. 2018; Martinello et al. 2017;

R. Wang et al. 2019; O'Byrne et al. 2022). We know, from the theory of statistical physics, that near critical points we can find long-range correlations, scale-invariant spatio-temporal patterns, maximal sensitivity to perturbations, enhanced dynamical range, etc. (Binney et al. 1992; Muñoz 2018). Therefore, it was conjectured that the brain, by being near criticality, could extract crucial advantages from such a plethora of spontaneously-generated collective properties (Shew et al. 2013; Chialvo 2010).

From the theoretical side, a lot of effort has been devoted to trying to discern what type of criticality is the most pertinent to describe brain activity (Mora et al. 2011; Muñoz 2018). Some of the different possible scenarios that have been explored include:

1. Avalanche criticality: the first empirical evidence supporting the critical hypothesis raised from the observation of *neural avalanches*, which are irregular outbursts of neural activity interspersed by quiescent periods, detected both *in vitro* (Sanchez-Vives et al. 2000; Eytan et al. 2006; Segev et al. 2001) and in *in vivo* (Meister et al. 1991; Steriade et al. 1993) experiments. Remarkably, in their seminal paper 20 years ago, Beggs and Plenz showed the existence of scale-invariant statistics in both the size and duration these avalanches, with power-law exponents that matched those of a mean-field critical branching processes (Beggs et al. 2003; Petermann et al. 2009; Mazzoni et al. 2007). Thus, this type of criticality is associated to a state in which the instability occurs at a population level, with non-zero average correlations between neurons and a single dominant slow time-scale (Beggs et al. 2003; Chialvo 2010; Cocchi et al. 2017; Moretti et al. 2013; Dahmen et al. 2019; Corral López et al. 2022).

2. The edge of synchronization: the previously introduced resting-state of noisy and variable neural activity, can be understood as an *asynchronous state*, in which excitatory and inhibitory inputs to any given neuron typically cancel with each other, so that the mean input is below the activation threshold. Within this picture, one can consider a neural network composed of excitatory and inhibitory units, so that its overall state can be shifted (by varying, e.g., the intensity of the synaptic strengths) from the previously discussed asynchronous phase to a synchronous one, where collective oscillations emerge (Liang et al. 2020). It has been argued that the brain could operate at the edge of a synchronization phase transition, jointly exploiting the advantages of both asynchronous noisy states (e.g., for information processing) and synchronous ones (e.g., for coherent information transmission through oscillations) (Yang et al. 2012; Villegas et al. 2014; Poil et al. 2012; Liang et al. 2020; J. Li et al. 2020; Santo et al. 2018; Buendía et al.

2021).

3. The edge of chaos: it has been hypothesized that the brain could be operating close to an order-to-chaos phase transition (Steyn-Ross et al. 2010; Magnasco et al. 2009; Solovey et al. 2015), a regime that has been shown to maximize information transmission, memory capacity and input representations, among other properties (Crutchfield et al. 1988; Langton 1990; Melanie 1993; Boedecker et al. 2011; Morales et al. 2021b). Experimental evidence supporting this hypothesis was found in high temporal-resolution electrocorticography data from human brains. In L. Alonso et al. 2014, the authors showed how the maximum Lyapunov exponent of the dynamics (which, when positive, characterizes the presence of chaos, as we will see in Chapter 3), fluctuated around the threshold of instability in awake subjects, indicating that brain dynamics is finely tuned to the edge of a phase transition; but became more negative (i.e., stable) in anesthetized subjects. This finding suggests that the edge of stability could be associated with a functional brain, whereas deviations from this edge are indicative of a loss of consciousness or changes in functional dynamics.

In more generic terms, one can say that criticality emerges whenever the baseline phase loses its stability and shifts continuously to another phase. Thus, when talking about general features of criticality and to avoid leaning towards a particular hypothesis regarding the type of underlying phase transition, we will more generally refer to it as the *edge of instability* (Morales et al. 2023).

Recently, with the advent of modern techniques enabling simultaneous recordings of thousands of neurons, complementary experimental evidence in favor of the critical hypothesis have emerged from unexpected angles (Tkačik et al. 2015; Fagerholm et al. 2021; Meshulam et al. 2019; Dahmen et al. 2019; Morales et al. 2023). In the following chapter, we will combine theoretical aspects of linear-response theory and the renormalization group to further study the validity and implications of this attractive hypothesis.

# The critical brain

We adore chaos because we love to
produce order.

M. C. Escher, 1958

[...] vi la circulación de mi oscura
sangre, vi el engranaje del amor y la
modificación de la muerte, vi el Aleph,
desde todos los puntos, vi en el Aleph la
tierra, y en la tierra otra vez el Aleph y
en el Aleph la tierra.

J. L. Borges, *El Aleph*

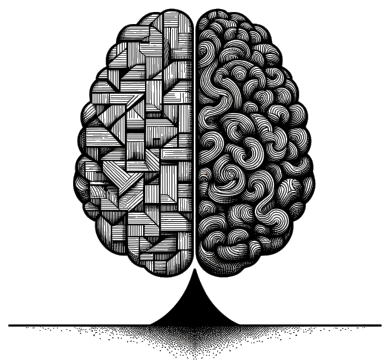

## 2.1 Introduction: a new light into the criticality hypothesis.

Built upon the foundations of statistical mechanics, the so-called "criticality hypothesis" has revealed itself as an elegant solution in the search for a mathematical characterization of the brain dynamical regime (see Plenz et al. 2021; Cocchi et al. 2017; Massobrio et al. 2015; Wilting et al. 2019; O'Byrne et al. 2022; Muñoz 2018 for some recent reviews). Nevertheless, despite its conceptual appeal and thrilling implications, the validity of this hypothesis as an overarching principle of dynamical brain organization remains a controversial issue (Touboul et al. 2017; Beggs et al. 2012; Cubero et al. 2019).

From the empirical side, evidence of putative brain criticality often relies on the observation and characterization of neuronal avalanches and their power-law statistics in size and duration (Beggs et al. 2003; Petermann et al. 2009; Plenz et al. 2021; Z. Ma et al. 2019; Liang et al. 2020). However, it has been suggested (among other possibilities), that the observed power laws: (i) could be a side-effect of experimental artifacts (Bédard et al. 2006; Bédard et al. 2009); (ii) might be better fit by exponential distributions (Touboul et al. 2010; Touboul et al. 2017; Dehghani et al. 2012); (iii) may not be the fingerprint of any critical process, but arise from a combination of exponential distributions (Reed et al. 2003), or from an underlying multiplicative noise (Sornette 1998).

Over the past decade, other approaches have emerged that compare empirical properties of neural systems with those expected on a system poised near a critical point. This includes, for instance, the study of long-range spatio-temporal correlations (Palva et al. 2013; Poil et al. 2012); the search for statistical critical-like patterns of activity (Tkačik et al. 2015; Mora et al. 2011); or the analysis of whole-brain critical models fitted to match empirically observed correlations (Cabral et al. 2017; Tagliazucchi et al. 2012; R. Wang et al. 2019). Despite this extensive effort, novel theoretical frameworks and more-stringent experimental tests are still much needed to either prove or disprove this conjecture.

As we saw in Chapter 1, systems at or near criticality exhibit scale invariance, meaning that their properties remain unchanged under rescaling, so that their behavior is characterized by universal features independent of the specific details or size of the system (Muñoz 2018). The presence of scale invariance is often captured by powerlaw —or scale-free— distributions of diverse quantities (Sornette 2006), and can be understood because of the divergence of the system's correlation length, which correlates the system's variables over larger and larger scales as we approach the critical point. Renormalization Group (RG) was also introduced in Chapter 1 as a way of capturing these universal features that persist across different scales as we zoom out from a micro- to a macroscopic description of the system (Kenneth G Wilson 1983; Binney et al. 1992; Efrati et al. 2014). In the original real-space RG, as proposed by Kadanoff for spin-systems (Kadanoff 1966), one has to "coarse-grain" neighboring spins into blocks, which are then clustered into new blocks at successive steps of the RG transformation, thus constructing effective descriptions of the microscopic system at progressively larger spatial scales.

The problem one encounters when trying to adapt this methodology to neural data is the absence of a lattice geometry or, more generally, the lack of an embedding metric space in which the notions of *scale* and *distance* are well-defined, so that we can understand what it means to coarse-grain the activity of "neighboring" neurons. Extensive efforts have been devoted to work around this problem, specially in trying to develop algorithms that could generalize the notion of RG to complex networks (Kim 2004; Gfeller et al. 2007; Serrano et al. 2008; C. Song et al. 2005; García-Pérez et al. 2018; Villegas et al. 2023). Direct application of the former RG schemes to the realm of neural networks remains elusive nonetheless. At larger —already macroscopic— scales, a hierarchical coarse-graining of the anatomical regions of the brain unveiled the existence of self-similarity across human connectomes using a geometric RG that embedded the network structure into an underlying hidden metric space (Zheng et al. 2020; García-Pérez et al. 2018). On a more theoretical side, but still within the realm of macroscopic field equations, a renormalized theory for a simplified version of the stochastic Wilson-Cowan model was recently derived, allowing to uncover the structure of non-linear interactions across scales (Tiberi et al. 2022).

Delving deep into the microscopic level, it is unfeasible however to try to map the millions of individual synapses that make up the connectivity patterns between neurons. How could we then construct effective descriptions of a system at progressively larger scales, when we lack any information about the underlying structure among its components? An out-of-the-box solution to this problem was proposed in Meshulam et al. 2019, in what they termed a Phenomenological Renormalization Group (PRG) approach. In the following section we present the main ideas behind the PRG approach, which will later serve us to look for scale-invariant properties of neural data.

## 2.2 Renormalization Group for neural data.

To understand the context and results of the PRG devised by Meshulam *et al.*, let us consider a set of $N$ neurons, such that the empirically-measured activity of the $i$-th neuron is discretized on a set of $T$ non-overlapping time bins of width $\Delta t$, with the average firing rate over the time bin $t_l$ represented as $x_i(t_l)$, with $l \in [1,T]$. The coarse-graining scheme then proceeds as follows: instead of grouping neighboring neurons using a criterion of spatial vicinity, we can cluster them on the basis of maximal pairwise Pearson's correlation:

$$C_{ij} = \frac{\langle \delta x_i \delta x_j \rangle}{\sqrt{\langle (\delta x_i)^2 \rangle \langle (\delta x_j)^2 \rangle}}, \tag{2.1}$$

where averages, $\langle \cdot \rangle$, are computed across the available discrete time steps and $\delta x_i = x_i - \langle x_i \rangle$. More specifically, at the $k$-th coarse-graining step, one would select the two most correlated variables, $i$ and $j_{*i}$, and combine their activities into a new coarse-grained, "block-neuron" variable given by:

$$x_i^{(k)} = z_i^{(k)} \left( x_i^{(k-1)} + x_{j_{*i}}^{(k-1)} \right), \tag{2.2}$$

where $k = \{1, ..., n_{steps}\}$ indexes the RG step —and one takes $x_i^{(0)}$ to be the original timeseries of the $i$-th neuron as extracted from the data— whereas the normalization factor, $z_i^{(k)}$, is chosen such that the average non-zero activity of the new variables $x_i^{(k)}$ is equal to one. The procedure is then iterated by considering the second most correlated pair of neurons, and so on, until a set of $N_k = N/2^k$ block-neurons is obtained, where each block-neuron is a cluster containing the summed activity of $K = 2^k$ original units (see Fig. 2.1 for a schematic representation).

Once the coarse-graining procedure has been completed, the goal is to study how the statistical properties of the system change as one moves progressively across scales. Following Meshulam et al. 2019, here we focus on four quantities that have a clear interpretation in the classical theory of critical phenomena (Binney et al. 1992):

1. The variance of the non-normalized coarse-grained variables:

$$M_2(K) = \frac{1}{N_k} \sum_{i=1}^{N_k} \left[ \left\langle \left( \tilde{x}_i^{(k)} \right)^2 \right\rangle - \left\langle \left( \tilde{x}_i^{(k)} \right) \right\rangle^2 \right], \tag{2.3}$$

where $\tilde{x}_i^{(k)}$ is the summed activity of the original variables inside cluster $i$ and $N_k$ is the number of clusters at step $k$ of the PRG. Notice that, for totally independent random variables, one would expect the variance to grow linearly in $K$ (i.e., $M_2(K) \propto K$), while for perfectly

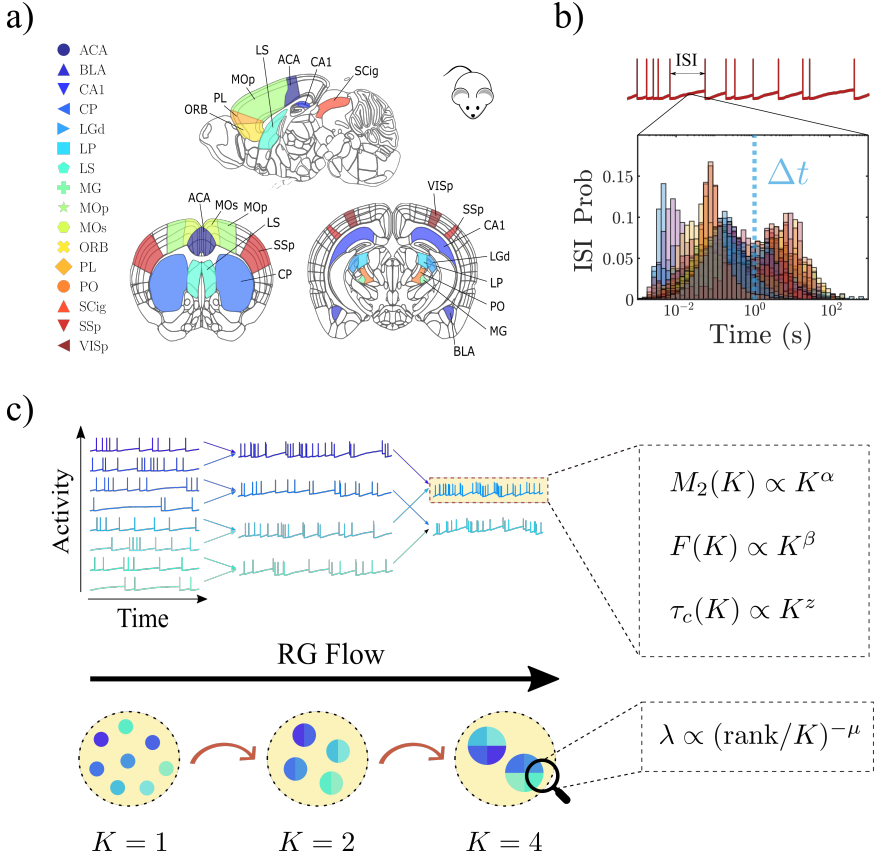

Figure 2.1: **Workflow of the phenomenological renormalization group analysis**. **(a)** The raw data is filtered for resting-state activity in regions with at least $N = 128$ recorded neurons. **(b)** For each region, activity is discretized in bins of width $\Delta t$ equal to the geometric mean of the inter-spike interval (ISI) distribution. **(c)** The original system is progressively coarse-grained by clustering together pairs of most-correlated neurons, so that at step $k$ each new variable contains the summed activity of $K = 2^k$ original neurons. By changing the scale $K$, different properties are analyzed as we move towards more coarse-grained descriptions of the system.

correlated variables $M_2(K) \propto K^2$. Non-trivial scaling is therefore characterized by:

$$M_2(K) \propto K^\alpha, \qquad (2.4)$$

with a certain intermediate value of the exponent $1 < \alpha < 2$.

2. The "free-energy" for the coarse-grained variables:

$$F(K) = -\log(S_K), \qquad (2.5)$$

where, $S_K$ is the probability that a given coarse-grained neuron is silent at any time step. As more and more of the original neurons,

$\sigma_i$, are grouped into cluster variables $x_i^{(k)}$, one would expect that the probability of having "silent" block-neurons (i.e., the probability that all neurons inside a cluster are silent) decreases rapidly with the size $K$ of the clusters, leading to:

$$F(K) \propto K^\beta, \tag{2.6}$$

where, in particular, one expects an exponential decay (i.e., $\beta = 1$) for initially independent variables, while $\beta \neq 1$ reflects non-trivial scaling.

3. The autocorrelation function of the coarse-grained variables:

$$C^{(k)}(t) = \frac{1}{N_k} \sum_{i=1}^{N_k} \frac{\langle x_i^{(k)}(t_0) x_i^{(k)}(t_0+t) \rangle - \langle x_i^{(k)} \rangle^2}{\langle (x_i^{(k)})^2 \rangle - \langle x_i^{(k)} \rangle^2} \tag{2.7}$$

which, at the steady state, is independent of $t_0$. Given that, commonly, fluctuations at larger spatial scales relax with a slower characteristic time scale, one should expect the autocorrelation function to decay more slowly as one averages over more neurons. Assuming that correlations decay exponentially in time with a characteristic time scale, $\tau_c^{(k)}$, at each coarse-graining step (i.e., $C^{(k)}(t) \sim e^{-t/\tau_c^{(k)}}$), dynamical scaling implies that the average correlation function collapses into a single curve when time is re-scaled so that:

$$C^{(k)}(t) \equiv C(t/\tau_c^{(k)}), \tag{2.8}$$

and that such a characteristic time obeys and scaling law with the cluster size:

$$\tau_c(K) \propto K^z, \tag{2.9}$$

where $z$ is the dynamical scaling exponent.

4. The covariance matrix for all the neurons inside clusters at step $k$ of the coarse-graining:

$$C_{i,j}^{(k)} = \frac{1}{N_k} \sum_{l=1}^{N_k} \langle \delta\sigma_i^{(l)} \delta\sigma_j^{(l)} \rangle, \tag{2.10}$$

where $\delta\sigma_i^{(l)} = \sigma_i^{(l)} - \langle \sigma_i^{(l)} \rangle$ is the activity of neuron $i$ belonging to cluster $l$, and the sum is over all $N_k$ clusters of size $K = 2^k$. As argued in Meshulam et al. 2019, if correlations are self-similar across scales, then we should see this by looking inside the clusters, as they are analogous to spatially contiguous regions in a system with local interactions. In particular, at a fixed point of the renormalization group flow, the eigenvalues of the covariance matrix must obey a powerlaw dependence on the fractional rank:

$$\lambda \propto \left(\frac{\text{rank}}{K}\right)^{-\mu}. \tag{2.11}$$

At the light of the above scaling relations, in what follows we will employ the introduced exponents $\alpha$, $\beta$, $z$, and $\mu$ to characterize the potential scale-invariant properties of neural activity in the mouse brain.

## 2.3 Quasi-universality across brain regions

By applying the previously introduced PRG to the activity of over one thousand neurons in the CA1 region of the mouse hippocampus, Meshulam *et al.* observed the emergence of non-trivial scaling of the variance, "free-energy" and autocorrelation times; as well as spatial scaling reflected in the powerlaw dependence with the fractional rank for the eigenspectrum of the covariance inside clusters (Meshulam et al. 2019). All these results, which together evidenced the scale-invariant nature of the underlying neural dynamics, gave suddenly a new life to the theory of criticality in real neural networks.

To assess the robustness or possible universality of these findings, we decided to use the PRG framework to analyze the activity of different regions in the mouse brain. In particular, most of the forthcoming analyses rely on the empirical electrophysiological data presented in Steinmetz et al. 2019, where the activities of thousands of individuals neurons in the mouse brain were simultaneously recorded at a high temporal resolution (200Hz). During the experiments, mice were trained to perform a perceptual task that required vision, choice, action and behavioral engagement, but activity was also recorded during periods of resting-state or spontaneous dynamics. Thus, in order to conduct a comprehensive analysis, we initially segregated the original time series into "resting-state" and "task-induced" activity, focusing on the former for the purpose of our study (nevertheless, a comparison between the observed scaling exponents for the two conditions can be found in Appendix B.IV.§). In all cases, we restricted our analyses to areas with at least $N = 128$ simultaneously recorded neurons, as reported in Fig. 2.1.

Before delving any further into the results, let us briefly mention one important aspect that needs to be (and it is rather often not) carefully considered when dealing with correlation measures over time series. Typically, studies involving neural recordings of activity, both at the single-neuron level and at lower spatial resolution scales (EEG, fMRI, etc.), involve the discretization of the time series into bins of a certain width $\Delta t$, so that the activity $x_i(t_k)$ of neuron (or macroscopic region) $i$ at bin $k$ can be defined as $x_i(t_k) = \int_{t_k}^{t_k+\Delta t} x_i(t)dt$. Regardless of whether the bin size is chosen to match the time resolution of the recording technique, a relevant timescale of the input task, or it is just arbitrarily selected, the fact is that correlations and correlation-based properties of the system may depend drastically on the choice of $\Delta t$ (Neto et al. 2022; Cessac et al. 2017). This is especially true for single-neuron activity, given that neurons within the same population can operate at broadly different time-scales, with rate frequencies ranging from milliseconds to tens of seconds (see Appendix B.II and Fig. B.1 therein).

To cope with such a heterogeneity, in the forthcoming analyses the relevant timescale for each brain region (i.e., our choice of $\Delta t$) is defined as the *geometric mean* of the inter-spike interval (ISI) distribution across all neurons (see Fig. B.1). As a sanity check, we observed that this choice of $\Delta t$ reproduces to great extent the scaling exponents found in Meshulam et al. 2019 for the mouse hippocampus[1] when using Steinmetz *et al.*'s recordings for this very same area. Moreover, to ensure the consistency of the results, we additionally confirmed that the documented exponent values show only small variations when the time-discretization bin is changed, with the exception of $\mu$, which displays an increase during longer time-scales beyond the population activity's typical inter-spike-interval (see Appendix B.IV.† and Fig. B.6 therein).

Fig. 2.2 summarizes the analysis presented in Morales et al. 2023 for the emergence of scale-invariant properties across RG steps. For the scaling of the variance, free-energy and autocorrelation time, the goodness of each powerlaw fit was estimated by means of its R-squared value and comparing it with an equivalent exponential fit (see Appendix B.IV.∗, Fig. B.5). For the probability density of eigenvalues —which is directly related to the observed scaling in the rank-ordered plot (see Appendix A.II)—, we computed the log-likelihood ratios (Clauset et al. 2009) between the estimated powerlaw and alternative exponential and lognormal distributions (see Appendix B.IV.∗, and Tables B.1 and B.2). Each exponent for a particular region can then be expressed as $e = \bar{e} + \text{MAE} + \sigma$, where $\bar{e}$ is the average across different experiments (measured over different mice), MAE is the mean-absolute-error, computed as the average across experiments of the individual experimental errors, which are in turn measured over split-quarters of data, and $\sigma$ is the standard deviation across experiments (see Tables B.1 and B.2 for values of all measured exponents and their errors in every region).

---

[1] In the original work by Meshulam *et al.* spike trains were constructed using a bin width of $\Delta t \approx 33$ms, matching the sampling rate of 30Hz for the recording technique.

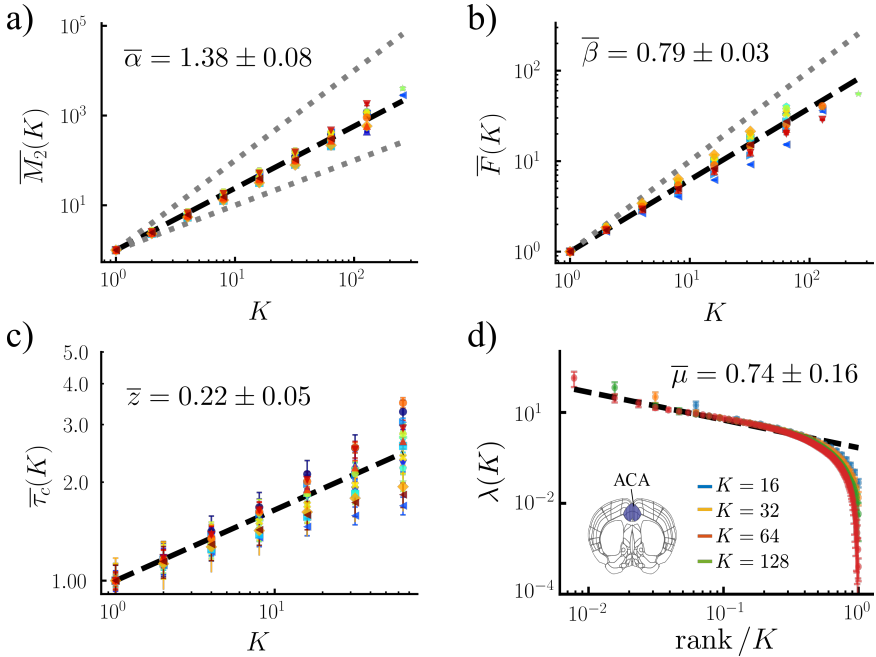

Figure 2.2: **Phenomenological RG analyses unveils the existence of quasi-universal scaling over 16 different regions of the mouse brain**. **(a)** Variance of the non-normalized activity as a function of the cluster size, $K$ (dotted lines with slopes 2 and 1, mark the fully correlated and independent limit cases, respectively). **(b)** Scaling of the free energy, as defined in Eq. (2.5) (dotted line corresponds to the independent-variables limit case). **(c)** Scaling of the characteristic autocorrelation time for coarse-grained variables. **(d)** Scaling of the average covariance matrix over clusters of size $K = 16$ (blue), 32 (yellow), 64 (green) and 128 (red) units, in one of the 16 considered brain regions. The rank-ordered eigenvalues decay as a power law of the fractional rank (rank/$K$) and, even more remarkably, the curves cut-offs at different values of the clusters size, $K$, collapse into the same curve. To facilitate the comparison between regions with different number of neurons, the first three measures were normalized by the value of each quantity at $K = 1$ (e.g., $\overline{M_2}(K) = M_2(K)/M_2(K = 1)$. Errorbars are computed as the standard deviation across split-quarters of data, with lengths typically smaller than the marker size.

An almost perfect scaling for the variance, $M_2(K)$, of the non-normalized activity of block-neurons is observed (Fig. 2.2a), with an average exponent $\overline{\alpha} = 1.38 \pm 0.08$ across all regions. In particular, we measure $\alpha_{CA1} = 1.37 \pm 0.03 \pm 0.02$ for the CA1 region in the mouse hippocampus, which is within errorbars of the value $\alpha = 1.56 \pm 0.07 \pm 0.16$ reported in Meshulam et al. 2018 for this same area. Notice how the exponent values for all the regions are always between the expected ones for uncorrelated ($\alpha = 1$) and fully-correlated ($\alpha = 2$) variables, revealing consistently the existence of non-trivial scale-invariant correlations.

Meanwhile, the "free-energy" as defined in Eq. (2.5) also exhibits a clear scaling with the cluster size (Fig. 2.2b), with an average exponent $\overline{\beta} = 0.79 \pm 0.03$ across regions ($\beta_{CA1} = 0.78 \pm 0.04 \pm 0.05$ for the hippocampus, to be compared with the value $\beta = 0.87 \pm 0.014 \pm 0.015$ reported in Meshulam et al. 2018).

Finally, temporal and spatial scaling are manifested in the curves shown in Fig. 2.2c-d, respectively. The former is reflected in the observed dynamical scaling for coarse-grained neuron's autocorrelation times $\tau_c(K)$, with an average exponent $\overline{z} = 0.22 \pm 0.05$ across regions (also $z_{CA1} = 0.18 \pm 0.03 \pm 0.01$ for the CA1 area, in perfect agreement with the one reported in Meshulam et al. 2018 for this region ($z = 0.22 \pm 0.08 \pm 0.10$), but also compatible with the exponent values reported in Fagerholm et al. 2021 using a different approach). As an additional test for dynamical scaling, we show in Appendix B.III.∗ (Fig. B.2) how the curves for the auto-correlation functions at different coarse-graining levels collapse when time is appropriately re-scaled.

On the other hand, spatial scaling manifests itself in the collapse of the covariance eingenspectra at different levels of coarse-graining when plotted against the fractional rank (rank$/K$). For clarity, results are only plotted for one of the regions (ACA), but we observed a clear powerlaw scaling of the eigenvalues with the rank in all the analyzed brain areas (see Fig. B.4 in Appendix B.III.†), with an average exponent across regions $\overline{\mu} = 0.84 \pm 0.14$. Likewise, the value reported in Meshulam et al. 2019 for CA1 ($\mu = 0.76 \pm 0.05 \pm 0.06$) is in excellent agreement with our measured value for the same region, $\mu_{CA1} = 0.78 \pm 0.08 \pm 0.02$. Since the spectrum of covariance eigenvalues is known to be strongly sensitive to the samples-to-neurons ratio, $a_0 = T/N$ (with $N$ the number of recorded neurons and $T$ the number of samples, see Kong et al. 2017), we further verified in Appendix B.IV.‡ that our choice of timebin width yielded a sufficient ratio, $a_0$, as to ensure the robustness of the exponents estimates (see Fig. B.7).

In addition to the above results, similar signatures of scale invariance to those observed for resting-state activity emerge in PRG analyses of neural activity recorded while mice are performing a task (see Fig. B.8 in Appendix B.IV.§). In particular, we showed how the dispersion and mean values of the scaling exponents across regions are not significantly altered ($p > 0.1$ on a two-sample $t$-test for each exponent) when comparing the resting-state and task-induced activity. Nevertheless, such a comparison must be taken with a grain of salt, as for non-resting-state activity we are far-off from the equilibrium steady-state assumptions that underlie any Renormalization Group analysis.

As a control test, we further verified in Appendix B.IV.¶ that the non-trivial scaling features revealed by the PRG analyses are lost for all areas when the correlation structure of the data is broken either by: (i) reshuffling the times of individual spikes in the time series (Fig. B.9b); (ii) shifting each individual time series by a random time span while keeping the sequence of spikes (Fig. B.9c); or (iii) shuffling spikes across neurons (Fig. B.9c).

At the light of the above results, we can confidently state that strong signatures of scale invariance with quasi-universal exponents are indeed observed across brain regions.

Is it however right to assume that such scale-invariant properties naturally follow from an underlying critical dynamics? The answer to this question is not straightforward as, for instance, an alternative explanation to the phenomenology observed in Meshulam et al. 2019 was recently derived in terms of latent fields in a model with no critical point in its dynamics (Morrell et al. 2021). In the next section we will tackle this problem not only by looking for signatures of criticality, but actually inferring from the data how far the underlying dynamics lies from a critical point.

## 2.4   How far from the critical point?

Given an empirical system showing some degree of scale invariance, as in the previously discussed experimental recordings of neural activity, can we quantify how close it is to the edge of a phase transition or critical point?

We begin the following section presenting the simplest neuronal network model that exhibits a phase transition between two different regimes. Working upon analytical results derived from this simple linear model, we will then introduce two different methods that allow us to infer the distance to criticality from neural recordings of activity, such as the ones analyzed above for the dataset in Steinmetz et al. 2019.

### 2.4.1   The linear rate model

Let us consider a very simple linear-rate model (LRM) in which the time evolution of $N$ linearly interacting units is described by a continuous variable $x_i(t)$ representing its activity or "firing rate" at instant $t$ (Sompolinsky et al. 1988; Hu et al. 2020). Neurons in this model interact with each other through a synaptic connectivity matrix, $J$, and their dynamics is described by the following continuous-time differential equation:

$$\tau \dot{x}_i(t) = -x_i(t) + \sum_{j=1}^{N} J_{ij} x_j(t) + \xi_i(t) \ , \qquad (2.12)$$

where, for each neuron, $\xi_i(t)$ is a external input modeled for simplicity as a zero-mean white noise with $\langle \xi_i(t)\xi_j(t') \rangle = \delta_{ij}\delta(t - t')$; and $\tau$ is the characteristic time scale for the firing rates changes. The synaptic connectivity matrix is assumed to be random, with elements $J_{ij}$ sampled from a Gaussian distribution with zero mean and variance $\sigma^2 = g^2/N$ (i.e., $J \sim \mathcal{N}(0, g^2/N)$).

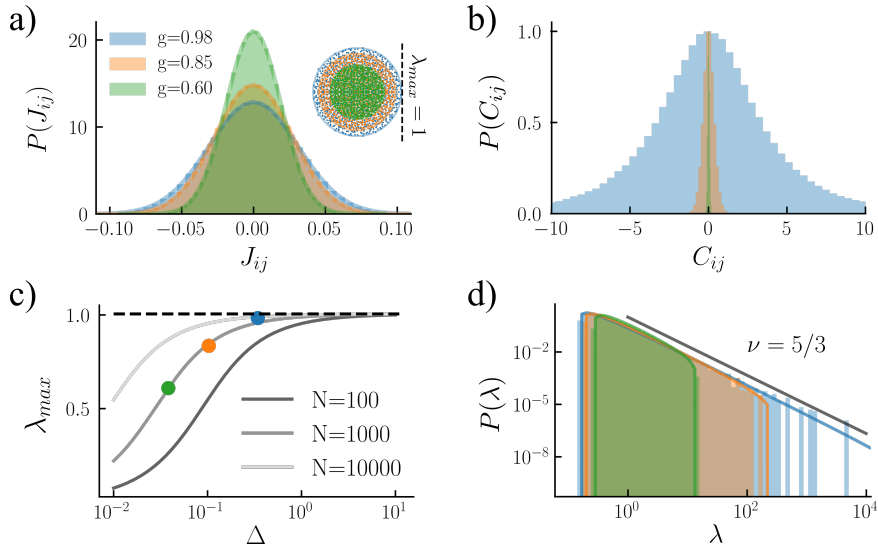

Figure 2.3: **Statistics of pairwise covariances provide two different measures of distance to the edge of instability.** For each panel, we compare simulated results for the linear-rate model (LRM) using 3 different values of $g$. **(a)** Entries of the connectivity matrix $J_{ij}$ in the LRM follow a Gaussian distribution of zero mean and variance $g^2/N$. The spectrum of the matrix is contained in a circle of radius $g$ in the complex plane (inset), with $\lambda_{max} = 1$ marking the edge of instability (black dashed line). **(b)** Entries of the sampled spike-count covariance matrix $C_{ij}$ for $N = 10^3$. Histograms are re-scaled to their maximum value for visualization purposes. **(c)** Maximum eigenvalue of the connectivity matrix $J$ as a function of the relative width of the distribution for the covariance matrix entries. Full lines are drawn using Eq. (2.16) for different values of $N$, whereas circles are obtained through simulations of the LRM for the 3 considered values of $g$ (black dashed line marks the edge of instability). **(d)** Distribution of the covariance matrix eigenvalues in the LRM. Full lines are plotted using the analytical expression in Eq. (2.17), whereas histograms are obtained from simulations. As $g \to 1$ the tail of the distribution approaches a power law with exponent $\nu = 5/3$ (black continuous line).

A well-known result from random-matrix theory, the circular law, asserts that for $N \to \infty$ a random matrix with entries sampled from a Gaussian distribution $\mathcal{N}(0, g^2/N)$ has a spectrum of eigenvalues uniformly distributed over a disc of radius $\lambda_{max} = g$ in the complex plane (Ginibre 1965; M. Mehta 1967; Girko 2005; Tao et al. 2010) (see Fig. 2.3a). Thus, denoting by $I$ the identity matrix, it is straightforward to see that the steady-state solution for the system described by $N$ coupled equations following Eq. (2.12) is stable provided all the eigenvalues of the matrix $(-I + J)$ are negative; this is, when $\lambda_{max} = g < 1$. In other words, the distance to the edge of instability in this simple linear rate model can be fully characterized by the maximum eigenvalue of the interaction matrix.

Let us note that the connectivity pattern described by a random matrix $J$ is far from being realistic in a biological sense. In particular, real neural networks are expected to obey Dale's law (Shu et al. 2003; Xue et al. 2014; Liang et al. 2020), meaning that each presynaptic neuron form either only excitatory or inhibitory connections with all connected postsynaptic neurons (i.e., all elements along a column of the connectivity matrix should have the same sign). Nevertheless, although the spectrum of such matrices does not obey the circular law (Rajan et al. 2006), the results presented below can be easily generalized to connectivity matrices where excitation and inhibition are properly considered (Dahmen et al. 2019; Hu et al. 2020).

## 2.4.2 Using empirical correlations: a measure from the first and second moments

In principle, one could possibly infer a connectivity matrix from experimental recordings of activity by constructing maximum entropy models that aim to match the observed mean and variance of the empirical correlations (see, e.g., Cocco et al. 2022 and references therein). However, it has been shown that activity-based estimates of connectivity tend to be biased towards inferring links between unconnected but highly correlated neurons (Das et al. 2020). Moreover, this approach involves an estimation of the $N^2$ entries of the connectivity matrix while, as we just saw, the dynamical stability within the simple LRM approximation would be completely determined with only information about the largest eigenvalue of the connectivity matrix.

Remarkably, we will see that an estimation of $\lambda_{max}$ can be obtained from a measure of the pairwise covariance of the activity across neurons (Dahmen et al. 2019). Notice that, compared to the synaptic efficacies that would constitute the weights of a connectivity matrix, correlations can be measured experimentally in a much simpler and robust way, since large-scale simultaneous recordings of neurons have become more and more attainable in recent years, thanks to the development of high-performance neural probes (Steinmetz et al. 2021; Demas et al. 2021; Zong et al. 2022).

To find a relation between neural activity correlations and structure, one first need to compute the so-called spike-count covariance matrix (sometimes also called noise covariance or long-time-window covariance). The elements of this matrix are pairwise covariances of the time-integrated activity across many samples or trials, measuring the degree to which trial-to-trial fluctuations from the average response are shared by a pair of neurons (M. R. Cohen et al. 2011), i.e.:

$$C_{ij} = \lim_{\Delta t \to \infty} \frac{1}{\Delta t} \langle \Delta s_i(t) \Delta s_j(t) \rangle \, , \qquad (2.13)$$

where averages are taken over samples and $\Delta s_i(t)$ is computed as:

$$\Delta s_i(t) = \int_t^{t+\Delta t} x_i(t') - \langle x_i(t') \rangle dt' \ .$$  (2.14)

Notably, this covariance matrix is linked in a rather model-independent way to the underlying effective connectivity matrix $J$ [2] through the following relationship (Dahmen et al. 2019; Hu et al. 2020; Pernice et al. 2011; Trousdale et al. 2012):

$$C = (I - J)^{-1}(I - J^T)^{-1} \ ,$$  (2.15)

under the main assumption of uncorrelated inputs. A simple proof of this equation for a more general case in which the white noise input has non-unitary variance is presented for the sake of completeness in Appendix A.I.

Despite the seeming simplicity of the above expression, one cannot naively calculate $J$ by inverting Eq. (2.15) for an empirically measured $C$ matrix, because the recorded neurons represent only a small subsample of the whole local network. To circumvent this difficulty, Dahmen *et al.* resorted to ideas and tools from the physics of disordered systems, such as spin glasses, to propose an estimator, $\hat{g}_d$, for the maximum eigenvalue $\lambda_{max}$ of the effective connectivity matrix based on the relative dispersion of the elements in $C$ (Dahmen et al. 2019):

$$\hat{g}_d = \sqrt{1 - \sqrt{\frac{1}{1 + N\Delta^2}}} \ ,$$  (2.16)

where $\Delta = \delta c / \bar{c}$, being $\delta c$ the dispersion of the entries of the spike-count covariance matrix, calculated as the standard deviation of the out-of-diagonal terms of $C$; and $\bar{c}$ is the mean variance (i.e., the mean of the diagonal terms).

Thus, this rather elegant result provides us with an overall measure of the network stability which depends only on the relative dispersion of pairwise covariance values and the system size. Moreover, the authors showed that this result is insensitive to the details of the underlying dynamics and connectivity, so that Eq. (2.16) still holds (as a linear-response approximation) when considering spiking-neurons dynamics or excitatory-inhibitory connectivities (see Dahmen et al. 2019 and Ocker et al. 2017 for more details). As we will see in the following section, an alternative estimation of $\lambda_{max}$ can be obtained using the full eigenspectrum of the empirical covariance matrix.

---

[2] By *effective connectivity matrix* we will refer to the product of the anatomical connectivity, weighted by synaptic efficacies, and the intrinsic excitability of individual neurons. Thus, $J_{ij}$ can be understood as the change in firing probability of a postsynaptic neuron $i$ due to a single spike of a presynaptic neuron $j$ (Dahmen et al. 2019).

### 2.4.3 Using empirical correlations: a measure from the spectral distribution

In a mathematical tour-de-force, Hu and Sompolinsky took the theory of linear rate models one step further, deriving an analytical form for the full probability density of the long-time-window covariance eigenvalues in the systems described by Eq. (2.12) (Hu et al. 2020):

$$p_{lrm}(\lambda) = \frac{3^{\frac{1}{6}}}{2\pi g^2 \lambda^2} \left[ \sum_{\xi=1,-1} \xi \left( \lambda(1 + \frac{g^2}{2}) - \frac{1}{9} \right. \right.$$
$$\left. \left. + \xi \sqrt{\frac{(1-g^2)^3 \lambda(\lambda_+ - \lambda)(\lambda - \lambda_-)}{3}} \right)^{\frac{1}{3}} \right] , \quad (2.17)$$

for $\lambda_- \leq \lambda \leq \lambda_+$, with:

$$\lambda_{\pm} = \frac{2 + 5g^2 - \frac{g^4}{4} \pm \frac{1}{4}g(8 + g^2)^{\frac{3}{2}}}{2(1-g^2)^3} , \quad (2.18)$$

while $p_{lrm}(\lambda) = 0$ for values of $\lambda$ out of the support ($\lambda > \lambda_+$ and $\lambda < \lambda_-$). Observe, in particular, how the upper limit of the support, $\lambda_+$, diverges in the limit of $g \to 1$ (i.e., close to the edge-of-instability), when the distribution develops a long (power law) tail of large eigenvalues:

$$\lim_{g \to 1} p_{lrm}(\lambda) \sim \frac{\sqrt{3}}{2\pi} \lambda^{-\frac{5}{3}} . \quad (2.19)$$

While the above analytical result was derived in the limit of $N \to \infty$, Fig. 2.3 shows that Eq. (2.17) match accurately the spectrum obtained numerically in simulations of the linear-rate model with $N = 10^3$ units. Thus, in practice, provided one has simultaneously recorded enough neurons (on the order of hundreds) and has enough samples as to meaningfully compute their covariance matrix (Kong et al. 2017), it is possible to infer the dynamical regime of the recorded population simply by fitting the sampled covariance matrix eigenvalue spectrum to the theoretical distribution given by Eq. (2.17), using $g$ as the fitting parameter. More concretely, we will be earching for the value $\hat{g}_s$ that minimizes the $L^2$-norm using the Cramer-von Mises statistics between the empirical cumulative distribution, $F_n(\lambda)$, and the theoretical one $F_{lrm}(\lambda) = \int_{-\infty}^{\lambda} p_{lrm}(\lambda) d\lambda$ (Hu et al. 2020):

$$D_{CvM}^2 = \int (F_{lrm}(\lambda) - F_n(\lambda))^2 dF_n(\lambda) \quad (2.20)$$

$$= \frac{1}{12n^2} + \frac{1}{n} \sum_{i=1}^{n} \left( F(\lambda_i) - \frac{2i-1}{2n} \right) , \quad (2.21)$$

where $n$ is the total number of samples and $\lambda_i$ are the eigenvalues of the empirical long-time-window covariance matrix. This will constitute our second method to estimate the distance to the edge-of-instability from data.

We remark that, although the above expression for the density of eigenvalues was derived assuming a linear-rate model of recurrently connected neurons, Hu and Sompolinsky showed that it provides also an excellent fit to the numerical spectrum of a network of nonlinear rate neurons driven by external noise, as well as for networks of generalized leaky integrate-and-fire (LIF) neurons. Let us caution, in any case, that nonlinearities may introduce a bias in the inferred value of $g$ (see Hu et al. 2020 for more details).

### 2.4.4 All regions lie close to criticality

In this section, we will be applying the two different methods of estimating the maximum eigenvalue of the effective connectivity matrix, as discussed above, to the previously introduced dataset of Steinmetz *et al.* (Steinmetz et al. 2019). To avoid confusion, we will denote by $\hat{g}_d$ and $\hat{g}_s$ the empirical estimates of the distance to criticality obtained using the actual data from the relative dispersion of covariances (first method, as in Dahmen et al. 2019) and their eigenvalue spectrum (second method, following Hu et al. 2020), respectively.

First of all, since both methods rely on the computation of the long-time window covariance matrix (Eq. 2.13), one first needs to split the original time series of spiking activity for each neuron into $T$ samples of width $\Delta t$. In practice, the condition $\Delta t \to \infty$ can be approximated by choosing a time window large enough for autocorrelations to decay (Fig. 2.4.a for one particular region). For this purpose, a characteristic time, $\tau$, for each region was obtained from an exponential fit to the average autocorrelation decay time window. On the basis of these results (see Table B.1), a time bin $\Delta t = 1s$ was chosen so that $\Delta t > \tau$ while maximizing the number of samples available.

The spike-count covariance, $C$, can then be computed for the activity in each region after integrating the rates of individual neurons within the previously defined time window (Fig. 2.4b shows the distribution of entries for this matrix in one of the regions). We remark that, for this spike-count or noise correlations to be meaningful, neural activity across samples must be stationary, a condition that typically holds during recordings of resting-state type of activity. Nevertheless, an augmented Dickey–Fuller test (Dickey et al. 1979) was performed on each neuron time series to test this assumption, showing that only a small percentage of all neurons within a region present non-stationary spiking statistics (see Table B.1).

A first estimate of the distance to criticality in each region was ob-

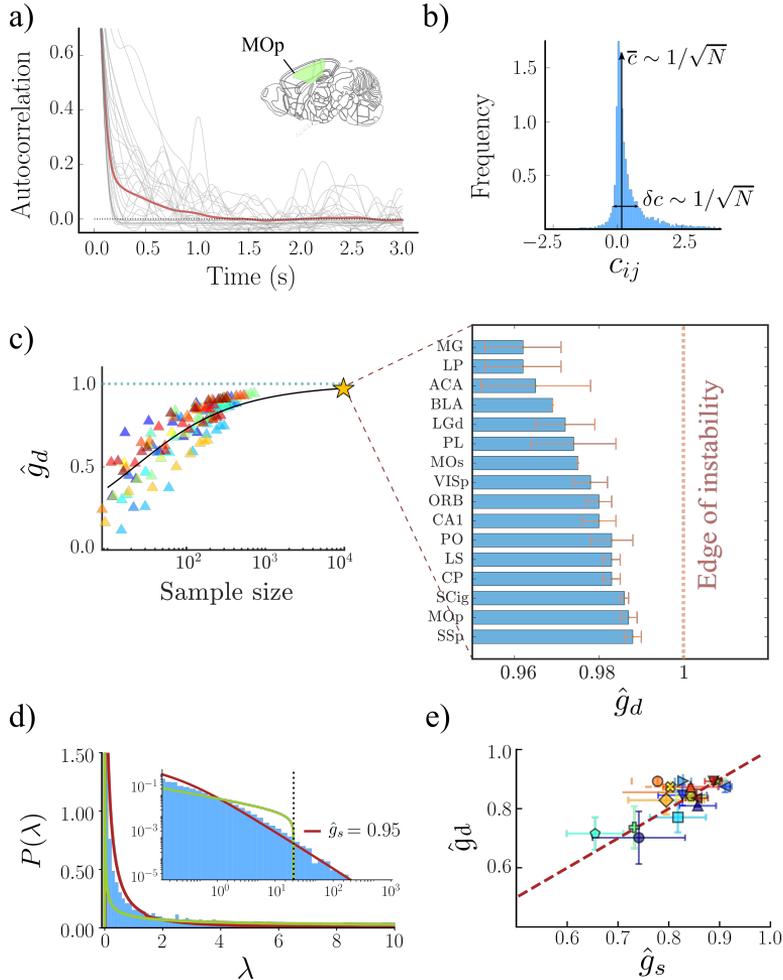

Figure 2.4: **Distance-to-criticality analysis locates the activity in 16 regions of the brain near the edge-of-instability.** **(a)** Decay of autocorrelations in the MOp region (grey lines: 30 randomly chosen neurons; red line: average over neurons). **(b)** Distribution of pairwise spike-count covariances in the MOp region. **(c)** For each region (color coded), we computed $\hat{g}_d(N)$ at different system sizes by subsampling the original data. Points were then fitted to Eq. (2.16) to obtain an empirical estimate $\hat{g}_d$ at a common number of neurons $N = 10^4$. Only an average over such fitted curves (black line) is shown for visualization purposes. **(d)** Covariance eigenvalues distribution for the MOp region, together with the best-fitting MP distribution (green line) and the best-fitting eigenvalue distribution for the LRM (red line, $\hat{g}_s = 0.95$). Inset: close-up in log-log scale. **(e)** Estimated values of $\hat{g}_d$ (dispersion method, Dahmen et al. 2019) and $\hat{g}_s$ (spectral method, Hu et al. 2020) for each of the 16 regions using the original number of neurons in each area. In all cases, errors are computed as the standard deviation over different recordings of the same region.

tained using Eq. (2.16) following the method in Dahmen et al. 2019 (see Fig. 2.4c). Notice, however, that our maximum eigenvalue estimator, $\hat{g}_d$, strongly depends on the number of neurons being recorded and, since the available empirical data heavily subsamples each region, direct application Eq. (2.16) would actually underestimate the real value of $\lambda_{max}$. To avoid this problem, but also to make a sensible comparison between regions with different number of recorded neurons, we took random subsamples of the recorded neurons to generate a curve, $\hat{g}_d(N)$, of the estimated maximal eigenvalue as a function of the network size. We then extrapolated the expected value of the estimator for a common (biologically realistic) number of neurons $N = 10^4$ in each region. Fig. 2.4c shows the distance to the critical point thus estimated, with errorbars computed as the standard deviation across experiments in each region. Notice that most values lie on a very narrow window between $\hat{g}_d \approx 0.96$ and $\hat{g}_d = 0.99$, with a mean value $\hat{g}_d = 0.978 \pm 0.009$, very close to the edge of instability

Alternatively, we saw that one can infer $g$ from the best-fitting parameter, $\hat{g}_s$, minimizing the distance between the empirical and theoretical eigenvalue spectra of the covariance matrix (Hu et al. 2020). To illustrate this, in Fig. 2.4d (left) we plotted the empirical distribution of covariance eigenvalues for an example region, together with the best fitting distribution, $p_{lrm}(\lambda)$, which was obtained for $\hat{g}_s = 0.95$, very close to the critical value. Moreover, we report for comparison the best-fitting Marchenko-Pastur eigenvalue distribution (i.e., the one expected for uncorrelated random variables), which clearly fails to capture the long tail of the empirical eigenvalue distribution.

To conclude, Fig. 2.4d (right) presents a comparison of the two alternative estimates of the distance to criticality, $\hat{g}_d$ and $\hat{g}_s$, for all the 16 mouse brain regions analyzed when the original number of neurons in each experiment is considered, revealing that —within errorbars computed across different recordings of the same region— they are typically in excellent agreement. Values for both estimates in each region can be found in Table B.1.

Thus, the above results let us conclude that all the analyzed regions are, to greater or lesser extent, close to the edge of instability. Could one then infer as well that the observed scale invariance has its roots on this underlying, close-to-critical dynamics? Remember that the presented estimates of the distance to the edge of instability are based on theoretical results over a LRM. Could this simple model, if poised near its critical point, show any scale invariance at all?

While it has been shown that similar scale-invariant properties can emerge in non-equilibrium systems like the Contact Process (Nicoletti et al. 2020), we do not have any evidence that a simple LRM can give rise to any kind of scaling behavior under renormalization. Investigating this potential link using the already introduced PRG framework will be the motivation of our next section.

## 2.5 Closing the loop: from critical models to scale invariance

In this section we will be applying the PRG analysis proposed in Meshulam et al. 2019 to the LRM described by Eq. (2.12), for which we can control the distance to the edge of instability through the coupling parameter $g$. *A priori*, one would expect to find non-trivial scaling for values of $g$ close to the critical coupling (i.e., $g \simeq 1$), but not when $g \ll 1$. However, there is no simple reason as to why the scaling exponents should be in principle similar to the empirically measured ones. We also remark that, since we are dealing with a stochastic rate model with a Gaussian external noise, neurons are never really "silent" and so there is no well-defined free energy, $F(K)$.

The following simulations were ran for networks of size $N = 1024$, choosing an integration step $h = 0.01$ and setting the white-noise variance, $\sigma^2$, and characteristic time scale, $\tau$, to one. The resulting time series were then binned with a $\Delta t = 0.1$ sampling window, before performing a PRG analysis at different values of the control parameter $g$. Results for these analyses are illustrated in Fig. 2.5; in particular they reveal that:

- The variance of the coarse-grained activity (Fig. 2.5a) scales with cluster size with a trivial exponent $\alpha \simeq 1$ for $g \ll 1$, as expected for independent units. However, for $g = 0.95$, —i.e., sufficiently close to the edge of instability— it shows a non-trivial exponent $\alpha = 1.38$, in surprisingly good agreement with the average exponent measured for the spiking neurons ($\bar{\alpha} = 1.38 \pm 0.08$, Fig. 2.2b).

- The characteristic time for the autocorrelation functions (Fig. 2.5b) of the coarse-grained variables shows a disruption of the scaling for $g = 0.2$ and $g = 0.6$ at large cluster sizes, and a relatively small powerlaw exponent $z$ even when only the first few steps of PRG are considered. Conversely, for $g = 0.95$ the scaling holds for the all the considered cluster sizes and the measured dynamical scaling exponent is very close to average experimental value ($\bar{z} = 0.22 \pm 0.05$, Fig. 2.2c).

- The rank-ordered eigenvalue spectrum (Fig. 2.5c) of the block-neurons activity covariance matrix has a cut-off that drastically depends on the cluster size for small values of the coupling parameter. On the contrary, for $g = 0.95$ we observe a —not perfect but much better— collapse of the curves at different levels of coarse-graining and a power-law trend with an exponent close to the value measured across actual brain regions ($\bar{\mu} = 0.84 \pm 0.14$ across all regions, Fig. 2.2d shows the collapse for one example region).

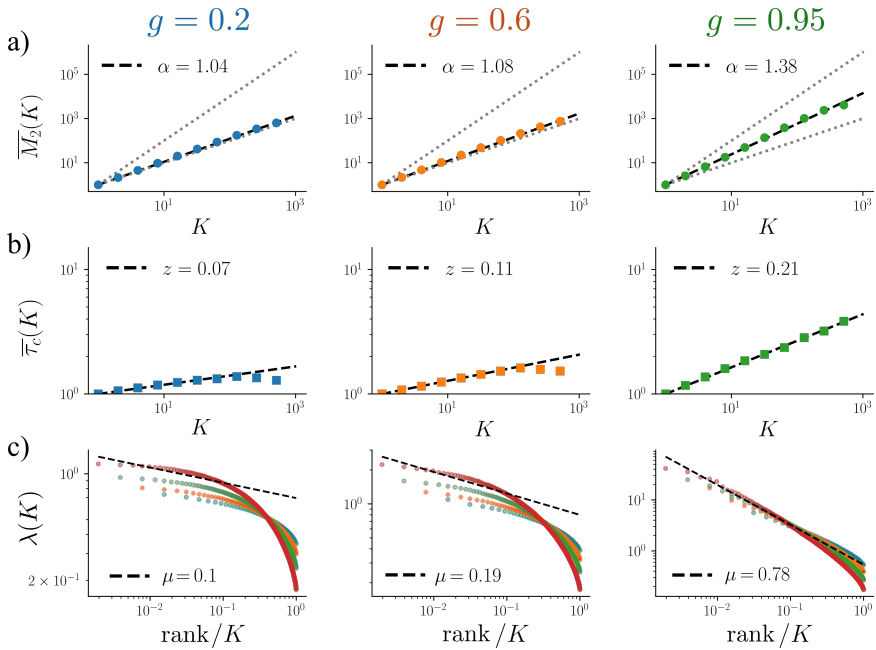

Figure 2.5: **A phenomenological RG analysis over a simple linear rate model shows signatures of scale-invariance when the system is close to the edge-of-instability.** In each column, we consider a different value of the overall coupling strength $g$ of the model. After a PRG analysis we then show **(a)** the variance of the non-normalized coarse-grained activity; **(b)** the scaling of the characteristic autocorrelation time for coarse-grained variables; and **(c)** the covariance matrix spectrum for the activity in clusters of size $K$. Only the close-to-critical case ($g = 0.95$) do the simulations show scaling behavior and power-law exponents similar to the ones observed in real data. Parameter values for the LRM are taken as $N = 1024$, $\sigma = 1$ and $\tau = 1$, for the number of units, variance of the external white noise and membrane time-scale, respectively.
.

The above results suggest that a simple LRM is indeed capable of capturing a great part of the arising scale-invariant properties observed in biological neural networks under a PRG flow. Interestingly, the fact that the observed scaling in the covariance eigenvalue spectra shows "better" power-law scaling for the empirical data (Fig. 2.2d and Fig. B.4 in Appendix B.III.†) as compared to the model (Fig. 2.5c), probably hints at some missing ingredient in our simple model to fully capture the observed spatial scale invariance. It will be the scope of future work to analyze the effects of including further features in the model such as non-linear activation functions, heavy-tailed distribution of weights or heterogeneous characteristic times scales for the neurons.

In the next section, however, instead of adding modifications to the original LRM and try to link the observed behavior with an existing closet-to-critical phenomenology in the brain, we will be taking a more exciting path: we will look at the effects over the dynamical regime of the neurons, as well as the possible emergence of scale-invariant properties when we modify directly the connectivity structure of *real* neural networks.

## 2.6 Linking dynamics and structure

In the previous section we went over a minimal dynamical model that showed scale-invariant properties only when poised close to a phase transition. Despite the success of this simple linear model in explaining most of the phenomenology arising from a PRG analysis, little has been discussed on the structural aspects of the neural network that allows for such behavior. This question is particularly important, as it has been extensively shown how the connectivity structure among neurons can affect the dynamical regime of neural networks (Millán et al. 2018; Marinazzo et al. 2014; Yamamoto et al. 2018; Montalà-Flaquer et al. 2022). Could, for instance, homogeneous networks of randomly-connected real neurons show the same scale-invariant properties, as a simple LRM would suggest?

To address this question, we resorted to experiments on topographically designed neuronal networks carried out by the group of J. Soriano, at the University of Barcelona, consisting of *in vitro* neural cultures grown over a printed circuit board (see Fig. 2.6 and Appendix I.† for a brief description of the experimental methods). Neural cultures show spontaneous activation (Orlandi et al. 2013) and can even generate various forms of collective spatiotemporal patterns, depending upon network connectivity features and the excitation-inhibition balance (Tibau et al. 2013; Okujeni et al. 2017; Yamamoto et al. 2018; Montalà-Flaquer et al. 2022). Recently, it has been shown that not only the dynamical properties of the cultures can be modified (for instance, by blocking inhibitory synapses, thus altering the excitation-inhibition balance in the network), but also the *structural* patterns that these continuously-evolving networks generate can be conditioned by the topographical design of the surface on which they grow (Montalà-Flaquer et al. 2022; Yamamoto et al. 2023).

For all the experiments presented below, neurons were grown on a poly-dimethylsiloxane (PDMS) substrate (Fig. 2.6a) composed of valleys and slits that, in each case, were distributed following one out of four possible topographical designs: (i) a completely flat surface in which neurons grew isotropically, thus serving as a control case; (ii) evenly-spaced, linear, parallel tracks; (iii) randomly positioned squares; and (iv) a fourth-order Sierpinsky square fractal pattern (see Fig. 2.6b).

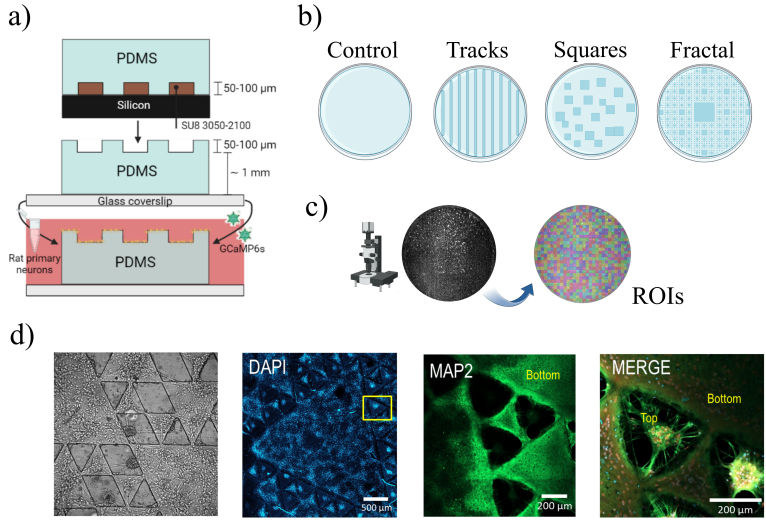

Figure 2.6: **Experimental setup.** **(a)** A silicon wafer with a topographic relief of 50 or 100 $\mu$m height was used as mould for pouring and curing PDMS. The resulting design corresponds to a reverse-relief replication of the original mould. **(b)** Topographical designs considered for the substrate, including a flat, control case; linear parallel tracks; randomly placed squares and a fourth-order Sierpinsky square fractal pattern. **(c)** Electron microscope images are parcelled into 732 region of interests (ROIs) in which we perform all the analysis. **(d)** From left to right: bright-field image of the culture with connectivity directed by a topographical pattern; three immunohistochemical images at different levels of increasing magnification. Figure partly adapted, upon private communication, from the still unpublished master thesis work of M. Olives Verger at J. Soriano's lab.

In each experiment, 732 regions of interest (ROI) were defined within the neural culture (Fig. 2.6c) and activity in each ROI, as measured by average fluorescence intensity, was recorded over a 15-minute period (see Appendix I.† for more details on the experimental protocol) . These traces of fluorescence activity were then translated into spike trains using the Schmitt trigger method, which considers a sharp change in fluorescence as a "spike" in a ROI whenever the fluorescence signal passes a high threshold and then stays elevated above a second lower threshold for at least 100ms. Once we have access to the spike times of each neuron in the culture for the length the recorded period, we constructed discretized trains of spikes using a timebin $\Delta t = 10$ms. In this way, all the information for a given experiment on one the four topographical designs is collected in a matrix $X \in \mathbb{R}^{N \times T}$, where $N$ is the number of ROIs and $T$ the number of time bins contained in the 15 minutes of recording.

To get some intuition on the emergent spontaneous dynamics of these neural cultures for the different topographical surfaces considered, we show in Fig. 2.7a raster plots for the spiking activity of individual ROIs to-gether with the autocorrelation functions of the mean population activity (Fig. 2.7b) and the ISI distributions considering all ROIs (Fig. 2.7c). For the isotropic case, neural cultures show large outburst of synchronous activity followed by long quiescent periods, leading to fast-decaying autocorrelations and average ISIs in the range of tens of seconds. Once heterogeneity in the connections is induced, we observe a decrease in the rate of decay of the autocorrelation functions, as well as changes in the shape of the ISI distribution towards smaller average values. Interestingly, the most complex dynamics in terms of temporal patterns of spiking activity seems to emerge for neurons growing on a fractal substrate, concomitant with a broader distribution of ISIs.

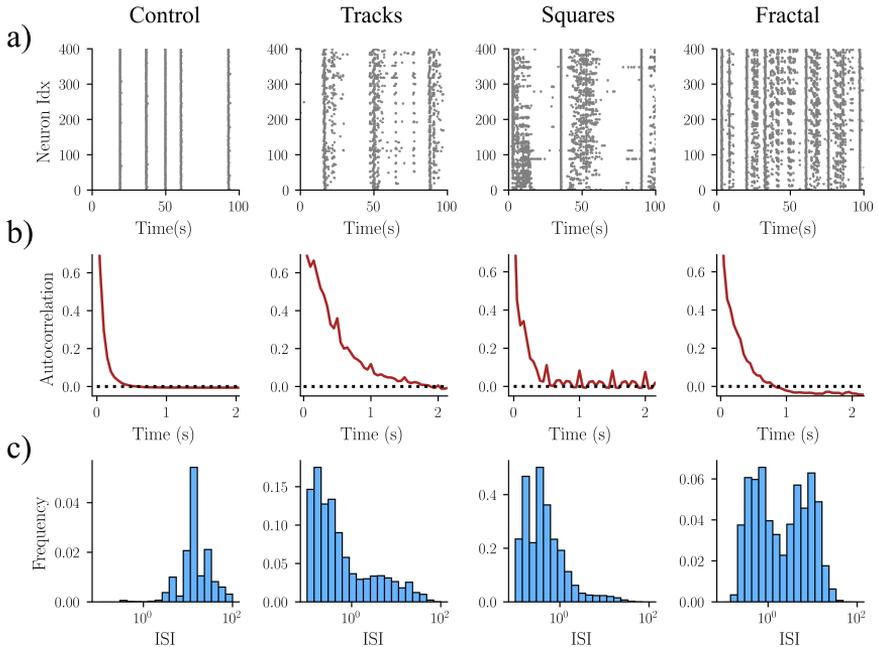

Figure 2.7: **Heterogeneous connections leads to heterogeneous firing.** For each of the four topographical substrates considered (in columns), we show: **(a)** 100ms of a raster plot for all recorded ROIs; **(b)** autocorrelation functions for the average population activity; **(c)** ISIs distributions considering all recorded ROIs. To obtain the above results, spike trains were constructed averaging the activity of each ROI across bins of width $\Delta t = 10$ms.

We can finally apply the same PRG analysis presented in section 2.2 to the matrix $X$ containing the recorded activity for all neurons, looking for signatures of scale invariance on each culture characterized by a given underlying topographical substrate. Fig. 2.8 shows results of these analysis across 8 coarse-graining steps for the variance, $M_2(K)$ (Eq. 2.4), free-energy, $F(K)$ (Eq. 2.5), and characteristic autocorrelation time, $\tau_c(K)$ (Eq. 2.9), of the coarse-grained variables against the size, $K$, of the clusters; together with the rank-ordered spectrum, $\lambda_n$ (Eq. 2.11), of the activity covariance matrix inside the clusters. Notice how, for the isotropic (control) case, we observe a trivial scaling for the variance, close to the fully correlated case, and no scale-invariance for the temporal and spatial correlations. Once we add anisotropy into the neural substrates, the heterogeneity of connections can give rise to exponents and scaling relations closer to the ones we observed for different regions of the mouse brain in Section 2.3. However, notice how only in the case in which neurons are grown over a fractal surface we find the autocorrelation time of the coarse-grained variables, $\tau_c$, to grow as a power law with the size $K$ of the clusters.

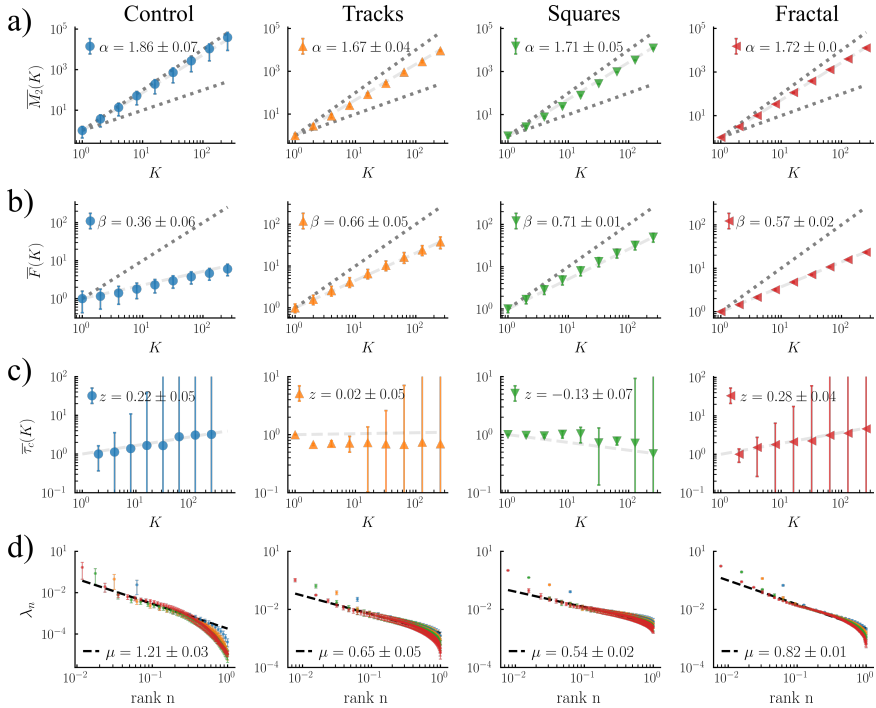

Figure 2.8: **Non-trivial scale-invariance arises in neural cultures with a high degree of network heterogeneity.** Results are shown for four different topographies of the circuit boards (in columns). At each step $k$ of the PRG, in which the original ROIs are grouped into clusters of size $K = 2^k$, we plot: **(a)** the normalized variance, **(b)** the normalized free-energy, and **(c)** the normalized autocorrelation decay times of the coarse-grained variables. In **(d)** we show the rank-ordered spectrum of the covariance matrix at different steps of the coarse-graining, with clusters of size $K = 16$ (blue); $K = 32$ (yellow); $K = 64$ (green) and $K = 128$ (red).

The above results shed some light into the role of the structure on the emergence of scale-invariant properties of the dynamics for real neural networks. Although a careful study of the actual connectivity patterns induced by the underlying topographical surface is still much needed, we hypothesize that neurons growing over a fractal surface not only develop heterogeneous connections among each other, but also an increased hierarchical-modular architecture for the overall network, a key ingredient that has been observed in biological, *in vivo* neural networks (Meunier et al. 2010; Sporns 2010; Diez et al. 2015). Interestingly, it has been shown that extended critical-like regions, known as Griffiths phases, could emerge from such an existent structural heterogeneity (Moretti et al. 2013). Unfortunately, given the limited amount of samples in the recordings —on the order of $N/T \sim 1$ for any measure of long-time-window correlations— it is not possible to provide a reliable estimate of the distance to criticality as we did in Section 2.4.4 using the available experimental data.

So far in this chapter, we were able to exploit the potential of large-scale single-cell recording techniques in both, *in vivo* and *in vitro* biological neural networks, to unveil some of the fascinating phenomenology that, we hypothesize, steams from an underlying critical dynamics. Unfortunately, this type of microscopic measures —when performed *in vivo*— require invasive techniques (see Appendix B.I) and are therefore typically unviable for human subjects.[3] When studying the human brain, neuroimaging techniques such as EEG, fMRI or MEG can only provide spatially coarse-grain measures of the neural activity (or a related quantity, in the case of fMRI). Could we still investigate the underpinnings of the critical hypothesis from these types of measures in the human brain? The last part of this chapter is devoted to some preliminary, yet fascinating, connections between frequency-dependent criticality and neurological disorders.

## 2.7 Beyond zero-frequency criticality in the human brain

Our estimates for the distance to a phase transition of a given recorded neural dynamics, as employed in Section 2.4.4, have been counting so far on spike-count or noise correlations. This type of long-time-window measures integrate out all potential correlations that could emerge at shorter time scales, reflecting only the tendency of neuron pairs to co-fluctuate around their average rate. To capture temporal correlations beyond this slow time-scale, one can generalize the notion of covariance to the frequency domain, defining a frequency covariance matrix (also known as *coherence* matrix)

---

[3]To this day, current existent measures of human brain single-cell activity come mostly from invasive epilepsy monitoring using the so-called "Behnke-Fried" electrodes, or recordings of deep brain stimulation electrodes in Parkinson's disease.

as:

$$S(\omega) = \mathcal{F}[C(T)] = \langle \mathbf{x}(\omega)\,\mathbf{x}^\dagger(\omega) \rangle \ , \qquad (2.22)$$

where $\mathbf{x}(\omega)$ is the Fourier transform of the neural signal $\mathbf{x}(t)$, and $\mathbf{x}^\dagger(\omega)$ is its complex conjugate. Hence, $S(\omega)$ generalizes the notion of covariance matrix measuring the strength of the dependency between nodes at a given frequency $\omega$.

By the Wiener-Khintchine theorem (Wiener 1930; Khintchine 1934), $S(\omega)$ can also be computed as the Fourier transform of the time-lagged cross-correlation function, which is given by:

$$C(\tau) = \langle \mathbf{x}(t)\,\mathbf{x}^T(t+\tau) \rangle \ . \qquad (2.23)$$

In fact, the long-time-window covariance used for the distance-to-criticality estimation arises as a particular case of the coherence matrix at zero frequency, i.e., $S(\omega = 0) = C_\infty$. For a stochastic linear model driven by external white-noise, as the one described by Eq. (2.12), an analytical expression for the time-lagged correlation can be obtained:

$$C(\tau) = \begin{cases} C_0\,e^{-J^T\tau}, & \tau > 0 \\ e^{J\tau}C_0, & \tau < 0 \end{cases} \qquad (2.24)$$

Using again the Wiener-Khintchine theorem, the coherence matrix for the stochastic linear model can then be written as:

$$S(\omega) = \mathcal{F}\left[\sigma(T)\right] = (J + i\omega\mathbb{I})^{-1}\left(J^T - i\omega\mathbb{I}\right)^{-1}. \qquad (2.25)$$

Consequently, $S(\omega)$ is an Hermitian, positive definite matrix, with an spectral distribution defined on the real line, $\mathbb{R}$, making it suitable to interpret their eigenvalues and corresponding eigenvectors as principal components (in the PCA language) at the given frequency. Eq. (2.25) can be written in a more convenient form:

$$S(\omega) = |a(\omega)|^2(\mathbb{I} - a(\omega)gW)^{-1}(\mathbb{I} - a^\dagger(\omega)gW)^{-T} \ , \qquad (2.26)$$

where $a(\omega) = (1 + i\omega)^{-1}$ and we defined $J \equiv gW$. Note that $|a(\omega)|^2$ is nothing but the power-spectrum of an Ornstein-Uhlenbeck noise. Hence, Eq. (2.25) takes the form of the expression for the long-time-window correlation (Eq. 2.15), with an effective coupling parameter, $\tilde{g}(\omega) = a(\omega)g$, and a pre-factor given by the coefficient $|a(\omega)|^2$ of the power-spectrum. As a word of caution, we remark that it is convenient to work with the normalized coherence matrix, divided by the power-spectrum coefficient $|a(\omega)|^2$, in order to compare in a consistent way the shape of $S(\omega)$ for different values of $\omega$.

In the simplest case, for which the elements of $J$ are drawn as independent and identically distributed (iid) Gaussian variables, $J \sim \mathcal{N}(0, g^2/N)$, the spectral density of $S(\omega)$ follows the distribution given by Eq. (2.17) with an effective coupling parameter:

$$\hat{g}(\omega) = \frac{g}{\sqrt{1 + \omega^2}} \ .\tag{2.27}$$

We remark that, although coherence is perhaps the most common approach as a frequency-domain version of the covariance matrix, it is still based on second-order statistics (i.e., power spectra) and is therefore only sufficient when the time series being considered are Gaussian processes. Considering that neural signals are almost certainly not Gaussian (Malladi et al. 2018), full statistical dependence will not be quantified by coherence when used on such signals.

Nevertheless, knowing that on larger spatial scales synchronization effects can give rise to collective modes of brain dynamics at particular frequencies, we set out to quantify a frequency-dependent distance to criticality on magnetoencephalography (MEG) neural data of resting-state patients from the OMEGA dataset (Niso et al. 2016), which includes recordings from healthy, control subjects, and patients that suffer from Parkinson's disease. In particular, within each group we choose those patients with at least 20 minutes of total recorded time, which left us with 6 subjects on the control group and 8 subjects in the Parkinson's group (see Appendix B.I for more details on the structure and pre-processing of the data). Once again, to ensure a sufficient samples-to-size ratio we divided the original length of the recordings over 480 chunks or samples, each of 2.5s of duration, over which we then computed the frequency-dependent correlations using Eq. (2.22).

Fitting now, for each considered frequency $\omega$ on the range between 0 and 50Hz, the spectrum of the estimated coherence matrix $S(\omega)$ using the theoretical expression given by Eq. (2.17), we obtained an estimate $\hat{g}_s(w)$ of the distance to criticality at the given measured frequency. Since we are already taking the theoretical results derived in Hu et al. 2020 very far from their original context of a simple LRM, for the results presented below we will not focus our attention on the exact value of the estimator $\hat{g}_s(w)$, but rather on its relative magnitude with respect to the long-time-window or zero-frequency covariance case, $\hat{g}_s(0)$.

Interestingly, for the healthy patients, these analysis revealed the existence of close-to-critical frequency modes beyond the slow time scale associated to $\omega = 0$ (see Fig. 2.9a). In particular, we observed how the relative distance to the edge of instability, as given by $\hat{g}_s(w)/\hat{g}_s(0)$, peaked for frequencies near the alpha band (i.e., $\sim 8$ to 12 Hz) in this control group.

This peak, however, was not present for the neural time series of patients with Parkinson disease, for which the distance to criticality was rather consistently high across a large range of frequencies before dropping signif-icantly. These results seem to be in agreement with experimental observations arguing that patients with Parkinson's disease had altered alpha and theta oscillations in the parietal region of the brain (Ye et al. 2022), unveiling the importance of brain dynamics and, in particular, of the relative distance to criticality in different frequency-dependent modes of activity, for the correct functionality of the brain.

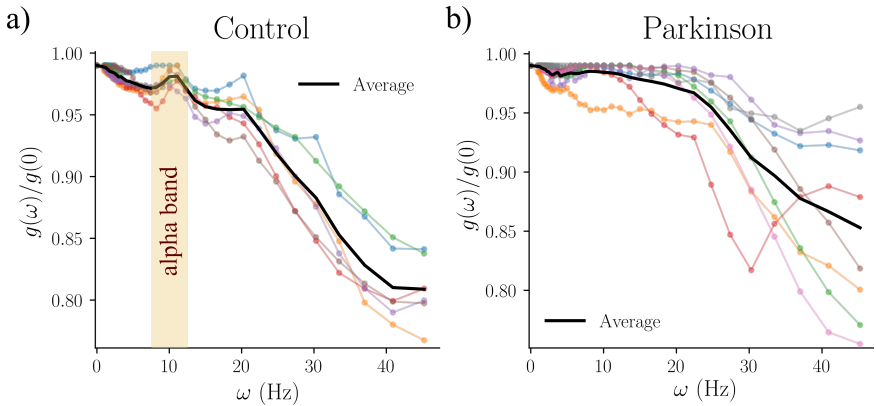

Figure 2.9: **Parkinson patients show a broader range of close-to-critical frequencies.** In both, the control **(a)**, and Parkinson **(b)** groups, each colored curve represents measures of the distance to criticality for one patient at different frequencies, while the black solid curve represents is an average over all available patients in each group. The alpha band, typically located between 8 and 12Hz, is highlighted in yellow for the control group.

## 2.8 Conclusions and perspectives

In this chapter we observed how the scale-invariant nature of neural activity in the mouse brain emerges under a phenomenological renormalization group flow, resembling closely what happens at critical points of continuous phase transitions. In fact, the so-determined scaling exponents exhibit little (yet some) variability across brain regions, so we referred to them as been "quasi-universal", in seeming analogy with universality in critical phenomena.

Setting out to quantify the actual distance of empirical data to the edge of instability, we found that all 16 regions analyzed turn out to be relatively close to the edge of instability. Although interpreting the reason behind the differences in the distance to criticality across regions remains an open and challenging goal, we speculate that this variability could be related to a hierarchy of information flow in the brain organization. Fig. 2.10a shows some preliminary results supporting this hypothesis, representing the distance to criticality against an estimated hierarchy score for each region

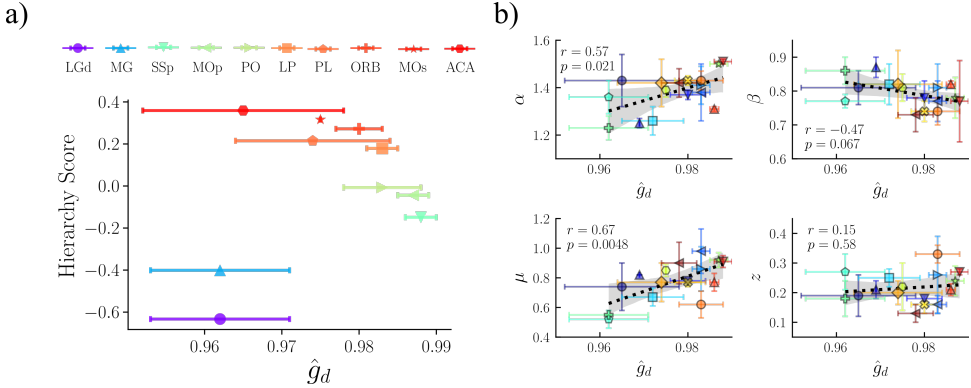

Figure 2.10: **Low-hierarchy regions show a dynamics closer to the edge of insta-bility. (a)** Hierarchy score, as quantified in Harris et al. 2019 against estimated distance to criticality. Regions are colored according to their score value. **(b)** Values of the average scaling exponents, $\alpha$, $\beta$, $z$ and $\mu$, observed in each region against the estimated distance to criticality. For each exponent, we fit a linear regression to the data points, showing the correlation coefficient, $r$, and $p$-value to reject the null hypothesis that there is no linear relationship between 2 variables.

—as defined in Harris et al. 2019— with low (high) scores corresponding to sensory (higher-level) areas. Notice how sensory areas such as primary cortices MOp and SSp (with low scores; see, e.g., Fig. 6 in Harris et al. 2019) tend to operate closer to the edge of instability than secondary cortices (such as MOs) or areas in the prefrontal cortex (such as ORB, PL, and ACA). This seems to suggest that there could exist a relationship between the dynamical regime of a given area and its hierarchical score, with low-score regions being more "critical". On the other hand, a clear trend is not observed in thalamic regions LGd, and MG (Fig. 2.10a)

During this chapter, we argued that the scale-free properties observed across all regions could arise from a close-to-critical dynamics. Is there, however, any relation between the variability in the scaling exponents and our estimates for the distance to criticality? Fig. 2.10b shows, for every region, each of the computed exponents against the estimated distance to the critical point. Interestingly, a statistically significant linear dependence seems to exist between the estimated distance, $\hat{g}_d$, and the exponents $\mu$ ($r = 0.67$; $p < 0.01$) and $\alpha$ ($r = 0.57$; $p < 0.01$), whereas such relation can not be ascertained with a 5% significance level for the exponents of the free-energy and autocorrelation times.

Following with the results, we found that a very simple linear-rate model tuned to the vicinity of its critical point generates scale-invariant patterns of activity, with associated critical exponents that match remarkably well those observed for the resting-state activity across brain regions. Understanding whether more elaborate models (including spiking neurons, non-linear interactions or fulfilling Dale's rule, for example) do behave in the same universal way or can even provide a better description of the scaling features of biological neural networks remains as an open task for the future.

Aiming at understanding the role of structure in the emergence of scale-invariant properties, we performed a PRG analysis over topographically-designed neural cultures. We showed that heterogeneity (and possibly a hierarchical-modular architecture) in the network structure are ingredients that can affect the dynamical regime of the network and, consequently, its scaling behavior. To what extent this type of architectures push the network towards an underlying close-to-critical dynamics, as well as a detailed analysis of the effective connectivity patterns emerging under different types of topographical substrates is the scope of ongoing work.

Finally, we moved away from the microscopic, single-neuron recordings to measures of MEG activity over macroscopic regions of interest (ROIs) in the brain. In particular, we studied whether neural systems could show close-to-critical dynamics in the temporal structures of correlations beyond the slow time scale given by the long-time-window (or zero-frequency) covariance. Remarkably, a clear peak in the relative distance to criticality (with respect to the zero-frequency mode) was observed in healthy patients for frequencies in the alpha band, while these peaks were consistently flattened out in the measures carried over patients with Parkinson's disease. We notice that this preliminary results seems to be in agreement with previous studies linking critical dynamics and Parkinson's disease Hohlefeld et al. 2012; West et al. 2016; Zimmern 2020. In particular, in West et al. 2016 the authors used a detrended fluctuation analysis to conclude that patients in more severe stages of Parkinson's disease presented a dynamics closer to the onset of a pathological synchronization, towards a possible hyper-synchronized critical state. Our results could indeed provide new light into this hypothesis, as they seem to reveal the specific frequency-modes of activity that show an "overly-critical" dynamics when compared to healthy brains, adding to the number of works that highlight the relevance of the critical hypothesis approach also at a clinical level (Zimmern 2020).

# Learning machines learn to learn

"In terms of evolutionary history, it was only yesterday that men learned to walk around on two legs and get in trouble thinking complicated thoughts.
So don't worry, you'll burn out."

HARUKI MURAKAMI, *The Wind-Up Bird Chronicle*

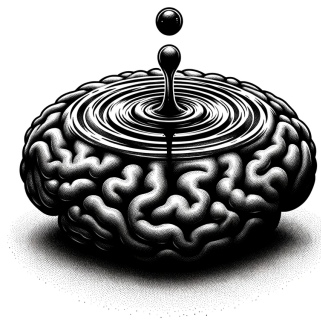

## 3.1 Introduction to Echo State Networks

Many of the ideas behind the theory of criticality presented in Chapter 1 and tested in Chapter 2 can be also extended to the realm of artificial neural networks. In fact, advances in the area of machine learning (ML) have explored since the eighties the idea of computation at the "edge-of-chaos" (Packard 1988; Langton 1990; Bertschinger et al. 2004). Over the past decade, new exciting works have emerged at the boundary of machine learning and statistical mechanics, which explored, for instance, the connection between the Renormalization Group (RG) and Deep Learning (Koch et al. 2020; P. Mehta et al. 2014; Oprisa et al. 2017), or the emergence of optimal computational capabilities near phase transitions in neuromorphic devices (Hochstetter et al. 2021). Nevertheless, the increasing complexity of current state-of-the-art machine learning models, together with the detachment of these models from biologically-plausible architectures, such as recurrent neural networks (RNNs), hinders our ability to fully characterize and control the dynamical regime of learning machines. Can we find a ML framework that can naturally exploit the virtues of systems operating near a critical point, while remaining biologically plausible and tractable from the perspective of their dynamics?

The answer to this question would not arrive until the beginning of this new millennium. While a growing ML community was still thawing out from the second AI winter (Fradkov 2020), a new approach to training RNNs, later coined under the common term of Reservoir Computing (RC), was discovered independently on two different architectures: the Echo State Network (ESN) in a rate-based, discrete-time model of non-linear units (Jaeger 2001b); and the Liquid State Machine (LSM) for a biologically-inspired model of continuous-time spiking neurons (Maass et al. 2002). Both paradigms were built upon the idea that, under certain conditions, the state of an RNN can be understood as a function of the input history presented to the network. Not surprisingly, results showing how these types of networks could exploit the combined advantages of stability (order) and responsiveness to inputs (disorder) when they operate at the borderline between order and "chaos", appeared only a few years after the invention of this machine learning paradigm (Bertschinger et al. 2004; Boedecker et al. 2011).

In the following two chapters, we will focus our attention into the discrete-time ESN. The simplicity of its architecture, consisting of an input and output layers connected through feed-forward weights to a recurrent network (the *reservoir*), together with a fast and computationally much less expensive training compared to previous RNN models, make these networks especially well-suited for learning dynamical systems, even when those display chaotic or complex spatiotemporal behavior (Pathak et al. 2017; Pathak

et al. 2018).

As we will see in the next section, the simplicity of the training in the vanilla ESN emerges from the fact that only weights at the output layer are adjusted, while input-to-unit and unit-to-unit connections are left unchanged from their initial random values. Although very flexible, this approach also leaves the open question of how to design the reservoir connectivity so as to maximize the performance of the network in a given task. While most reservoir computing approaches consider a reservoir with fixed internal connection weights, plasticity was rediscovered as an unsupervised, biologically inspired adaptation to implement an adaptive reservoir. It appeared first as a type of Hebbian synaptic plasticity to modify the reservoir weights (Babinec et al. 2007), but soon the ideas of non-synaptic plasticity that inspired the first intrinsic plasticity (IP) rule (Triesch 2005) were also implemented in an ESN (Schrauwen et al. 2008). After that, several different models of plasticity have been implemented in RC networks with promising results (Steil 2007; Yusoff et al. 2016; X. Wang et al. 2019). Today, the fact that biologically meaningful learning algorithms have a place in these models, together with recent discoveries suggesting that biological neural networks display RCs' properties (Ju et al. 2015; Enel et al. 2016), make RC a field of machine learning in continuous growth.

From a functional perspective, ESNs have shown to successfully perform in a wide number of tasks, ranging from speech recognition (Skowronski et al. 2007) and channel equalization (Jaeger et al. 2004), to robot control (Hertzberg et al. 2002) or stock data mining (Lin et al. 2008). Here, we will focus in the challenging problem of chaotic time series forecasting. This type of task has been addressed extensively in the scientific literature (Babinec et al. 2009; Lin et al. 2009; Yusoff et al. 2016; H. Wang et al. 2019), and ESNs implementing plasticity rules to improve time series forecasting have also been treated before (Yusoff et al. 2016; H. Wang et al. 2019; Babinec et al. 2007). However, a real understanding of the effects of this type of unsupervised learning over the dynamical state of the reservoir is still missing.

Despite the overall simplicity of these networks (and much like in any type of ML model), their performance over different types of tasks is still subject to the daunting problem of parameter selection. Therefore, in this section we first introduce the overall framework upon which ESNs are constructed, beginning with their architecture and following with the equations governing the reservoir dynamics, highlighting the role of each tunable parameter within the model.

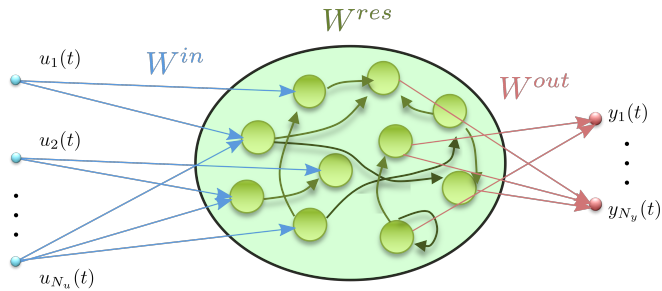

Figure 3.1: **Architecture of an archetypal Echo State Network.** Inputs are filtered through feed-forward connections $W^{in}$ into a reservoir consisting of $N$ recurrently-connected non-linear units. The internal states of these units are then projected for read out through a second feed-forward connectivity matrix $W^{out}$, generating the output of the network.

### 3.1.1 Architecture

Fig. 3.1 shows a schematic representation of the three fundamental ingredients that constitute a vanilla Echo State Network, but that are also present in any more complex version of the original model (Lukoševičius 2007; Jaeger 2008; Sussillo et al. 2009; Büsing et al. 2010; DePasquale et al. 2018; Chen et al. 2020).

The core of an ESN lies in its *reservoir*, where a number $N_x$ of time-discrete computational units[1] communicate with each other through a sparse recurrent connectivity matrix $W^{res} \in \mathbb{R}^{N_x \times N_x}$. Typically, these connections are randomly initialized according to a uniform or Gaussian distribution , although extensive work has been devoted to find optimal initializations for this connectivity matrix (Rad et al. 2010; Rodan et al. 2011; H.-T. Fan et al. 2017; Aceituno et al. 2020; Saadat et al. 2023), as well as plasticity rules that could evolve the recurrent weights to learn some characteristic features of the input distribution (Babinec et al. 2007; Yusoff et al. 2016; H. Wang et al. 2019). As a thrilling new line of research at the boundary of ML and neuroscience, we highlight the recent work of Damicelli et al. 2022, which aims at integrating real brain connectomes into the underlying reservoir networks, allowing to probe the influence of brain connectivity patterns and modularity over the reservoir performance. From a more computational perspective, the recently uncovered equivalence of ESNs to a nonlinear vector autoregression, which require no random matrices and fewer hyperparameters, seems to open the door to a whole new generation of reservoir computers (Gauthier et al. 2021).

---

[1]Although we will commonly refer to the individual elements of the reservoir as "units", in some contexts we may use the term "neuron" (in a very generic sense) to highlight the parallelism with biological neural networks.

Following the diagram depicted in Fig. 3.1, at each time step every unit inside the reservoir updates its state according to the activity of other presynaptic units and the incoming external signal from the $N_u$ input units. Synaptic connections with these input units take place through a feed-forward weight matrix $W^{in} \in \mathbb{R}^{N_x \times N_u}$ —typically dense and drawn randomly from a uniform distribution. Finally, a set of $N_y$ output units connected to the reservoir through a matrix $W^{out} \in \mathbb{R}^{N_y \times N_x}$ reads out the state of the reservoir at each time step. Although this output matrix is typically initialized following the same distribution as $W^{in}$, it has been shown that pruning and regularization of the output weights can lead to an increase in the generalization capability of the network (Dutoit et al. 2008).

In practice, we remark that the number of units inside the reservoir is much larger than the dimensionality of the input space, so that the reservoir effectively acts as a non-linear expansion of the inputs (Lukoševičius 2012). In fact, reservoirs can be understood as general spatiotemporal kernels, capable of computing a broad set of nonlinear mappings of the input data, on which linear regression or classification can be easily performed (Hermans et al. 2012). Moreover, as we will see in the following section, within a broad region of the parameter space the states of the reservoir units can hold a "memory" of the inputs, providing a temporal context that is well-suited for time series prediction tasks. We now introduce the discrete-time equations that govern the dynamical state of the units inside the reservoir, working towards a deeper insight of the properties that make ESNs great learners of temporal dependencies.

### 3.1.2 Internal dynamics and the echo state property

Let $\mathbf{u}(t) \in \mathbb{R}^{N_u}$ be a multivariate input signal of dimension $N_u$ defined at discrete time steps $t = 1, ..., T$, where $T$ is the total length of the input time series (see Fig. 3.1) . At each time step, the input, previously filtered through a weight matrix $W^{in} \in \mathbb{R}^{N_x \times N_u}$, is fed to the $N_x$ internal units of the reservoir, whose states $\mathbf{x}(t) \in \mathbb{R}^{N_x}$ evolve according to the following discrete-time equation:

$$\mathbf{x}(t) = \tanh(\varepsilon W^{in}\mathbf{u}(t) + \rho W^{res}\mathbf{x}(t-1)) \ , \qquad (3.1)$$

where the internal connections between units in the reservoir are given by $W^{res} \in \mathbb{R}^{N_x \times N_x}$, defined as a sparse matrix with unitary maximum eigenvalue (i.e., $W^{res} \equiv (1/\tilde{\rho})\tilde{W}^{res}$, where $\tilde{W}^{res}$ is a sparse matrix with maximum eigenvalue $\tilde{\rho}$ ). In the above equation, the *input scaling factor*, $\varepsilon$, and the *spectral radius* or largest eigenvalue of the effective connectivity matrix, $\rho$, are the two fundamental parameters that will determine the dynamical regime of our networks.[2] For the rest of this chapter, we chose the hyperbolic tangent as our activation function, but other nonlinearities have been shown to be equally valid (Lukoševičius 2012).

The spectral radius determines, under linear approximation, the dynamical stability inside the reservoir when no input is fed into the network. Thus, a spectral radius exceeding unity has been often regarded as a source of instability in ESNs due to the loss of the so-called **echo state property**, a mathematical condition ensuring that the effect of initial conditions on the reservoir states fades away asymptotically in time (Jaeger 2001c; Jaeger 2001a; Yildiz et al. 2012). Nevertheless, later studies have shown that the echo state property can be actually maintained over a unitary spectral radius, and different sufficient conditions have been proposed (Buehner et al. 2006; Yildiz et al. 2012; Gallicchio 2018) (see in particular Manjunath et al. 2013, where the authors analyze the problem from the lens of non-autonomous dynamical systems, deriving a sufficient condition for the echo state property with regard to a given input). Intuitively, the larger the spectral radius, the more weight is given to the "history" of the system through the contribution of the past states of the reservoir.

On the contrary, the input scaling factor often shows an opposite role. For larger values of $\varepsilon$, the inputs push the activity arriving to each unit towards the saturating ends of the non-linearity, which can make the units effectively act in a binary switching manner, potentially converting an initially expanding mapping into a contracting dynamics (Lukoševičius 2012).

In the next section we will introduce the main ideas behind the supervised training and testing protocols used along with ESNs. For the purpose of this chapter we will focus on the typical setup for a time-series prediction task, but a training protocol for an image classification task will be presented in Chapter 4, when studying the reservoirs high-dimensional internal representations.

## 3.2 Training and testing for a time-series prediction task

Let us first begin providing some general rules regarding the pre-processing of the training set. To facilitate the comparison between the performances of two networks with the same set of hyper-parameters over two different training sets, it is first advisable to standardize the input time series so that they always have zero mean and unitary standard deviation. Alternatively, one can divide the series by their maximum absolute value, effectively forcing them to lie in the $[-1, 1]$ interval. Moreover, if $\mathbf{u}(t)$ is unbounded and one expects the presence of strong outliers, it is also a good practice to bound the training and test data (for instance, squashing it with a tanh function). Otherwise, the reservoir dynamics could respond to these outliers by jumping into points of its phase space that are not well covered by the usual trajectories of $\mathbf{x}(t)$, leading to unpredictable outputs or saturating activities that "erase" the memory stored by the echo state property (Lukoševičius 2012).

---

[2]In the literature, one typically finds Eq. (3.1) written in terms of an effective reservoir connectivity matrix $W_{eff}^{res} := \rho W^{res}$. Here, we decided to factor out the contribution of the spectral radius to highlight the confronting roles of $\varepsilon$ and $\rho$.

The basic training protocol for a time-series prediction task is schematically depicted In Fig. 3.2. Once the training data has been correctly preprocessed, the input series is filtered at each time step by a weight matrix $W^{in} \in \mathbb{R}^{N_x \times N_u}$, which is typically dense and drawn from a uniform distribution in the [-1, 1] range, and the states of the units inside the reservoir are updated following Eq. (3.1). The above formulation, with a common scaling factor for all inputs, is considered the standard practice to avoid increasing the number of tunable hyperparameters. However, in some cases it might be also useful to define a vector $\boldsymbol{\varepsilon} \in \mathbb{R}^{N_u}$ if, for instance, we think that some dimensions of the input should be weighted more strongly (Lukoševičius 2012).

On the other hand, as already introduced in the previous section, an initial sparse connectivity matrix, $\tilde{W}^{res}$, is typically drawn from a uniform distribution with symmetric range around zero (although zero-mean Gaussian distributions are also commonly employed). Then, the width of the non-zero weights distribution, given by the matrix largest eigenvalue, is rescaled so that the effective connectivity matrix have a spectral radius $\rho$: $W^{res}_{eff} = \rho W^{res} = (\rho/\tilde{\rho})\tilde{W}^{res}$, where $\tilde{\rho}$ is the maximum eigenvalue of $\tilde{W}^{res}$.

At each time step during the presentation of the training series, the states of the neurons in the reservoir produce the final output after being linearly filtered by a random matrix $\tilde{W}^{out} \in \mathbb{R}^{N_y \times N_x}$ (whose elements, for simplicity, follow the same distribution as those in $W^{in}$), so that the dynamics of the readout states is given by:

$$\tilde{\mathbf{y}}(t) = \tilde{W}^{out}\mathbf{x}(t) \ , \tag{3.2}$$

where $\tilde{\mathbf{y}}(t) \in \mathbb{R}^{N_y}$ is the $N_y$-dimensional output. Using a supervised learning scheme, the goal of the training is to find the output weights $W^{out}$ so that $\mathbf{y}(t)$ not only matches as closely as possible a desired target series, $\mathbf{y}^{target}(t) \in \mathbb{R}^{N_y}$, but can also generalize to unseen data. Because large output weights are commonly associated to overfitting of the training data (Lukoševičius 2012), it is a common practice to keep their values low by adding a regularization term to the error in the target reconstruction. Although several regularization methods have been proposed (Jaeger 2001b;

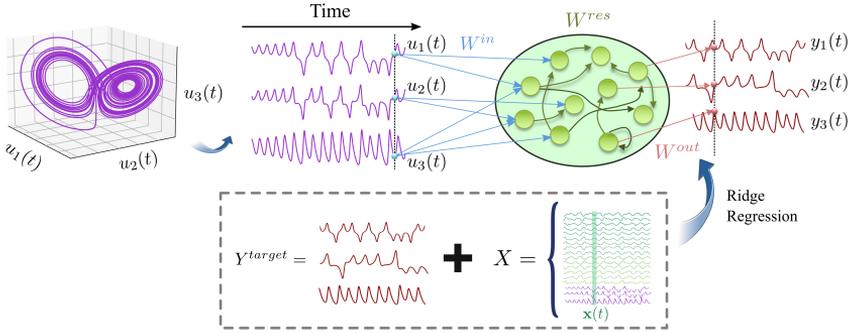

Figure 3.2: **Training protocol for a time-series forecasting task.** At each time step a point of a (possibly multivariate) time series $\mathbf{u}(t)$ is projected through a matrix $W^{in}$ into the $N$ units of the reservoir. For the length $T$ of the training series, reservoir states are collected into a matrix $X \in \mathbb{R}^{N \times T}$. Output weights $W^{out}$ are then computed offline through a ridge regression over the reservoir states for a given target series, typically involving a one-step-ahead prediction of the input (i.e., $\mathbf{y}_{target}(t) = \mathbf{u}(t+1)$).

Dutoit et al. 2008; Reinhart et al. 2010; Reinhart et al. 2011), here we use the ridge regression method, for which the error is defined as:

$$E_{ridge} = \frac{1}{N_y} \sum_{j=1}^{N_y} \left( \frac{1}{T} \sum_{t=1}^{T} \left( y_j^{target}(t) - y_j(t) \right)^2 + \beta \left\| \mathbf{w}_j^{out} \right\|^2 \right), \qquad (3.3)$$

where $\mathbf{w}_j^{out}$ are the projection weights of all reservoir neurons to the $j$-th output (i.e., the $j$-th row of $W^{out}$), $\|\cdot\|$ stands for the Euclidean norm and $\beta$ is the regularization coefficient, which sets a penalty for large values of the output weights. Notice that choosing $\beta = 0$ removes the regularization, turning the ridge regression into a generalized linear regression problem.

Defining $Y^{target} \in \mathbb{R}^{N_y \times T}$ as the matrix whose columns are the output targets $\mathbf{y}^{target}(t)$ at each time $t$, and $X \in \mathbb{R}^{(N_u+N_x) \times T}$ the *design matrix* consisting of all concatenated vectors $[\mathbf{x}(t); \mathbf{u}(t)]$, the expression for the optimal readout weights $W^{out}$ that minimizes the above error can be easily obtained (Lukoševičius 2012) as:

$$W^{out} = Y^{target} X^T \left( X X^T + \beta I \right)^{-1}, \qquad (3.4)$$

where $I$ is the identity matrix. The above solution, also known as regression with Tikhonov regularization (McDonald 2009), is highly stable and, more importantly, because the terms $Y^{target} X^T \in \mathbb{R}^{N_y \times (N_u+N_x)}$ and $X X^T \in \mathbb{R}^{(N_u+N_x) \times (N_u+N_x)}$ do not depend on the length $T$ of the training series, they can be easily updated *online* during the presentation of the inputs for arbitrary large training data. Nevertheless, there exists other solutions that

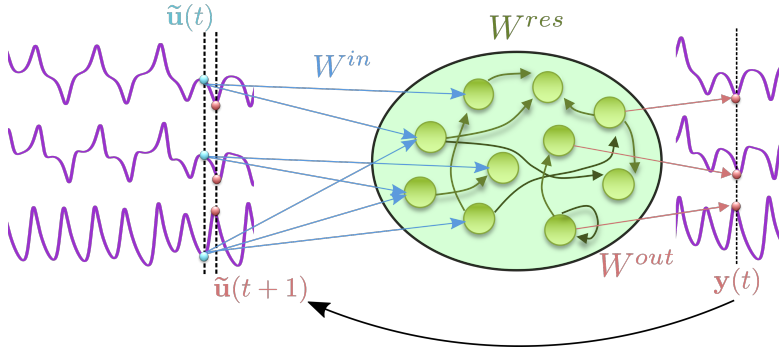

Figure 3.3: **Test protocol for a time-series forecasting task.** The network runs in generative mode for $T_{test}$ time steps. Beginning with an initial input seed $\mathbf{u}(0)$, at each time step $t$ the network output $\mathbf{y}(t)$ serves as input, $\tilde{\mathbf{u}}(t+1)$, at the next time step.

are widely used, such as:

$$W^{out} = Y^{target} X^+ \; , \tag{3.5}$$

where $X^+$ is the Moore-Penrose pseudoinverse of $X$ (Penrose 1955). Alternatively, it has been shown that adding a simple scaled white noise to Eq. (3.1) for the reservoir states update produces a similar regularization effect to ridge regression (Lukoševičius 2012), while being at the same time more biologically grounded.

In most cases, the output weights are trained in a one-step-ahead prediction task, in which the network output at time $t$ is required to be the next point $\mathbf{u}(t+1)$ of the input time series (i.e., $\mathbf{y}_{target}(t) = \mathbf{u}(t+1)$). The predictive ability of the network is then assessed by asking the network to continue from the end of the training series in *generative* (also known as autonomous) mode, meaning that the reservoir takes its former output at time $t$ as input at time $t+1$ (see Fig. 3.3). Different measures have been designed to quantify the goodness of predictions in time-series forecasting tasks, including the standard root mean-square error (RMSE):

$$\text{RMSE} = \frac{1}{N_y} \sum_{j=1}^{N_y} \left( \frac{1}{T} \sum_{t=1}^{T} \left( y_j^{test}(t) - y_j(t) \right)^2 \right) \; , \tag{3.6}$$

and the furthest predicted point (FPP) (Morales et al. 2021a), defined as:

$$FPP := \max\{k : |\mathbf{y}^{test}(k') - \mathbf{y}(k')| \leq \epsilon \quad \forall \, k' \leq k\} \; , \tag{3.7}$$

where $\epsilon$ is some arbitrary small threshold value.

## 3.3 Plasticity models in ESNs

Since there exists a plethora of plasticity mechanisms that biological neural networks can employ to learn from the incoming stimuli (Abbott et al. 2000; Mozzachiodi et al. 2010; Berlucchi et al. 2008; Watt et al. 2010), it is useful, when trying to translate this type of learning rules into a RC framework, to categorize them into two distinct families: rules that aim at modifying the recurrent weights $W^{res}$ of the reservoir (*synaptic plasticity*), and rules which alter the inner excitability of the reservoir units (*intrinsic plasticity*). In this manner, changes are based on the activity stimulated by the input, so that information carried by the training signal is partly embedded in the structure of reservoir.

As we saw in the previous section, one of the advantages of the RC paradigm is its simple and efficient —since typically offline— learning framework, which trains only a set of output weights, but let the reservoir connectivity matrix take completely random values, only bounded by the chosen spectral radius. This obviously leaves a lot of room for improvement, as not just any connectivity matrix will generate an optimal mappings of the input space into the high-dimensional space of reservoir states. Although there is empirical evidence of non-Hebbian forms of synaptic plasticity (A. Alonso et al. 1990; H. K. Kato et al. 2009), most rules modifying the synaptic strength among neurons fall into the category of Hebbian learning. The Hebbian rule, as originally proposed by Hebb (Hebb 1949), can be described mathematically as a change in the synaptic strength between a presynaptic and a postsynaptic neuron that is proportional to the product of their activities. In the discrete-time language of ESNs, this would translate into:

$$w_{kj}(t+1) = w_{kj}(t) + \eta x_j(t) x_k(t+1) \ , \tag{3.8}$$

where $w_{kj}$ is the weight element of $W^{res}$ connecting the "postsynaptic" unit $k$ with the "presynaptic" unit $j$ . In the above equation, all weights in the reservoir are updated in parallel at each discrete time step, with a parameter $\eta$ that sets the learning rate. An obvious flaw follows from Eq. (3.8): as the weights grow in the direction of the correlations between pre and postsynaptic units, the connections get stronger following Hebb's postulate, until activity eventually spread uncontrollably throughout the network. To avoid this undesired property, one possibility is to normalize the weights arriving to each postsynaptic neuron $k$, so that $\sqrt{\sum_j w_{kj}^2(t)} = 1, \ \forall t \in [0, T]$. One can achieve this by rewriting the update rule in Eq. (3.8) as:

$$w_{kj}(t+1) = \frac{w_{kj}(t) + \eta x_k(t+1) x_j(t)}{\sqrt{\sum_{i=1}^N \left( w_{ki}(t) + \eta x_k(t+1) x_i(t) \right)^2}} \ . \tag{3.9}$$

Note, however, that Eq. (3.9) is no longer local, meaning that a modification in a given weight $w_{kj}$ also depends on the weights of unit $k$ with all other units. This makes the update not only computationally very expensive, but also biologically less realistic. Nevertheless, if one assumes a small learning rate $\eta$ and close-to-linear activation functions in the absence of external inputs, it is possible to derive a local approximation to Eq. (3.9), known today as Oja's rule (Oja 1982):

$$w_{kj}(t+1) = w_{kj}(t) + \eta x_k(t+1)\left(x_j(t) + x_k(t+1)w_{kj}(t)\right) \ . \qquad (3.10)$$

A complete derivation of this rule can be found for the sake of completeness in Appendix B.II. Generally, all local Hebbian rules take the form $w_{kj}(t+1) = w_{kj}(t) + \Delta w_{kj}(t)$. However, it has been suggested that a change in the sign of Hebbian plasticity rules (so that $w_{kj}(t+1) = w_{kj}(t) - \Delta w_{kj}(t)$) may be advantageous in making an effective use of the dynamical range of cortical neurons (Barlow 1989), while also promoting decorrelation between the activity induced by different inputs. For the purpose of this chapter we will work with a form of this so-called anti-Hebbian learning known as the *anti-Oja* rule, obtained simply by changing the sign of the weight update in Eq. (3.10) (see Babinec et al. 2007; Yusoff et al. 2016 for applications of this rule to ESNs).

Having defined a rule that adapts the internal weights between reservoir units, we move now to non-synaptic forms of plasticity, which adjusts the neurons' internal excitability instead of the individual synapses. Based on the idea that every single neuron tries to maximize its information transmission while minimizing its energy consumption, Jochen Triesch proposed a mathematical learning rule that leads to maximum entropy distributions for the neurons activity with certain fixed moments (Triesch 2005). Although the original derivation of Triesch applied to Fermi activation functions and exponential desired distributions, the work was soon extended in Schrauwen et al. 2008 to account for neurons with hyperbolic tangent functions. In this case, each neuron updates its state through the following expression, which is an extension of Eq. (3.1):

$$x_k(t) = \tanh(a_k z_k(t) + b_k) \ , \qquad (3.11)$$

where $a_k$ and $b_k$ are gain and bias terms for the postsynaptic neuron, and:

$$z_k(t) = \varepsilon \sum_{i=1}^{N_u} w_{ki}^{in} u_i(t) + \rho \sum_{j=1}^{N_x} w_{kj}^{res} x_j(t-1) \ , \qquad (3.12)$$

is the total input arriving to unit $k$ at time $t$ . Online learning rules for the gain, $a_k$, and bias, $b_k$, can then be derived from the minimization of the Kullback-Leibler divergence with respect to a desired Gaussian output distribution (Schrauwen et al. 2008), so that at each time step:

$$a_k(t+1) = a_k(t) + \Delta a_k(t) \ , \tag{3.13}$$

$$b_k(t+1) = b_k(t) + \Delta b_k(t) \ , \tag{3.14}$$

with

$$\Delta b_k(t) = -\eta \left( -\frac{\mu}{\sigma^2} + \frac{x_k(t)}{\sigma^2} \left( 2\sigma^2 + 1 - x_k^2(t) + \mu x_k(t) \right) \right) \ , \tag{3.15}$$

$$\Delta a_k(t) = \frac{\eta}{a_k(t)} + \Delta b_k(t) z_k(t) \ , \tag{3.16}$$

where $\eta$ is the learning rate, while $\mu$ and $\sigma^2$ are the mean and variance of the targeted Gaussian distribution for the neurons activity, respectively. The above update equations constitute what is known today as the intrinsic plasticity (IP) rule.

We remark that the above plasticity mechanisms, although often regarded as a form of unsupervised learning, are completely oblivious to the task at hand. Their aim is that, by increasing the dynamical range of the networks (anti-Hebbian synaptic rules) or maximizing the information carried by the neurons output (intrinsic plasticity), the resulting network will improve its ability to learn how to perform some task over the presented inputs. Thus, within the overall training process of the networks, we like to refer to the stage mediated by input-induced plastic changes as a *learning-to-learn* phase. In the following section we will test whether these biologically-inspired rules can indeed improve the performance of ESNs on a time-series prediction task.

## 3.4 Forecasting chaos

For our purposes, the task at hand will consist on a one-step-ahead forecasting problem of a well-known chaotic timeseries: the Mackey-Glass dynamical system (Mackey et al. 1977). The Mackey-Glass (MG) series is a classical benchmarking dataset in time-series forecasting, generated from the time-delay differential equation:

$$\frac{dx}{dt} = \left[ \frac{\alpha x(t-\tau)}{1 + x(t-\tau)^\beta} - \gamma x(t) \right] \ , \tag{3.17}$$

where $\tau$ represents the delay and the parameters are set to $\alpha = 0.2$, $\beta = 10$ and $\gamma = 0.1$, a common choice for this type of tasks (Yusoff et al. 2016; Ortín et al. 2015). This dynamical system is known to exhibit a chaotic behavior for time delays $\tau > 16.8$, so we constructed a series of length $T = 4000$ as training set with $\tau = 17$ (MG-17), often used as an example of a weakly chaotic series. We refer the reader to our work in Morales et al. 2021a for a comparison of the performance with a series showing stronger chaotic behavior for $\tau = 30$ (MG-30). More details regarding the integration of the training and test series can be found in Appendix B.I.

When implementing any of the already introduced plasticity rules, we ran the corresponding algorithms over the reservoir weights (anti-Oja rule), or the units' gain and bias (IP rule) using all $T$ points of the training series as input, and then passing over the full training set a number $n_{ep}$ of times (known as *epochs* in the ML jargon). At the end of this learning-to-learn phase, the reservoir weights —as well as the gain and biases for the units in the case of intrinsic plasticity— are kept fixed to their last configuration.

The evaluated prediction task consisted on the continuation of the series from the last input of the training set. Accordingly, for an input $\mathbf{u} = [u_1, u_2, ..., u_T]$ the target series in the supervised training was defined as the one-step-ahead sequence $\mathbf{y}^{target} = [u_2, u_3, ..., u_{T+1}]$. For the off-line computation of the output weights, $W^{out}$, we kept all internal reservoir states of the ESN and only after passing all the input training set we applied Eq. (3.4) (see Fig. 3.2 in the previous section).

Right after training of the output weights, the ESNs were asked to continue the series in generative mode for a number of steps $F = 400$, beginning with the last input of the training set,. To quantify the error between the points of the predicted and target test series, we used two different magnitudes as described in Section 3.2: the root mean-square error (RMSE) (Eq. 3.6) and the furthest predicted point (FPP) (Eq. 3.7), with a tolerance value for significant deviation, $\epsilon = 0.01$, representing around 1% of the maximum distance between any two points in the training series.

### 3.4.1 Hyperparameter selection

One of the biggest drawbacks of Echo State Networks is their high sensitivity to hyper-parameters choice (see Lukoševičius 2012 for a detailed review on their effects over the network performance). For the purpose of this chapter, weights in the reservoir, input and output layers are initialized randomly according to a uniform distribution between -1 and 1. Sparseness in the reservoir matrix is set to 90%, while input and output connections are initialized as dense matrices. When incorporating plasticity rules, an extra tunable hyperparameter $\eta$ is included, describing the learning rate in the update rules. When IP is implemented, we find that best results are obtained when using $\mu = 0$ and $\sigma = 0.5$ as the mean and variance of the targeted distribution of the neuron states. For the sake of comparison between different models, we choose a com-mon non-optimal —but generally well-performing— set of hyper-parameters $\{N = 300, \rho = 0.95, \varepsilon = 1, \beta = 10^{-7}, \eta = 10^{-6}\}$ for all simulations.

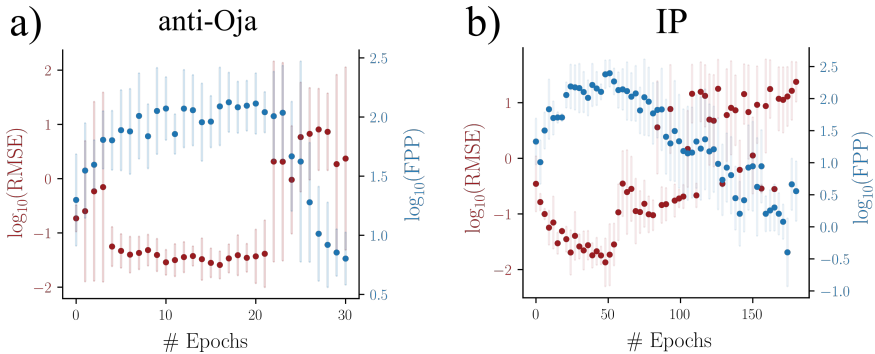

Figure 3.4: **Performance shows local maximum with the length of training.** For each plasticity rule, performance measures were computed across $F = 400$ predicted points in a MG-17 chaotic time series forecasting task, and plotted as a function of the number of training episodes (epochs). At each epoch, averages were computed over 20 independent realizations of a 300 units ESN.

### 3.4.2 Assessing forecasting performance

To analyze the effects of plasticity over the ESN performance on a one-step-ahead prediction task over the MG-17 series , we compute the evolution of the RMSE and FPP across epochs for both the synaptic (Fig. 3.4a) and non-synaptic (Fig. 3.4b) plasticity rules considered. Notice how, in both cases, ESNs show optimal performance at a certain number of epochs (as measured by a minimum of the RMSE, and maximum of the FPP), followed by a decline in the network's predictive capabilities as the training continues.

The above behavior reveals that both synaptic and intrinsic plasticity rules can increase the performance of ESNs during the first epochs of learning, but then a regime follows in which the networks lose their forecasting capabilities. In the following section we will delve into the "guts" of the reservoirs as they undergo plasticity changes, aiming at understanding the effects of these rules over the dynamics of the network and, ultimately, the observed shift in performance.

## 3.5 Plasticity rules shape the dynamical regime

### 3.5.1 Changes at the population level

We will first take a look at properties of the overall population dynamics under the effects of plasticity. In particular, because we expect correlations among units to decay under the effect of anti-Hebbian rules, we will measure the average equal-time cross-correlation among neurons as:

$$C = \frac{1}{n_{trials}N(N-1)} \sum_{n=1}^{n_{trials}} \sum_{i \neq j}^{N} c_{ij}^{(n)} \ , \qquad (3.18)$$

where $c_{ij}^{(n)}$ is the Pearson's correlation coefficient between neurons $i$ and $j$ at trial $n$:

$$c_{ij}^{(n)} = \frac{\sum_{t=1}^{T} \left( x_i(t) - \overline{x}_i \right) \left( x_j(t) - \overline{x}_k \right)}{\sqrt{\sum_{t=1}^{T} \left( x_i(t) - \overline{x}_i \right)^2} \sqrt{\sum_{t=1}^{T} \left( x_j(t) - \overline{x}_j \right)^2}} \ . \qquad (3.19)$$

Moreover, since it was suggested that information transfer and storage in ESNs are maximized at the edge between a stable and an unstable (chaotic) dynamical regime (Boedecker et al. 2011), we will check whether plasticity rules could be driving the network dynamics near such a phase transition when optimal performance is observed, using the maximum Lyapunov exponent (MLE) as a control parameter.

Let's assume that a given reservoir $\mathcal{R} \equiv \{\varepsilon W^{in}, \rho W^{res}\}$ is in a state $\mathbf{x}(t_0)$ of its phase space at an initial time step $t_0$, and that a second, identical reservoir $\tilde{\mathcal{R}} = \mathcal{R}$ is defined, but with an infinitesimally small perturbation over the initial state such that $\tilde{\mathbf{x}}(t_0) = \mathbf{x}(t_0) + \delta \mathbf{x}$. If, for a given set of inputs $\mathbf{u}(t)$ the networks were poised in a dynamically-stable regime, one would expect that $\exists k > 0 : \mathbf{x}(t'_k) = \tilde{\mathbf{x}}(t'_k) \ \forall k' > k$, meaning that both trajectories eventually converge over the phase space. On the other hand, if the reservoirs show an unstable or chaotic dynamics, we would expect both trajectories to further diverge in time. The MLE is a typical measure of the system sensitivity to such perturbations (and, by extension, a way of quantifying chaos), which is defined as:

$$\text{MLE} := \lim_{k \to \infty} \frac{1}{k} log \left( \frac{\gamma_k}{\gamma_0} \right) \ , \qquad (3.20)$$

where $\gamma_0 \equiv \|\tilde{\mathbf{x}}(t_0) - \mathbf{x}(t_0)\|$ is the initial distance between the perturbed and unperturbed trajectories, and $\gamma_k \equiv \|\tilde{\mathbf{x}}(t_k) - \mathbf{x}(t_k)\|$ is the distance between the trajectories at time-step $k$ (Boedecker et al. 2011; Sprott 2003). Thus, a positive value of the MLE is characteristic of chaotic dynamics, whereas the system is said to be stable to local perturbations provided MLE $< 0$.

Moreover, for the case of synaptic plasticity, we will also track the plasticity-induced changes in the spectral radius, $\rho$, of the reservoir matrix, given the discussed relevance of this parameter in determining the existence of the echo state property (see Section 3.1.2).

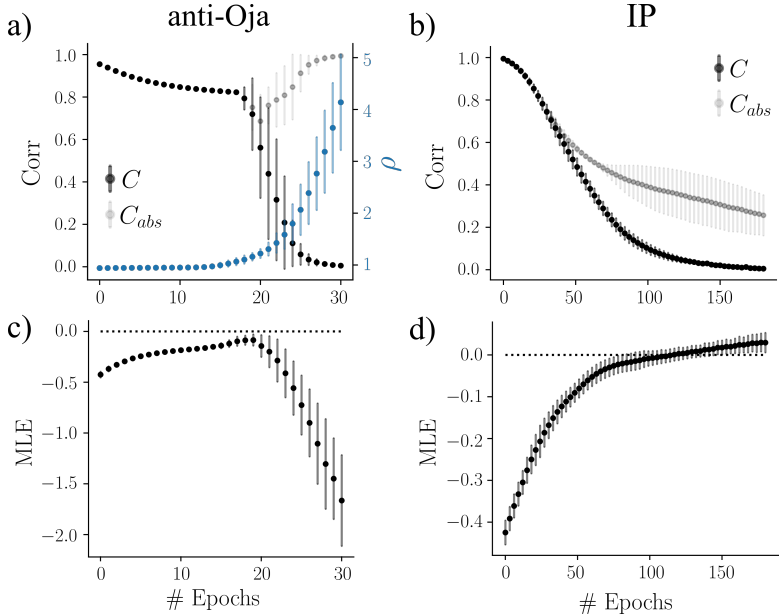

Figure 3.5: **Synaptic and intrinsic learning rules show different types of transitions.** **(a)** Evolution of the average Pearson's correlation coefficient (black curve) and average *absolute* Pearson's correlation coefficient (gray curve) between reservoir states, together with the reservoir connectivity spectral radius (blue curve) against the number of epochs during anti-Oja training. **(b)** Same as in (a) for reservoirs trained using the IP rule. **(c)** Evolution of the MLE for the anti-Oja rule. **(d)** Same as in (c) for reservoirs trained using the IP rule. For each epoch, error bars are computed as the standard deviation over 20 independent realizations of an ESN.

Fig. 3.5 shows how, for the initial stages of the plasticity learning phase, both the anti-Oja and IP rules induce decorrelations inside the reservoir, as reflected by a decrease on the average Pearson's correlation coefficient. Nevertheless, while this decrease happens in a continuous, smooth manner for the intrinsic plasticity (Fig. 3.5b), the anti-Oja rule shows a sharp transition from a high-correlated to low-correlated state happening at a critical number of epochs $n_{crit}^{AO} \sim 20$, which is concomitant with a steep increase in the spectral radius (Fig. 3.5a). Interestingly, while the average correlation goes to zero in both rules, the average *absolute* correlation between unit pairs increases past $n_{crit}^{AO}$ in the synaptic plasticity rule, coinciding with the drop in forecasting performance observed in Fig. 3.4a. We will get a better understanding of this behavior when looking at the single-unit activity in the next section, but for now, let us stress that this difference in the nature of the transition for the anti-Oja and IP rules can be further highlighted by measures of the maximum Lyapunov exponent (MLE).

In the synaptic case, the MLE increases towards zero during the first stage of the learning and then, coinciding with the sharp drop in correla-tions, decreases again at a much faster rate (Fig. 3.5c). However, the fact that the MLE remains negative across all epochs of the learning, shows that the transition is not of the order-to-chaos type. On the contrary, excitability changes due to intrinsic plasticity can drive the network from an originally stable regime to a chaotic regime with MLE $> 0$ at a critical number of epochs $n_{crit}^{IP} \sim 100$ (Fig. 3.5d), effectively inducing the loss of the echo state property and, consequently, the loss of any forecasting capability of the network (Fig. 3.4b). Notably, the results presented here for the IP rule seem to agree with the observations in Boedecker et al. 2011, where it was suggested that information transfer and storage in ESNs are maximized at the edge between a stable and a chaotic dynamical regime. In our case, however, two different behaviors on the error performance emerge across IP training epochs (Fig. 3.4b). A first deterioration in performance takes place after $\sim 50$ epochs, when the reservoir is approaching the edge of instability from the stable regime (MLE $< 0$). At this point, the network is still able to predict part of the test series, and the error remains bounded. Then, a second steep increase of the RMSE follows when the network crosses the boundary MLE $= 0$. Past this limit, the outputs generated by the network can span values outside of the support of the original input, effectively translating into a potentially diverging error.

Since a poor performance is usually associated to the loss of the echo state property, we will now take a look at the evolution of the "memory" of the networks at different stages of the training. The task of memory capacity (MC), as introduced in Jaeger 2001a, is based on the network's ability to retrieve past information from the reservoir using the linear combinations of reservoir unit activations. In particular, given $T$ points of a single-variable random training series, $u(t-d)$, drawn from a uniform probability distribution in the interval $[-1, 1]$, the goal is to generate $L$-dimensional outputs $\mathbf{y}(t)$, such that, at each time $t$, the element $d$ of the output vector is a reconstruction, $\tilde{u}(t-d)$, of the input $d$ steps *before* the current time:

$$y_d(t) = \sum_{k=1}^{N} w_{d,k}^{out} x_k(t) = \tilde{u}(t-d) \ . \tag{3.21}$$

For that, each of the $L$ output units is independently trained to approximate past inputs with a different value of $d$. To assess the ability of each ESN model in restoring previous inputs fed into the network, we compute the (short-term) MC as introduced by Jaeger in Jaeger 2001a:

$$\text{MC} = \sum_{d=1}^{\infty} \text{MC}_d = \sum_{d=1}^{\infty} \left( \frac{\langle (u(t-d) - \overline{u})(y_d(t) - \overline{y}_d) \rangle}{\sigma_u \cdot \sigma_{y_d}} \right)^2 \ , \tag{3.22}$$

where $\text{MC}_d$ is just the squared Pearson's correlation coefficient between the actual input $d$ steps before, $u(t-d)$, and its reconstruction at time

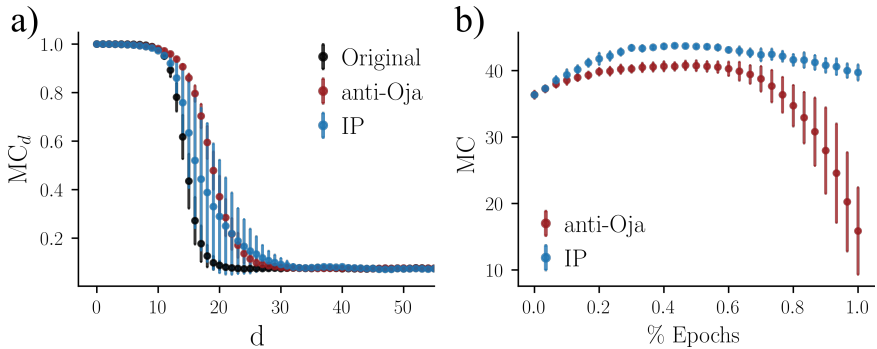

Figure 3.6: **Memory capacity can be enhanced through plastic learning. (a)** Memory curves, $MC_d$, as a function of the delay $d$ for an ESN with no plastic training (black), and the same network after been optimally trained with the anti-Oja (red) and IP rules (blue). **(b)** Evolution of the average memory capacity of the network with the number of learning epochs (normalized, for visualization purposes, by the maximum number of epochs in each rule). Errors in all cases are computed as the standard deviation over 20 independent realizations of the networks.

$t$: $y_d(t) = \tilde{u}(t - d)$. Thus, a value $MC_d \sim 1$ typically means that the system is able to accurately reconstruct the input fed to the network $d$ steps ago. In consequence, the sum of all $MC_d$ values can be understood as an estimation of the number of past inputs the ESN is able to recall. Although the sum runs to infinity in the original definition— accounting for the complete past of the input— in practice it was shown in Jaeger 2001a that a theoretical limit for the memory capacity on a reservoir of $N$ neurons is given by $MC_{max} \approx N - 1$, so we can set an upper threshold on the sum given by $d_{max} = N$.

Fig. 3.6a shows the memory curves for the original (i.e., non-plastic) ESN and for two networks trained with the optimal amount of epochs using the anti-Oja and IP rules. We notice how models with implemented plasticity outperform the original ESN, which shows a faster decaying memory. These results are in agreement with the average values presented in Boedecker et al. 2011, where the maximum memory was observed at the edge of instability for a random recurrent neural network. Moreover, we can see how during plastic learning, the average memory capacity increases up to a certain critical number of epochs for both types of rules (Fig. 3.6b). Nevertheless, while in the case of the anti-Oja rule the ability of the network to retrieve past inputs drops drastically around the critical number of epochs $n_{crit}^{AO} \sim 20$, for the IP rule the memory capacity remains fairly high even after plasticity has driven the reservoir into the chaotic regime.

### 3.5.2 Changes at the single-unit level

So far, we have focused on understanding the effects of plasticity at the network level, but nothing has been said about the way each individual unit "sees" or "reacts" to the input after implementation of the plastic rules. To shed some light into this question, we define the effective input of a unit $n$ at time $t$ as the presynaptic input arriving at unit $n$ once it has been filtered through the input mask, $u_n^{eff}(t) = \sum_{j=1}^{N_u} w_{nj}^{input} \cdot u_j(t)$. In this way, Eq. (3.1) for the state update of a single unit can be rewritten as:

$$x_n(t+1) = tanh\left(a_n\left(\varepsilon u_n^{eff}(t) + \rho\sum_j w_{nj}^{res}x_j(t)\right) + b_n\right) \ , \qquad (3.23)$$

where we also included the possibility of having gain, $a_n$, and bias, $b_n$, terms for the case or learning through the IP rule. In Fig. 3.7a we plot the response of two different units to this effective input before (black dots) and after the implementation of the anti-Oja and IP rules (red and blue dots, respectively) for an optimal, close-to-critical number of epochs in each case.

It can be clearly seen that plasticity has the effect of widening the activity range of the units, specially on those areas —as highlighted in yellow— in which the same point may lead to very different continuations of the series depending on its past history (see Fig. 3.7c, showing $\sim 500$ points of the effective input $u_n^{eff}(t)$ arriving to one of such neurons). In the case of the IP rule, a further displacement of their activity towards the center of the nonlinearity was observed, which should come as no surprise since we chose a zero-mean Gaussian as our target distribution.

As an end to the results presented in this chapter, we applied the same neuron-level framework to see if we can understand the effects of an overtrained anti-Oja and IP plasticity, using a total of $n_{AO} = 25$ and $n_{IP} = 175$ epochs, respectively. From the resulting plot of the activity as a function of the effective input, as shown in Fig. 3.7b, two different paths leading to the deterioration of performance can be identified. On the one hand, overtraining with the IP rule drives the network states to a seemingly blurred phase space representation at each unit of the reservoir, in which every effective input value can lead to very different responses of the same unit, with all units behaving similarly. On the other hand, an excess of synaptic learning through the anti-Oja rule induces the split of the original phase space representation into two disjoint regions. We observed that the instability in this case is associated with a self-sustained periodic dynamics of the reservoir states, leading to consecutive jumps from one region of the phase space to the other.

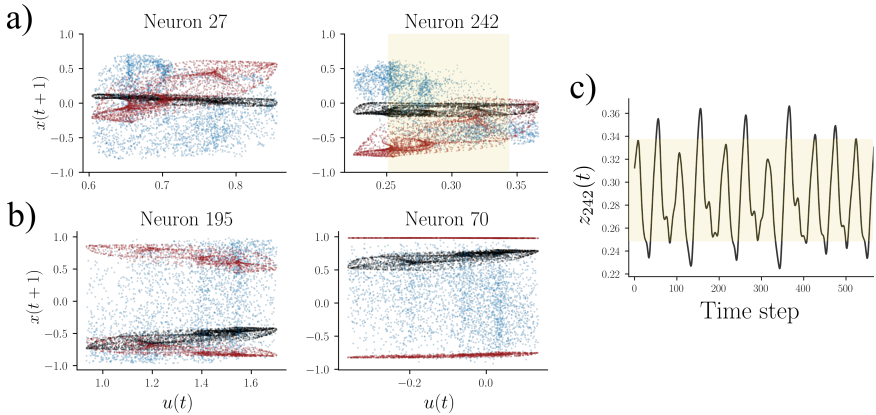

Figure 3.7: **Plasticity leads to an expansion in the units phase space. (a)** Activity of 2 different neurons as a function of the effective input in a non-plastic ESN (black) and in the same reservoir after training it with the anti-Oja rule (red) and the IP rule (blue) for an optimal number of epochs. **(b)** Same as in (a), but within a regime of overtrained plasticity. **(c)** Evolution of the effective input over a section of the training. We highlight in yellow the range of inputs for which the activity broadens more notably with respect to the non-plastic case, coinciding with the most variable part of the series.

## 3.6 Conclusions and perspectives

In this chapter, we showed how numerical implementation of unsupervised synaptic and non-synaptic plasticity rules can improve the performance of Echo State Networks (ESNs) in a chaotic-time-series forecasting task. Both anti-Oja and intrinsic plasticity (IP) rules induced decorrelations inside the reservoir states, effectively expanding the regions spanned by individual units in their activity phase space. In Morales et al. 2021a, the fact that mechanisms apparently so disparate exhibited similar effects at the neuron and network level motivated the idea of synergistic learning involving both, synaptic and nonsynaptic plasticity, a phenomenon that has been extensively backed up also in biological systems (Hanse 2008; Mozzachiodi et al. 2010).

At the network level, we showed for the anti-Oja rule how the sudden increase in the reservoir weight matrix spectral radius co-occurred with a sharp drop on the states pairwise correlations. More concretely, we observed that the optimal number of epochs happened just before the transition to a periodic self-sustained dynamics inside the reservoir, Similarly, continuous application of the IP rule also tended to decorrelate the states of the neurons within the reservoir, but the transition to a low-performance regime was of different nature: in this case the reservoir enters into an unstable regime, characterized by a positive value of the maximum Lyapunov exponent (MLE). This high-to-low performance transition with the number of epochs was also evinced by looking at the ability of the network to retrieve past inputs on a memory capacity task, showing that both types of plasticity rules can increase the memory of reservoirs up to a critical number of epochs, prior to the loss of the echo state property.

Interesting results emerged also by looking at the single-unit level before and after the implementation of plasticity rules. In particular, we saw how plasticity rules expanded each neuron activity space, adapting to the properties of the input and thus enhancing the dynamical range of the whole network. We used this framework to further analyze the differences in the regime of low performance for the anti-Oja and IP rules. In the synaptic case, once we reach the critical number of plasticity training epochs, the phase space region spanned by the activity of each single neuron splits into two disjoint regions, with states jumping from one region to the other at consecutive time steps as the overall dynamics falls into a stable fixed point. In the IP rule, on the other hand, we found that decorrelation of the states and expansion of their phase space continues progressively up to a point in which units loose their selectivity to particular parts of the input, becoming instead reactive to all inputs in a manner highly-dependent of the reservoir history.

Our findings also rise interesting questions that will hopefully stimulate future works. The computational paradigm of reservoir computing has been shown to be compatible with the implementation constraints of hardware systems (Van der Sande et al. 2017; Tanaka et al. 2019; Cramer et al. 2020). The finding that a physical substrate with non-optimized conditions can be used for computation has been exploited in the context of electronic and photonic implementations of reservoir computing (Appeltant et al. 2011; Brunner et al. 2013). Thus, the results presented in this chapter anticipate a potential advantage of considering such plasticity rules in physical systems, a thrilling line of research that has just set off with the first implementation of plasticity within a neuromorphic chip (Cramer et al. 2020).

# A representation of reality

Para llorar, dirija la imaginación hacia usted mismo, y si esto le resulta imposible por haber contraído el hábito de creer en el mundo exterior, piense en un pato cubierto de hormigas o en esos golfos del estrecho de Magallanes en los que no entra nadie, nunca.

JULIO CORTÁZAR, *Historias de Cronopios y Famas*

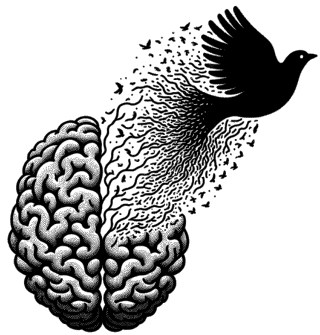

## Black Swan

And then we saw it.
On the very far side of a Bell's curve,
right where nights are born dawnless
and late buses overflow with time,
as countries unwound their borders against
flocks of purple martins,
and the sound of pines trimmed its wavelength,
shorter and shorter,
now a ripple carrying the color of the ocean.
We saw it.
The nights and flocks, the smell of pines,
                                        the waves.

And we knew it was real,
'cause we both saw it with our eyes,
although our eyes were never open.

# 4.1 Introduction: the problem of perception

For centuries, the epistemological problem of perception has been a fertile ground that attracted many of the greatest philosophers throughout history, giving rise to theories about whether reality exists outside of our minds, and if so, whether we can rely upon our senses to attain an objective observation of it. Skeptics, inspired by Descartes arguments, maintain that, since our only access to the external world is mediated by potentially misleading perceptual appearances, so the "veil of perception" refrains us from any possible knowledge of an external reality. Paraphrasing the words of David Hume: nothing is ever directly present to the mind in perception except perceptual appearances (Lyons 2023). Following this perspective, there is no reason to believe (nor disbelieve) in the existence of an objective external reality. Even if such a reality existed, the "veil of perception" would keep us from having any knowledge of it.

As a response to the skeptical argument, idealists like Kant would make a distinction between the *phenomenal* objects of perception —which are collections of appearances— and *noumenal* objects —which are things in themselves, with an existence that is independent of human perception. Thus, idealists apply Descartes' skepticism of the *noumenal* world, which is unknowable to us, but they can close the epistemological gap arguing that the *observable* world is, in fact, made out of the perceptual constructs in our minds, for which we can know something.

On the opposite side of the spectrum sit realists, for whom the perceptual constructs are, at least to some extent, representations of an existing physical reality. Furthermore, realists would argue that our sensory systems have evolved to capture meaningful aspects of the world, providing us with accurate information about it. Regardless of whether they assume that conscious sensory experiences have an almost 1:1 correspondence with external reality (naive or direct realism), or they deny direct-world involvement in perception —with the later taking place through the lens of a conceptual framework (representational or indirect realism)— all realists would agree on the fact that the *real* world consists of noumenal objects, with a physical existence (Lyons 2023). As we will see, the ideas presented in the following chapter lie at the basis of what I like to denote as a *two-fold realism*, in which experience with the external, *noumenal* world can be seen as taking place across two different phases.

The first part of experience, which I will call the **representational phase**, involves the direct translation of incoming stimuli into patterns of spiking activity across thousands or millions of neurons within the sensory regions. Given the degree of complexity of neural networks, it is within reason to assume that no two sensory inputs can elicit the exact same re-sponse in the brain (in fact, as we will see in Chapter 5, one fixed stimulus can evoke very different responses across time). Therefore, this first leg of the journey can be understood as a softened version of naive realism, in which sensory representations (but not yet perceptions) have an almost 1:1 correspondence with external reality.

The second part, denoted in what follows as the **perceptual phase**, takes place as higher cortical areas read-out the information in sensory regions, evoking what we understand as conscious experiences of the external world. These are indeed perceptual constructs of our minds in response to a physical reality (indirect realism), and —as any skeptical would claim— they cannot be proven to convey faithful information about it.

Nevertheless, I will argue that actual knowledge about the external world can be extracted from the physical responses of neurons during this representational phase of sensory experience. To do so, we first need to transform the classical *epistemological problem of perception* —that is, the problem of acquiring objective knowledge from reality by means of arising mental experiences— into the *ontological problem of representation*: how can the external reality be encoded into a language of millions of firing neurons?

To get some more intuition into the concept of neural representations, let us imagine we have recorded the activity of a set of $N$ neurons that we think are involved in the processing of a certain type of input (for instance, pyramidal neurons in the olfactory cortex responding to the presence of odor stimuli). One can then interpret the average response or firing rate of every neuron to one of such stimuli as a point in the $N$-dimensional phase space spanning all possible population states, which we will refer to as the *neural space*. Following this line of reasoning, a set of $T$ stimuli would then be mapped into $T$ points of the neural space (see Fig. 4.1). However, one would expect that correlations among neurons generate some redundancy in the encoding; i.e., there are many neurons that share the same type of response or tuning to a set of stimuli. Thus, population responses are not expected to be randomly distributed across the whole $N$-dimensional space, but to occupy a space of dimension $D \ll N$, which we will refer to as the *representation manifold* (Chung et al. 2021; Ganguli et al. 2012). In more mathematical terms, one can think of a representation as an isomorphism $f : \mathcal{U} \longrightarrow \mathcal{X}$ between the space of all possible inputs $\mathcal{U}$ to the space of all possible neural population states $\mathcal{X} \in \mathbb{R}^N$ .

In this chapter we will be looking at the mathematical properties that characterize the "optimality" of such internal representation, eventually unveiling the existence of a fascinating relation between the geometry of these neural manifolds and the dynamical regime in which the underlying neural network operates. To derive these results we will resort to the Echo State Network (ESN) (Jaeger 2001c; Lukoševičius 2012) framework already in-troduced in Chapter 3, drawing a parallel with the experimental evidence found by Stringer et al. for the representation of visual stimuli in mouse V1 cortex (Stringer et al. 2019a). The next section presents a brief overview on this seminal work, which will provide us with a link between the geometrical properties of neural manifolds and the cooperative effects emerging from neuronal dynamics. At the same time, we will also set up the necessary tools to tackle the representation problem from the lens of Reservoir Computing.

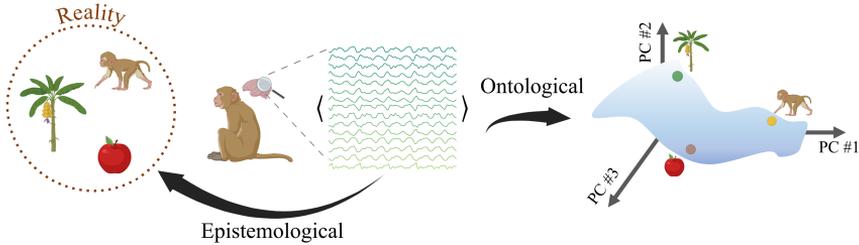

Figure 4.1: **A new arrow into the problem of perception.** In the brain, the "real", observable world is translated into patterns of spiking activity in sensory regions. The average of such responses to a given stimulus can be understood as a *neural representation* of the former in some high-dimensional manifold over the activity space. While the question of whether one can gain any knowledge about external reality from the study of the evoked neural activity is an epistemological one, the *physical* process of encoding reality into neural representations can be thought of as an ontological problem.

## 4.2 The tools of the trade

### 4.2.1 A smooth representation of external stimuli

Going back to the manifold picture as depicted in Fig. 4.1, one can intuitively hypothesize that the robustness of the response to a particular input will be likely conditioned by the "smoothness" of the neural manifold: if we want similar inputs to be encoded by similar patterns of activity, then small perturbations $d\mathbf{u}$ in the input space should account for small changes $df(\mathbf{u})$ in the representational space. Intuitively, this amounts to $f$ being continuous and differentiable on the set of all possible inputs, or, in other words, the representation manifolds should be continuous and differentiable (henceforth "smooth"). Indeed, within the field of ML, it has been shown that deep neural networks trained to classify images are more susceptible to *adversarial attacks* (subtly designed changes in some of the images pixels aiming at hindering the network's ability to classify images) when their internal representations of the input do not live in a smooth manifold (Madry et al. 2019; Nassar et al. 2020).

Remarkably, the authors in Stringer et al. 2019a presented a mathemati-cal proof that conditioned the smoothness of such manifolds to the spectrum of the covariance matrix of neuronal responses to a set of stimulus.  More specifically, let $X \in \mathbb{R}^{N \times M}$ be the matrix containing the average response (firing rate) of $N$ recorded neurons to a set of $M$ stimuli (see Fig. 4.2), corresponding to $M$ points in the neural space. Then, in the limit of $N, M \to \infty$, it is possible to derive a necessary condition for the smoothness of the manifold spanned by the $M$ population responses: the eigenvalues of the signal covariance matrix $C_X$, when ranked from the largest to the smallest, should decay with their rank according to a power law (i.e., $\lambda_n \sim n^{-\mu}$) with an exponent

$$\mu \geq 1 + 2/d , \tag{4.1}$$

where $d$ represents the embedding dimension of the inputs, and can be understood in the language of Principal Component Analysis (PCA) as the number of independent principal components needed to capture most of the variance of the stimuli.  In particular, the representation manifold is continuous (but not differentiable) if $1 + 2/d > \mu \geq 1$, and for $\mu < 1$ it cannot be continuous either.  Notice how, in any case, $\mu = 1$ serves as a lower bound of the exponent for complex, high-dimensional inputs, such as natural images.  Details of this proof are not presented here, but the interested reader is referred to Stringer et al. 2019a for an in-depth explanation of this smoothness-to-spectrum relation and its implications.

Remarkably, the above theoretical predictions were validated in experimental recordings of over 10000 individual neurons in the mouse visual cortex while the animals were exposed to a large sequence of images (Stringer et al. 2019a).  The authors showed that the brain is capable of generating internal representations which are optimal from an information-encoding point of view: they are as high-dimensional as possible (i.e., correlations show a slow-decaying spectrum, meaning that there are many relevant principal components capturing the signal variance of the representations), while respecting the aforementioned boundary for continuity and differentiability. In particular, for the encoding of natural images, this translates into measured exponents for the covariance eigenvalue spectrum $\mu \sim 1$ (see Fig. 4.2), but the same phenomenology (with $\mu \sim 1 + 2/d$) was observed when the original inputs were projected into lower-dimensional spaces before being presented to the mice.

We remark once again that the condition expressed by Eq. (4.1) is a property of the input-related or *signal* variance. However, when one measures the *empirical* covariance matrix $\hat{C}_X = \frac{1}{M-1} X X^T$ across a set of $M$ stimuli, the variance associated to each eigenvector (i.e., each principal component) will be a mixture of input-related and *noise* variance. Thus, before delving any further into the properties of neural manifolds, we first need a method that allows us to separate the directions of input-related variability

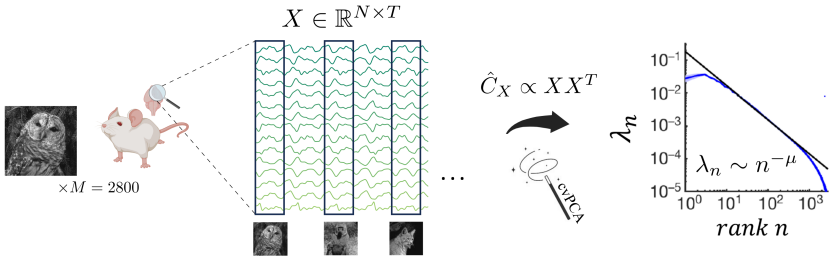

Figure 4.2: **Stringer *et al.*'s experimental setup.** A set of $M = 2800$ natural images where shown to mice while recording the activity of over 10,000 neurons in the V1 cortex of the animals. Responses to individual images were averaged over a time-window of 0.5s. The covariance matrix was then computed and filtered through a cross-validated Principal Component Analysis (cvPCA) to extract the signal variance, which showed a powerlaw decay with an exponent $\mu \sim 1$.

—which carry information about the represented external stimuli— from the directions of trial-to-trial variability (which, for practical purposes, we will identify here with noise, although it could emerge from the encoding of other unknown latent signals). Introducing such a method, known as cross-validated Principal Component Analysis (cvPCA), will be the scope of next section.

### 4.2.2 Disentangling input-related and background activity

When trying to make sense of high-dimensional data, as in the case of recordings of thousands of neurons, a common and very practical tool to reduce the dimensionality of the problem and find directions of interest is Principal Component Analysis (PCA). Mathematically speaking, given a matrix of observations $X \in \mathbb{R}^{N \times M}$, where $N$ is the number of variables and $M$ the number of samples or observations, PCA seeks to find a transformation or change of basis, $P$, such that $Y = PX$ is a better representation of the data, meaning that: i) the redundancy between the variables (i.e., the covariance) is minimized, and ii) the signal-to-noise ratio is maximized, this is, $P$ projects the original data into the directions of maximal variance. PCA finds a particularly elegant solution: in the new basis, the covariance matrix $C_Y$ must be diagonal, so the projector $P$ is just the matrix of eigenvectors of $C_X$ arranged in columns (Shlens 2014).

As mentioned in the previous section, if PCA was applied directly over the experimental recordings for the mouse V1 cortex in Stringer et al. 2019a, one would indeed find the directions of maximal variability of the neural responses, but nothing could be said about the origin of this *overall* variance. In fact, it is estimated that half of the variance of the visual-cortex activity is unrelated to stimulus-encoding activity (Stringer et al. 2019b). How could we then tell apart the variance steaming from the stimulus encoding (*signal* variance), from the intrinsic or trial-to-trial variability (*noise* variance)?

Stringer *et al.*. showed that the stimulus-related variance can indeed be extracted from empirical data by measuring the amount of shared variance between the neural responses to a repeated presentation of the same set of stimuli (Stringer et al. 2019a). We leave the mathematical details of this method, known as cross-validated Principal Component Analysis (cvPCA), to Appendix A.III.

Moreover, since cvPCA provides us with not only the eigenvalues (variance), but also the associated eigenvectors, we showed that the method can be easily extended to extract also the noise variance, assuming that signal and noise components span orthogonal sub-spaces within the $N$-dimensional neural space (Morales et al. 2023). One can then project the original activity into an input-related subspace $\Psi$, independent of the trial, and an orthogonal, trial-dependent subspace $\Sigma_k$, such that on trial $k$:

$$X_k = \Psi + \Sigma_k . \qquad (4.2)$$

Now that we know how to disentangle the input-related activity from the background fluctuations, we can begin to investigate how the dynamical properties of the neural network shape the geometry of the internal neural representations. To do so, we first introduce a suitable algorithm that can train the ESNs on an image-classification task.

### 4.2.3 Training protocol for an image-classification task

In order to adapt the ESN architecture —usually employed in time-series analyses— for an image classification task, we converted each input black and white image (consisting of $L_1 \times L_2$ pixels with a value in the $[0, 1]$ interval, representing a normalized gray-scale) into a multivariate time series where each image columns is a vector of $L_1$ elements or features that "evolve" along $T = L_2$ discrete "time" steps. One can then define a standard training protocol (Bianchi et al. 2021) in which, as illustrated in Fig. 4.3, at each time $t \in [0, T]$, vectors $\mathbf{u}(t) \in [0, 1]^{L_1}$ corresponding to columns of the image are fed as inputs to the ESN.

Using a supervised learning scheme, the goal of the ESN is to transform (map) the internal representations of the input into an output label $\mathbf{y} \in \mathbb{N}^F$ that correctly classifies each image in the test set as belonging to one of the $F$ existing categories or classes. This label consists of a vector in which every element is zero except for a value of one at the position corresponding to the assigned class (i.e., "one-hot-encoded" in the machine learning jargon). Several readout methods have been proposed in the literature to transform the information contained in the reservoir dynamics into the expected target output $\mathbf{y}^{target} \in \mathbb{N}^{\mathbb{F}}$, ranging from linear regressions methods over the reservoir states (Reinhart et al. 2010; Reinhart et al. 2011), to the use of Support Vector Machines (SVMs) or multilayer perceptrons as decoders (Babinec et al. 2007). Here, as with the timeseries prediction task, we will resort to a simple linear regression, but instead of using the activities of the units inside the reservoir, the readout will be carried over the *reservoir model space*, a method that has been recently proposed for the classification of multivariate-time series in Bianchi et al. 2021:

$$\mathbf{y} = W_{out}\theta_x + \mathbf{b}_{out} \; , \tag{4.3}$$

where $W_{out} \in \mathbb{R}^{F \times N(N+1)}$ and $\mathbf{b}_{out} \in \mathbb{R}^F$, defined as the output weights and biases, are determined through a ridge regression that minimizes the error between the produced and target label for all the presented images in the training set. In the above expression, $\theta_x = [\text{vec}(W_x); \mathbf{b}_x]$ is the reservoir model space, a set of parameters encoding the reservoir *dynamical* state for a given input (image), obtained from a linear regression to predict the next network state from the past one at discrete time steps:

$$\mathbf{x}(t+1) = W_x \mathbf{x}(t) + \mathbf{b}_x. \tag{4.4}$$

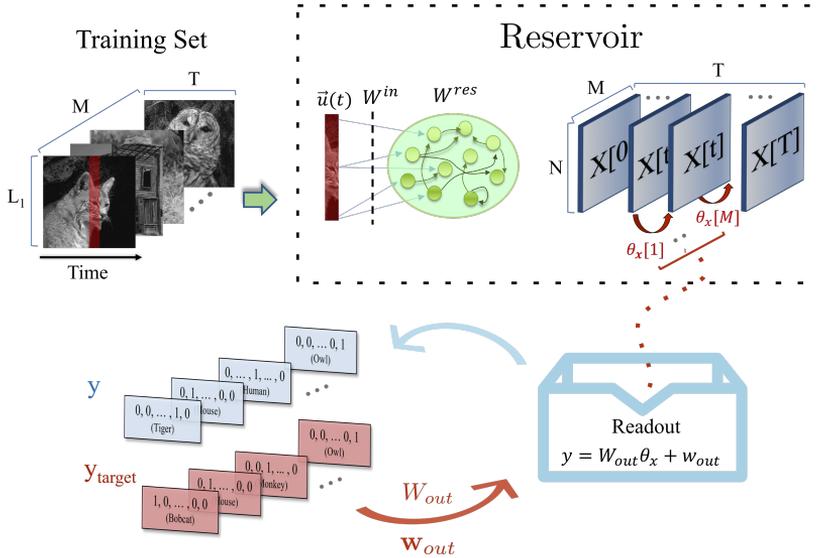

Figure 4.3: **Training process of an ESN for an image classification task.** Images are converted to multivariate time series and then fed into the reservoir. For each processed image a set of parameters $\theta_x$ is generated, which characterizes the high-dimensional state of the reservoir, i.e. the "*reservoir model space*". Those parameters are then fed into the readout module, that linearly transforms the information in the reservoir model space into an output label. Finally, output weights $W_{out}$ and biases $\mathbf{b}_{out}$ are generated by minimizing the error between the predicted and target labels. Red arrows indicate steps in which a ridge regression is performed.

In the next section, we will follow this protocol to mimic the experimental setup proposed in Stringer et al. 2019a using an ESN trained in an image classification task. Once trained, our ESN will be able to decode the identity of the presented images from the internal representations of such inputs inside the reservoir.

## 4.3 A three-sided coin: smooth representations and optimal performance meet at criticality

Since our goal is to mimic in our ML framework the experimental observations of Stringer *et al.*, there are certain questions that one can naturally ask even before delving into the simulations:

$\mathcal{A}$. Are ESNs able to generate "optimal" internal representations of $d$-dimensional inputs, which are as high-dimensional as possible while respecting the smoothness boundary given by $\mu > 1 + 2/d$?

$\mathcal{B}$. If so, what characterizes the dynamical regime in which these optimal representations emerge?

$\mathcal{C}$. Does the dimensionality and smoothness of the internal representations affects the network's ability to classify the input images?

In order to answer questions $\mathcal{A}$ and $\mathcal{B}$, we first need a way of tuning the dynamics of the ESNs. Recall from Chapter 3 that the evolution of the activity for the reservoir units follow a discrete-time equation that can be written as:

$$\mathbf{x}(t) = \tanh(\varepsilon W^{in}\mathbf{u}(t) + \rho W^{res}\mathbf{x}(t-1)), \tag{4.5}$$

where the values of $W^{in}$ are randomly drawn from a uniform distribution in the interval $[-1, 1]$, and $W^{res}$ is defined as a sparse matrix that has been rescaled so that it has unitary maximum eigenvalue (i.e., $W^{res} \equiv (1/\tilde{\rho})\tilde{W}^{res}$, where $\tilde{W}^{res}$ is a sparse matrix with maximum eigenvalue $\tilde{\rho}$ and elements drawn from a uniform distribution).

Thus, we will seek to analyze the reservoir input internal representations in terms of the trade-off between the two hyperparameters that drastically determines its dynamical regime and the existence of the echo state property: the *spectral radius* $\rho$, which controls the stability inside the reservoir when no input is fed into the network, and the *scaling factor* $\varepsilon$, which can turn an initially expanding mapping into a contracting dynamics, as stronger inputs tend to push the activities of the reservoir units towards the tails of the non-linearity (see Section 3.1.2). The number of units in the reservoir will be fixed to $N = 2000$ and the density of the reservoir weight-matrix elements (i.e., the percentage of non-zero connections) to 10%.

Having set the two knobs that tune the dynamical regime of the network, one just needs to define a good dial that serves as a control parameter to quantify how far is the network dynamics into the stable or unstable regimes (see Section 3.1.2). For this we will use the maximum Lyapunov exponent (MLE), as already introduced in Section 3.5, which serves as a measure of the system sensitivity to perturbations. We recall that negative values of the MLE characterize a system that is stable under local perturbations, whereas positive ones are associated to chaotic regimes in autonomous systems.[1]

Following the same methodology as in Stringer et al. 2019a, the ESN was first presented with a large set of high-dimensional, natural images, and the activity of the internal units in the reservoir was stored for each step of the training. Then, PCA was performed directly over the full set of neural activities $X \in \mathbb{R}^{N \times (T \times M)}$, where $T = 90$ is the number of pixels in the horizontal dimension of the images and $M = 2800$ is the total number of images. In this way, we obtained the variance along each principal component or eigenvector of the covariance matrix, which serves as a basis for the activity inside the reservoir. Notice that, because the dynamics of the system (as described by Eq. 4.5) is fully deterministic, all the variance in the reservoir states stems from input-related activity, and no cvPCA analysis is needed.

The surface in Fig. 4.4 shows the exponent of the covariance matrix eigenspectrum as a function of the hyperparameters $(\rho, \varepsilon)$ for an ESN presented with the same set of natural images as the ones used in Stringer et al. 2019a. As we can see, for a fixed value of the input scaling, $\varepsilon$, the spectrum of covariances decay slower as we increase the spectral radius, $\rho$. In other words, this means that the encoded representations become more high-dimensional as we increase the relative strength of the recurrent interactions with respect to the external input. Even more remarkably, notice how the boundary exponent for "smooth" high-dimensional input representations, $\mu \approx 1$, coincides almost exactly with the transition between the stable an unstable regimes (MLE = 0, color-coded). This means that the system is able to generate "optimal" representations of the input images, with properties akin to those observed in the visual cortex of mice, only when its dynamical regime is close to a critical point.

Although this emerging link between smooth, optimal neural representations of natural images and distance to criticality is quite a remarkable finding on itself, can we push it a bit further and test the validity of Eq. (4.1)

---

[1]Although this is obviously not the case in an ESN, for which the dynamics is input-driven, we will see that this quantity can still be informative as a control parameter for the underlying phase transition.

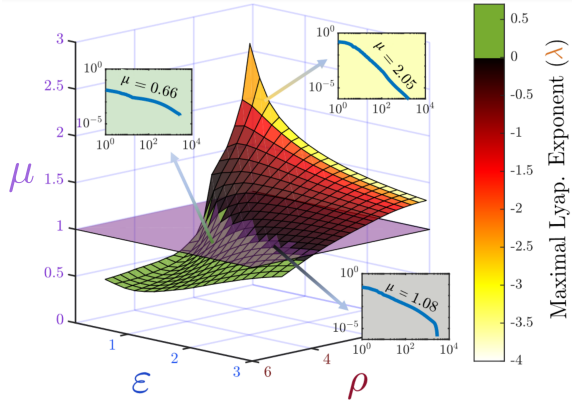

Figure 4.4: **Smooth, high-dimensional internal representations arise near the edge of instability.** Exponent for the powerlaw decay of the spectrum of the activity covariance matrix as a function of the spectral radius ($\rho$) and input scaling factor ($\varepsilon$) of the reservoir, plotted together with the maximum Lyapunov exponent (MLE) color-coded within the surface. The insets correspond to the activity covariance matrix eigenspectrum measured in three different points of the parameter space, where the variance in the $n$-th dimension scales as a powerlaw, $\lambda_n \sim n^{-\mu}$, of the rank. The purple plane marks the boundary $\mu = 1$ for smooth representations of high-dimensional inputs.

for inputs with a different embedding dimension $d$?

## 4.3.1 Matching visual cortex and ESNs

We will now be presenting the exact same sets of 8-dimensional and 4-dimensional images (constructed using a reduced-rank regression model from the natural ones, see Stringer et al. 2019a for more details) that were presented to mice in Stringer *et al.*'s paper. For each type of input, we tuned the network so it operated right before the onset of the unstable regime, i.e., for values in the parameter space $(\rho, \varepsilon)$ for which the MLE was near zero but still negative.

Fig. 4.5 shows the rank-ordered covariance eigenvalues of the reservoir states, together with the best powerlaw fit exponents, for natural and low-dimensional inputs. For the sake of comparison, we also reproduced in yellow the rank-ordered eigenvalue distributions obtained from the experimental data on mouse visual cortex (Stringer et al. 2019a). Remarkably, we found in all cases that the exponents observed in the mouse visual cortex activity could be well reproduced provided the reservoir dynamics was tuned close to the edge of instability. This finding suggests that one can set the network parameters in such a way that the neural activity manifold in which the input is represented is almost as high-dimensional as possible without loosing its "smoothness", and that such optimal solution is found near a phase transition.

At this point, it is probably pertinent to dig a bit deeper on the similarities and differences between the results presented in Stringer et al. 2019a for real, V1-cortex neurons in the mouse, and the powerlaw exponents obtained through our reservoir computing model.

1. First of all, as in the case of real neurons, the observed correlations between the internal units are not just a byproduct emerging from scale-free features of natural images (see second column in Fig. 4.5). In particular, one can see that the powerlaw decay of the eigenspectrum persists even in response to low-dimensional inputs whose embedding vector space can be spanned with just a few principal components (i.e. without a powerlaw decaying intrinsic spectrum).

2. In our model, images are processed sequentially in time along their horizontal dimension, so that for each image one can measure the activity of the $N = 2000$ internal units over $T = 90$ time steps. In contrast, activity of V1 neurons in Stringer et al. 2019a were averaged across the time span of each stimulus presentations (0.5 seconds).

3. The variance observed by Stringer *et al.* is not directly measured over the raw activity of the neurons, but filtered to extract only the input-related variance using the cvPCA method explained in detail in Appendix A.III. However, since our model is completely deterministic for a given initialization of an ESN, the stimulus-related variance computed through cvPCA trivially matches that of a standard PCA.

At the light of this last point, one could naturally wonder what would happen if we add a small additive noise term inside the activation function in Eq. (4.5), so that the dynamics now becomes stochastic:

$$\mathbf{x}(t) = \tanh(\varepsilon W^{in}\mathbf{u}(t) + \rho W^{res}\mathbf{x}(t-1) + \xi(t)) \ . \tag{4.6}$$

Are the powerlaw exponents robust to the introduction of such a noise? To answer this question, one can present this modified version of the original ESN with two repeats of the same input training set, collect at each time step the internal states of the noisy reservoir, and then apply a cvPCA analysis as proposed in Stringer et al. 2019a to filter for the signal variance. Once again, just as in the case of real V1 neurons, if we measure the covariance matrix over the raw, noisy activity, we find small exponents values that are below the critical threshold for continuity and differentiability of the neural manifold (Fig. 4.5, fourth column). Nevertheless, once a cvPCA has been performed over the internal states to filter out the noise correlations, we recover the expected exponents and their dependency with the dimension of the input (Fig. 4.5, fifth column).

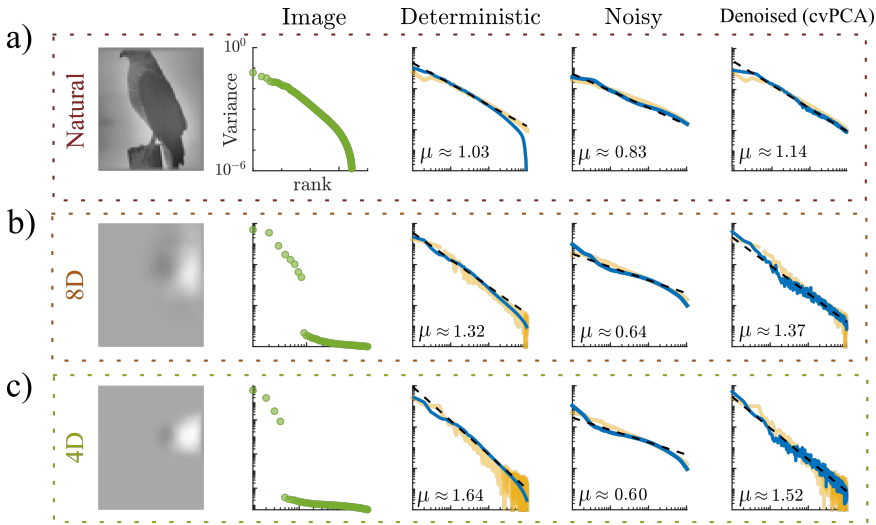

Figure 4.5: **ESNs near the edge of instability generate representations with the same eigenvalue distributions observed in real neurons.** From top to bottom: **(a)** results for natural, high-dimensional images; **(b)** results for 8-dimensional projections of the same set of natural images; **(c)** results for 4-dimensional projections of the same set of natural images. From left to right: one sample from the $M = 2800$ images in the training set; covariance eigenspectrum of the images pixel intensities; covariance eigenspectrum of the internal states of an ESN (blue line) and real, V1 mouse neurons (yellow line, plotted after Stringer et al. 2019a applying cvPCA) when subject to images of dimensionality $d$ ; same analysis, but now zero-mean white noise of amplitude $\epsilon = 0.4$ is added to the neuron dynamics (blue line), and no cvPCA is performed over the experimental recordings (yellow line); same analysis as in the previous panel, but now noise has been sustracted using cvPCA in both, simulations and experiments. To obtain the ESNs eigenspectra, parameters were chosen so that the networks operated near the edge of instability, with MLE $\sim -5 \cdot 10^{-3}$.

We will further comment on the possible implications of this result in the Section 4.4 at the end of the chapter, but for now, let us wrap up our findings tackling what we believe is a fundamental question from the perspective of machine learning: does working at the edge of instability, with its concomitant high-dimensional —yet smooth— neural representations, provide any functional advantage regarding the ability of the network to solve a given task?

### 4.3.2 Solving an image classification task

To assess the performance of ESNs, we relied on a classification task over the MNIST dataset —a common benchmark dataset consisting in 28x28 pixel pictures of hand-written digits (Deng 2012)—, following the protocol already introduced in Section 4.2.3. In particular, we trained the readout of the networks over one-third of the full MNIST training set (20,000 images), and then evaluated the classification error of the ESNs over the full test set (10,000 images).

Fig. 4.6 shows the percentage of digits that were correctly classified during the test as a function of the reservoir spectral radius for a fixed input scaling $\varepsilon = 0.6$, together with the MLE characterizing the dynamical regime of the network. The results highlight the fact that optimal performance ($\sim 2.2$ % error rate) is found just below the onset of chaos, when $\lambda \lesssim 0$. Most notably, the plot also evinces that the decay in performance is not only preceded by a positive MLE, but coincides too with exponents $\mu$ for the fit of the covariance matrix eigenspectrum that are below the limiting value $\mu_c \approx 1$, indicating a loss of continuity and differentiability of the neural representation manifold for high-dimensional inputs.

One could arguably say that using a reservoir computing approach, where images are processed in a time-sequential manner, is hardly comparable to the way in which visual stimuli are processed in the mouse V1 cortex. Nevertheless, we remark that Eq. (4.1) is a property of the *correlations* within the neurons when subject to stimuli with an embedding dimension $d$, but does not depend on the input structure, nor the observed powerlaw is inherited from the statistics of the natural images (see Fig. 4.5).

Should one then expect to find a similar match between edge-of-instability dynamics and optimal, smooth representations, for a different type of stimuli/inputs? In the next section we will present some interesting preliminary results that show how optimal internal representations can also arise in ESNs fed with complex time series.

### 4.3.3 The representation of a strange attractor

Since our purpose for this section is to assess the validity of Eq. (4.1) for the neural representations of input time-series, we will resort to the classical Lorenz dynamical system, described by the following set of differential equations:

$$\dot{x} = \sigma(y - x) \ , \tag{4.7}$$

$$\dot{y} = rx - y - xz \ , \tag{4.8}$$

$$\dot{z} = -bz + xy \ , \tag{4.9}$$

with $\sigma > 0$, $b > 0$ and $r > 0$. In particular, we will be using the original choice of parameters in Lorenz's paper ($\sigma = 10$, $r = 28$ and $b = 8/3$) for which the trajectories are known to generate an strange attractor with fractal structure (Strogatz 2000; Sprott 2003). This will be particularly useful, because the embedding dimension for the Lorenz system with this set of parameters has been already calculated, with a value $d_l \sim 2.1$ depending upon the algorithm being used (see Sprott 2003 for more details about the different definitions of attractor dimension).

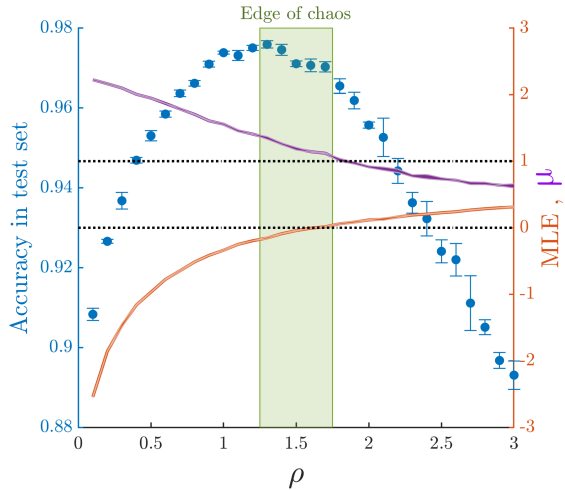

Figure 4.6: **Optimal performance and smooth representations meet near the edge of instability.** Accuracy in MNIST testset (blue dots), maximum Lyapunov exponent (orange line) and best fit exponent for the powerlaw spectrum of the activity covariance matrix (purple line). Errors were estimated as the standard deviation over ten different realizations of the ESN.

Therefore, in this section we will follow the training protocol presented in Section 3.2, using $\mathbf{u}(t) = \{x(t), y(t), z(t)\}$ as the input to the reservoir. More specifically, Lorenz equations were integrated (and sampled) with a time step $\Delta t = 0.005$ to retrieve $T_{train} = 10000$ points for the training set, then re-scaled so that all three variables of the time series lay within the $[0,1]$ range.

The reservoirs were constructed with $N = 800$ units, each connected on average to 10% of all existing units. Recurrent and input weights were drawn from a uniform distribution in the $[-1,1]$ range. As in the case of input images, the reservoir dynamical regime was modified by tuning the input scaling factor, $\varepsilon$, and the spectral radius of the internal connectivity matrix, $\rho$.

To see if there is a dependence between the smoothness of the representation manifolds and the dynamical regime of the network, we constructed ESNs with different sets of hyperparameters $\{\rho, \varepsilon\}$, then collected the internal states $\mathbf{x}(t)$ for each point of the input training set. Just as in Fig. 4.4 for the case of input images, we plotted in Fig. 4.7 the exponent $\mu$ for the powerlaw decay of the signal correlation eigenvalue spectrum against $\rho$ and $\varepsilon$, together with the MLE (color-coded) which will serve us as an order pa-rameter for the dynamical transition. Notice how, once again, the critical exponent $\mu_c = 1 + 2/d_l \approx 1.95$ (marked with a purple, continuous line) is reached right before the onset of the unstable regime.

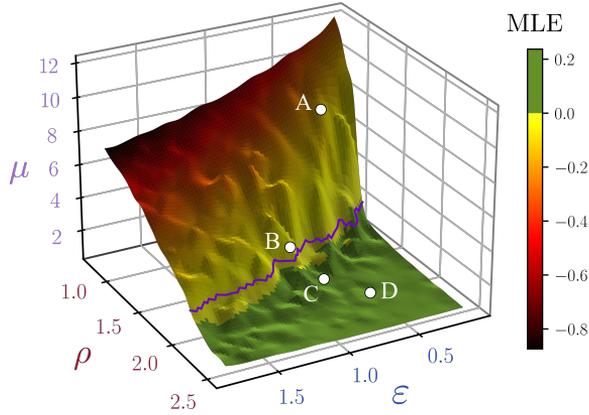

Figure 4.7: **Smooth, high-dimensional internal representations of time series arise near the edge of instability.** Exponent for the powerlaw decay of the spectrum of the activity covariance matrix as a function of the spectral radius ($\rho$) and input scaling factor ($\varepsilon$) of the reservoir, plotted together with the maximum Lyapunov exponent (MLE) color-coded within the surface. The purple continuous line marks the boundary $\mu = 1 + 2/d_l \approx 1.95$ for critically smooth representations of the Lorenz attractor.

Interestingly, using a time series with a low-dimensional attractor dynamics as input can provide us with a very illustrative picture of what it means to encode information in high-dimensional smooth manifolds. In Fig. 4.8 we plot the dynamics of the reservoir units (i.e., the internal representation of the input training set) projected into the first three principal directions, along with the eigenvalue spectrum for the signal covariance matrix for the four points A, B, C, and D marked in Fig. 4.7. We can see how, deep into the stable regime (point A, MLE $= -0.798$), the projected dynamics follow very closely the trajectories of the original Lorenz attractor, which means that the state of the system at each time step is fundamentally determined by its current input. As we approach the critical point (point B, MLE $= -0.0007$), the representations, when projected into the three directions of maximal variability, diverge considerably from the manifold spanned by Lorenz's attractor, but are still continuous and differentiable, as evinced by the smoothness of the curves. Close to the critical point but already into the unstable regime (point C, MLE $= 0.028$) the trajectories become ragged, reflecting the loss of differentiability in the manifold, as predicted by an exponent $1 < \mu < 1 + 2/d$. Finally, deep into the unstable regime (point D, MLE $= 0.122$) the internal representation becomes a "spiky" ball in a very high-dimensional space, that is not continuous nor differentiable.

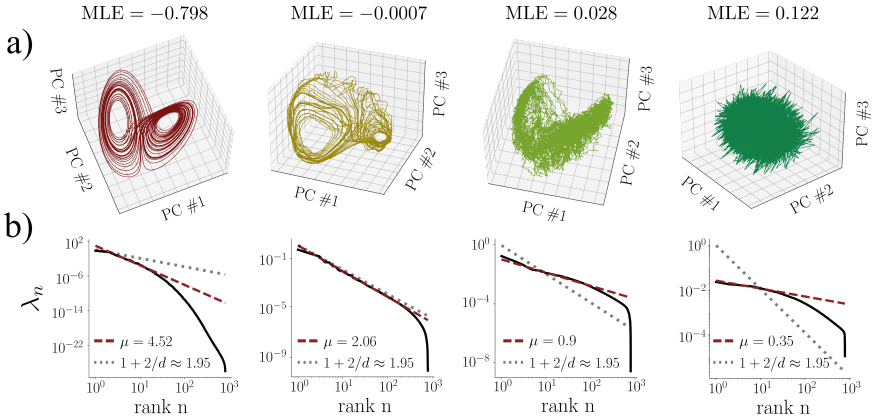

Figure 4.8: **Reservoir manifolds in PCA space. (a)** Projection of the reservoir activity during the training into its three first principal components. **(b)** Activity covariance matrix eigenspectrum, together with the best powerlaw fit (red dashed line) and the theoretical boundary for smoothness of the representation (gray dotted line). From left to right, columns correspond to points A, B, C and D in Fig. 4.7.

To further demonstrate that the representation in point B (i.e., close to the edge of instability) is able to capture finer details of the signal variance compared to the more "stable" encoding in point A, we projected the internal dynamics into the first 5 principal components and plotted the trajectories along each pair-wise combination of these directions, showing that lower-rank PCs encode relevant dynamics in the close-to-critical case, but account only for noisy fluctuations when the system dynamics is very stable (see Figs. D.1-D.4 in Appendix D).

Finally, in this last part of the chapter we will get back to the question of performance from a ML point of view, this time working on a time-series prediction task.

### 4.3.4 Solving a time-series prediction task

In this section, for each ESN characterized by a set of hyper-parameters $\{\varepsilon, \rho\}$, the network was trained on a one-step-ahead prediction task to infer points $\{x(t+1), y(t+1), z(t+1)\}$ of the Lorenz attractor from $\{x(t), y(t), z(t)\}$. For that purpose, the output weights were computed using a ridge regression over the responses of all units to the input training set, with regularization parameter $\beta = 10^{-7}$ (see Section 3.2). For testing, the ESN was asked to predict in autonomous mode (i.e., using the output at time $t$ as input in time $t + 1$) the next $T_{test} = 600$ points following the end of the training series. The performance was assessed using two different measures: the root mean-square error (RMSE) between the predicted and target series, as defined in Eq. (3.6), and the furthest predicted point (FPP), using a tolerance threshold of $\gamma = 0.02$ (see Section 3.2 in Chapter 3).

Fig. 4.9 shows maps akin to Fig. 4.7 for the RMSE (Fig. 4.9a) and FPP (Fig. 4.9b) over different dynamical regimes of the ESNs, as determined by the MLE (color coded). Once again, optimal performance (as given by a minimum of the RMSE and a maximum of the FPP) is achieved within the stable regime but right before the onset of instability. The results for a particular cross-section (marked with a white dashed line on the FPP-map) for a fixed value $\varepsilon = 0.73$ are plotted in Fig. 4.9c-d, where the error bars are computed as the standard deviation over 10 different of the ESN for a given set of parameters. Besides the MLE (Fig. 4.9e) and the fitted exponent $\mu$ for the covariance eigenvalues (Fig. 4.9f, blue curve), we also plot the participation dimension $D$ of the representational manifold (Fig. 4.9f, green curve), as defined by:

$$D = \frac{\left( \sum_i^N \lambda_i \right)^2}{\sum_i^N \lambda_i^2} \ , \qquad (4.10)$$

where $\lambda_i$ are the eigenvalues of the covariance matrix. Intuitively, if all $N$ units in the reservoir were independent of each other and had the same variance, then all eigenvalues would be equal and $D = N$ (Litwin-Kumar et al. 2017). Conversely, if the activity of the units is correlated so that responses are distributed equally in each direction of a $d$-dimensional subspace of the full $N$-dimensional neural space, then only $d$ eigenvalues will be nonzero and $D = d$. Fig. 4.9f (right) shows that not only optimal performance and smooth representations with $\mu \sim 1 + 2/d_l$ meet right before the critical point (green region, with MLE $\lesssim 0$ ), but also that the system undergoes a transition from a low-dimensional encoding (with $D \sim d_l$) to a high-dimensional encoding (with $D \gg d_l$) that is concomitant with the underlying dynamical transition.

Finally, to further visualize the predictive ability of the ESNs at different points of the hyperparameter space, we show in Fig. 4.10 the predicted and target multivariate time series in points A, B, C and D of Fig. 4.7, during a 1000-points prediction test. As we can see, the forecasting ability at either side of the critical region is considerably worse than that of an ESN trained in point C, whose dynamics present an average MLE very close to the critical value of zero. Interestingly, for a network trained in a strongly unstable regime (point D), the activity of the reservoir units can sometimes end up in regions of the phase space that are completely outside of the manifold spanned by typical responses to the input series, leading to predicted outputs that are not even contained within the training set (see, for instance, the existence of predicted values $z(t) > 1$ in the regime of point D, even when $\{x(t), y(t), z(t)\}_{train} \subset [0, 1]^3$).

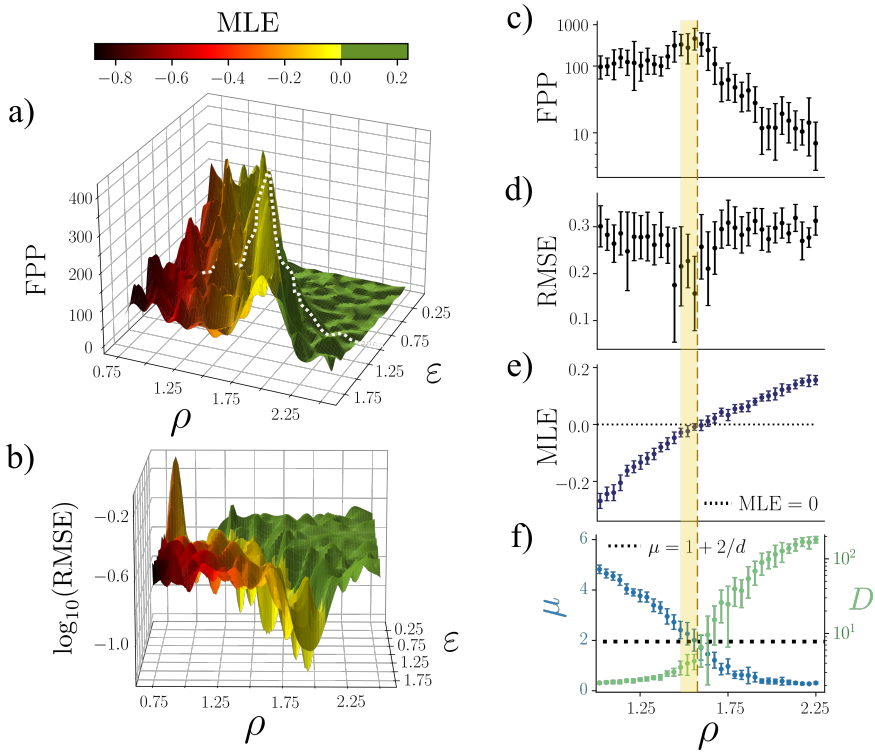

Figure 4.9: **Performance is enhanced right at the edge of instability.** **(a)** Surface plot for the furthest predicted point (FPP) in the test set. Discrete points in the surface were computed as the average across 10 different realizations of an ESN with parameters $\{\rho, \varepsilon\}$ at each point, covering a 400-points grid within the range shown for $\rho$ and $\varepsilon$. The surface was then constructed using a cubic polynomial interpolation. A trajectory of constant value $\varepsilon = 0.73$ across the surface is plotted in dashed white line. **(b)** Same as before, but now the logarithm of the root mean-square error (RMSE) between the predicted and target series during the test set is represented in the z-axis. Along the trajectory with constant $\varepsilon = 0.73$, we plot against the spectral radius: **(c)** the FPP (in logarithmic scale); **(d)** the RMSE; **(e)** the MLE; and **(f)** the powerlaw exponent $\mu$ (blue curve) and participation dimension $D$ (green curve).

## 4.4 Conclusions and perspectives

In Stringer et al. 2019a is was observed that neural encoding of different visual stimuli in the mouse V1-cortex was close to optimal, constrained by requirements of continuity and differentiability of the neural response manifold. In this chapter, we opened the door to the possibility that optimal, continuous and differentiable response manifolds emerge for neural dynamics laying close to a critical point. Indeed, we have shown that a simple ESN with randomly-connected units in the reservoir, when tuned close the edge of instability, is capable of reproducing powerlaw exponents similar to those found in mouse V1-cortex for the decay of the covariance matrix eigenspectrum, as well as the dependence of such exponents with the embedding dimension of the inputs. Intuitively, a slower decay of the eigenvalue spectrum (i.e., a smaller exponent $\mu$), translates into an enhanced ability of the network to encode finer details of the input-related variance in new relevant principal directions. However, if the decay is too slow, an excessive importance is given to such fine details at the cost of hampering the existence of a smooth manifold.

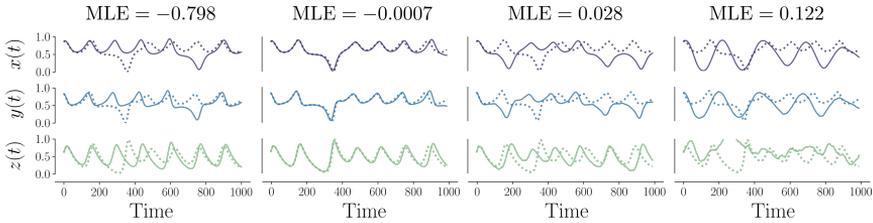

Figure 4.10: **Prediction of 1000 points of the Lorenz attractor.** From left to right, each column corresponds to points A, B, C and D of the hyper-parameter space in Fig. 4.7. Within each column, the three predicted coordinates $\{x(t), y(t), z(t)\}$ for a 1000-points test series of the Lorenz system are plotted (continuous line), together with the target series (dashed line).

On the other hand, adding stochasticity in the form of small-amplitude white noise naturally led to flatter eigenspectra, much like those found when PCA is performed over raw experimental data. Nevertheless, one can use the same cvPCA technique introduced also in Stringer et al. 2019a to extract the input-related variance of the activity, thus obtaining exponents similar to the fully deterministic case. This result suggests that the role of spontaneous activity and trial-to-trial variability on the representation of external inputs can be easily accounted for in our simple Echo State Network model.

We remark that, although giving a biological interpretation of the training algorithms and time-discrete dynamics of our simple model is out of question, there are still a lot of insights that can be drawn from simulations that try to mimic the experimental protocols. For instance, although we know that visual stimuli are not processed in a time-sequential manner across one of their spatial directions, we can argue that close to the critical point, because the echo-state property is enhanced and the reservoir states hold a fading memory of past inputs (Barancok et al. 2014; Manjunath et al. 2013; Bertschinger et al. 2004), the network units could effectively observe correlations between spatially distant points of the images.

Indeed, results obtained on image-classification and time-series prediction tasks suggest that input-representation manifolds that are critically high-dimensional may serve a bigger purpose than just being a mathematical curiosity, as ESNs poised near such a critical point show a better perfor-mance in forecasting and classification tasks, while the accuracy falls rapidly as soon as the representation manifold becomes discontinuous.

We also find important to clarify that further work needs to be done in order to fully characterize the stable-to-unstable transition that the dynamics of reservoirs undergo. Although the term edge of chaos as been widely employed in the literature to refer to this type of transition (Morales et al. 2021b; Boedecker et al. 2011; Barancok et al. 2014), this terminology —and the concept of chaos itself— should be taken with a grain of salt, as it is not devoid of criticism in this context. As pointed out in Manjunath et al. 2013, ESNs are an example of nonautonomous dynamical systems, for which typical concepts based in the theory of autonomous systems (e.g., "sensitivity to initial conditions", "attractor" and "deterministic chaos") do not directly apply (Clemson et al. 2014; Gandhi et al. 2012). Local perturbation experiments cannot therefore represent an ultimate evidence of chaotic dynamics in non-autonomous systems, since it might well be the case that the input drives the system towards and expanding dynamics for a certain time span, while the system shows on average a contracting, non-chaotic dynamics. Despite these caveats, at the light of the presented results it appears that there is indeed an actual dynamical phase transition occurring as the maximum Lyapunov exponent crosses from negative to positive values. Thus, in any case, it seems a sensible choice to use such a quantity as a control parameter when analyzing the dynamics of reservoir units.

Interestingly, leaving aside the challenges of determining the existence or absence of a chaotic regime, the effects of undergoing such a transition can be also visualized looking directly at the activities of neurons inside the reservoir against the incoming input. Fig. 4.11 shows that, if the network is deep into the stable state, where MLE $< 0$ but not close to the transition, neural responses are quite heterogeneous when compared among reservoir units, but also highly localized within each neuron phase space. This means that the individual neurons show a strongly Markovian dynamics, where the state at time $t$ is mostly determined by the current incoming input, with little influence from the history of the system. On the other hand, dynamical states characterized by MLE $> 0$ have neurons whose response extends across the full range of the non-linearity (with higher probability along the tails, reflecting a saturated behavior), but it is this same "phase space expansion" (akin to the one observed in Chapter 3 for reservoirs undergoing intrinsic plasticity changes for long periods) that makes units almost eventually indistinguishable from each other. It is only around the critical point, that we find a trade-off between dynamical richness in individual activity and variability across units.

Therefore, the presented results open the path for very exciting research avenues at the boundary of biology and machine learning, calling for theoret-ical formulations that can shed light into the fascinating properties of these input-representation neural manifolds and their relation with the criticality hypothesis.

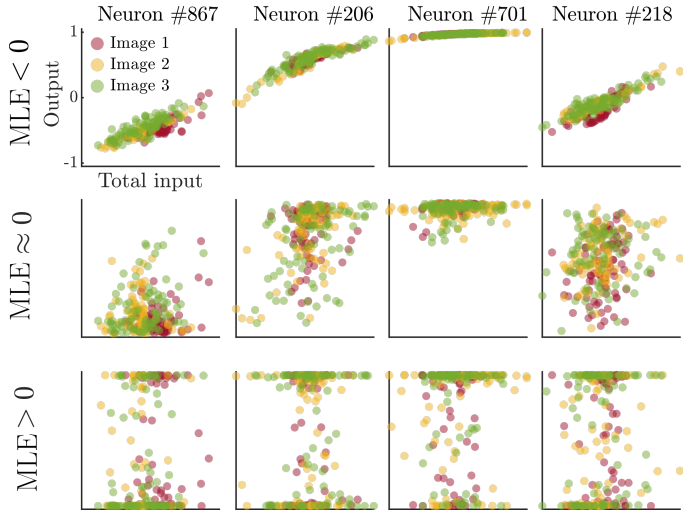

Figure 4.11: **Enhanced dynamical range of individual units near criticality.** For the different rows, ESNs were initialized using the exact same set of parameters $\{W^{res}, W^{in}, \rho = 2\}$, but tuning the dynamical regime through changes in the input scaling (from top to bottom: $\varepsilon = 2$, 0.5, and 0.01, respectively). Each column represents the activity of a particular unit of the reservoir against the incoming input to the unit for three different images (color-coded).

# Representational drift and the role of learning

"Liza! What was it yesterday, then?"
"It was what it was."
"That's impossible! That's cruel!"

FYODOR DOSTOEVSKY, *Demons*

I am made and remade continually.
Different people draw different words
from me.

VIRGINIA WOOLF, *The Waves*

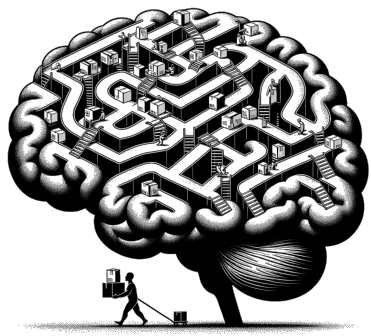

## 5.1 Introduction: an everchanging translation

Our brain is a universal encoding machine, trained over millions of years of evolution to process, represent, and interpret the thousands of incoming stimuli we are exposed to on a daily basis. As we saw in Chapter 4, an increasingly accepted hypothesis posits that the brain encodes the information about incoming stimuli in patterns of spiking activity in sensory areas, which constitute internal representation of the external world. We argued that this *representational* phase was then followed by a second, *perceptual* stage, in which conscious awareness of reality emerged as a read-out of this internal representations by higher cortical areas. Thus, within this framework, knowledge of an objective reality could be faithfully encoded in a "physical" representation of firing potentials within the neural space.

Over the past decade, however, the idea that these neural representations could be stable over time (i.e, the existence of a fixed translation from reality to neural codes) has been challenged by a number of experiments that simultaneously recorded the activity of hundreds of neurons in regions such as the hippocampus (A. Rubin et al. 2015; Ziv et al. 2013), the primary olfactory cortex (Schoonover et al. 2021), or the visual cortex (Marks et al. 2021; Deitch et al. 2021). In general terms, it is shown that the responses of neurons to a given set of stimuli is fairly stable at short time-scales (different trials within the same experiment), but can change over a span of days or weeks. This implies that neurons that were initially selective to a particular stimulus could stop being responsive after several days, and the other way around: neurons not involved initially in the representation of a given stimulus (i.e., unresponsive) could become selective to such an input within the same experimental conditions after a number of days. The empirical observation of what is known today as representational drift (RD) rapidly captured the interest of the scientific community, fostering a number of analytical and computational studies that aimed to explain this phenomenon.

In this chapter we will develop a biologically-realistic computational model to analyze the effects of spike-time dependent plasticity (STDP) as a potential mechanism behind the observed RD in olfactory cortex. In particular, we will show how STDP-mediated changes in the lateral olfactory track (LOT) connections under odor stimuli could account for the experimentally-observed slowing-down of the drift with the frequency of stimulus presentation.

Before delving any deeper into the particularities of odor encoding and the olfactory cortex structure, let us recapitulate first the observed empirical evidence for RD, as well as some of the already existing theoretical and computational approaches to the problem.

### 5.1.1 Experimental evidence of representational drift

Experimentally, RD has been found to affect spatial location representations in the hippocampus (Ziv et al. 2013; Khatib et al. 2022), task-information encoding in the posterior parietal cortex (Driscoll et al. 2017), odor encoding in olfactory cortex (Schoonover et al. 2021) and encoding of natural images in visual cortex (Marks et al. 2021; Deitch et al. 2021). On the other hand, population encoding in mouse motor cortex and in the forebrain of the zebra finch has been shown to remain stable across weeks (Katlowitz et al. 2018; Dhawale et al. 2017).

We remark that, when comparing the effects and properties of RD in different regions of the brain, it is not uncommon to find experimental results that are at odds with each other. For instance, despite the observed changes at the single-cell level, overall population statistics have been shown to remain invariant across weeks in the piriform cortex (Schoonover et al. 2021) and posterior parietal cortex (Driscoll et al. 2017). In the hippocampus, however, drift at relatively short time-scales of hours seems to be associated with an increased sparsification of the population response (Khatib et al. 2022). Moreover, neural population responses to drifting gratings in mouse visual cortex have been shown to be stable across weeks, while the same region showed drifting responses to natural movies across weeks, pointing to the existence of a stimulus-dependent drift in visual cortex (Marks et al. 2021). In contrast, drift rate in olfactory cortex was demonstrated to be fairly independent from the nature of the odor (Schoonover et al. 2021).

### 5.1.2 Existing models of representational drift

In order to study the overall mechanisms that could give rise to RD, computational approaches have typically resort to general models of leaky integrate-and-fire (LIF) or rate networks, leading to the formulation of a number of hypotheses regarding the origin of the drift. These include: i) symmetric STDP or synaptic turnover, in combination with homeostatic normalization of synaptic weights (Kossio et al. 2021); ii) noisy weight updates with additive (Qin et al. 2021) or correlated (Rule et al. 2022) noise; node or weight dropout (Aitken et al. 2021); and implicit regularization of the population activity (Ratzon et al. 2023).

Beyond studies whose aim is to find the underlying mechanism of RD, extensive work has also been done in trying to explain how the brain can extract consistent information of the external world from drifting representations. Possible explanations include the existence of an invariant, rotating submanifold in the representation space (Qin et al. 2021), a tracking process of the internal states by a "self-healing" readout (Rule et al. 2022), or the existence of a constant representational structure through drifting assem-blies (Kossio et al. 2021). As a different perspective into the problem, in Druckmann et al. 2012 the authors challenge the common assumption that unstable activity encodes time-varying properties of the system, showing how redundancy and sparsity can help to shape time-invariant representations from time-varying responses of individual neurons.

In this chapter, we will study the effects of RD on the population encoding of odors within the piriform cortex, as experimentally observed in Schoonover et al. 2021. In particular, we cast our attention into the empirically-observed dependency of the drift rate with the frequency of stimulus presentation (and hence, the role of learning), for which, at the time of writing, no theoretical explanation has been proposed to the best of my knowledge.

Notably, when talking about learning, one needs to be careful and distinguish between two phenomenologically different types of RD, often sheltered under the same term. On the one hand, there is experimental evidence of an input-evoked *fast* drift, observed in a time-scale of hours during representation learning of a new environment in the mouse hippocampus (Khatib et al. 2022). This type of drift has been recently associated to an increase in the sparsification of the population code during learning (Ratzon et al. 2023). On the other hand, there is evidence of a *slow* drift that takes place over longer periods of weeks, even in the absence of stimuli (Schoonover et al. 2021), for which the population statistics (average firing rate, sparsity of the responses, etc.) remain invariant.

To provide the reader with some more context into the problem, the next section presents an overview of the processes involved in the perception of odors, with special emphasis on the architecture and types of neurons that give rise to the odorants internal representations. With this knowledge of the underlying biological processes —and based upon further experimental results regarding odor encoding— we will then hypothesize on the possible mechanisms that could give rise to RD in the olfactory cortex before finally setting out to construct an operating model whereby we can mimic the experimental results in Schoonover et al. 2021.

## 5.2 Odor encoding in the olfactory cortex

### 5.2.1 A journey through perception: from latency to ensemble coding

In a time span of less than a second, inhalation unfolds a journey through perception as volatile molecules in the environment, known as odorants, bind to receptors in olfactory sensory neurons (OSNs) of the nasal ephitelium (see Fig. 5.1). In 1991, the pioneering work of Buck and Axell revealed that each OSN expressed just one out of $\sim 1000$ different odorant receptor genes (L. Buck et al. 1991). Although these receptors are each characterized by a high affinity to a particular odorant, they are still broadly tuned and can bind to other volatile compounds with a lower affinity. Nevertheless, it was shown that OSNs presenting the most affine receptors to a particular odorant at the given concentration always fire first (Malnic et al. 1999; Jiang et al. 2015).

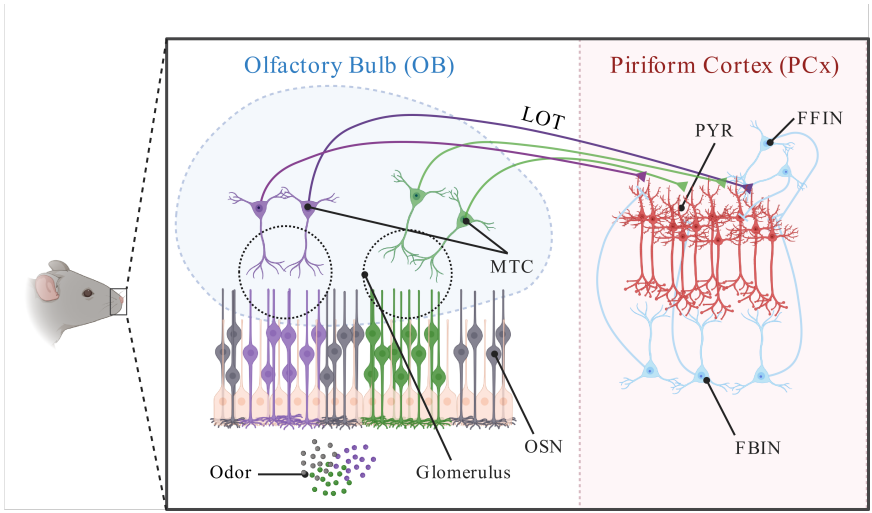

Figure 5.1: **Pathways of odor processing.** When a given odor reaches the nasal epithelium, the different odorants composing it (green, purple and gray molecules in the figure) bind to specific receptors in olfactory sensory neurons (OSNs). Next, (OSNs expressing the same type of odorant receptor project their axons into the same glomerulus, where they connect to the dendrites of mitral and tufted cells (MTCs). Random projections of MTCs axons into the piriform cortex (PCx) conform the lateral olfactory track (LOT). In the PCx pyramidal (PYR) neurons receive excitatory inputs from MTCs, as well as inhibitory connections from FFINs and FBINs. Illustration created using the BioRender software.

Following the next step of odor processing, all OSNs expressing the same type of receptor have shown to project their axons onto a unique pair of glomeruli in the olfactory bulb (OB) (Mombaerts et al. 1996; Hálasz et al. 1993). These spheroidal structures host the synaptic connections between the OSN axon terminals and the dendrites of the so-called secondary neurons: the mitral/tufted cells (MTCs) (see Fig. 5.1). Thus, for a given odor, although OSNs with less specific receptors will be eventually activated, bulb-cells associated to the most specific receptors will fire earliest, effectively transforming the initial receptor-specific encoding into a temporal encoding in the OB.

Once we cross from the OB into the olfactory cortex, things begin to get a bit more complicated. As the information encoded in spatio-temporal pat-terns of MTCs activity is conveyed through random and overlapping lateral olfactory track (LOT) connections into the piriform cortex (PCx), the original *temporal* encoding is translated into an *ensemble* code: odor identity in the PCx is determined by specific sets of principal neurons (pyramidal cells, mainly) recruited during the sniff, with no other information about the temporal profile of the incoming spikes (Stern et al. 2018). In fact, it has been shown that cortical odor responses are determined by the earliest-active glomeruli (the ones with higher specificity for the given odorant), thanks to the fast recruiting of inhibitory interneurons (feedforward inhibitory neurons (FFINs) and feedback inhibitory neurons (FBINs), see Fig. 5.1) that can suppress the cortical responses to later, less-specific OB inputs (Bolding et al. 2018; Stern et al. 2018; Bolding et al. 2020).

As we have seen in the introduction of this chapter, many different mechanisms have been proposed to explain the phenomenon of RD in neural networks. Before delving into the particular problem of simulating the drift in a model of the olfactory cortex, we will first hypothesize on the biological source of such a drift, assessing the validity of already existing models upon their ability to reproduce some of the experimental findings regarding RD in the olfactory cortex.

### 5.2.2 Pruning the hypothesis tree

Given the overall complexity of the odor encoding process from the arrival of odorant molecules in the nasal epithelium to the activation of pyramidal cells in the PCx, finding the link within this chain whence drift emerges might seem a daunting task. Let us, however, go through some of the experimental findings that could serve as a guide to construct our model.

In Bolding et al. 2020 it was observed that recurrent interactions at the PCx helped to rapidly recruit inhibitory feedback, thus halting the PCx response by discounting the effect of OB glomeruli with larger latency times (i.e., of those MTCs that got activated later on the sniff, and were therefore not so specific to the presented odorant). In fact, within the same study, the authors expressed the tetanus toxin in PCx neurons to block their ability to excite each other through recurrent interactions, which lead to an amplification of the PCx response, paired with an enhanced sensitivity of the encoding to odor concentration. These same findings were also validated in the computational model presented in Stern et al. 2018, showing that the identity of odors in PCx representations is fundamentally determined by the identity of those pyramidal neurons that are selectively activated by early-responding glomeruli. Thus, one can conclude that odor identity in PCx is mostly encoded in what we will call "primary" pyramidal neurons, i.e., neurons that are directly recruited through OB excitatory inputs.

On the other hand, it was reported in Schoonover et al. 2021 that only around 2.5% of the recorded pyramidal neurons in the PCx showed a stable response to a given set of odorants across the 32 days that lasted their experiment. Thus, given the accumulating evidence indicating that MTCs responses at the OB remain stable over long periods (of at least several months, Bhalla et al. 1997; Hiroyuki K. Kato et al. 2012), it follows that any proposed drift mechanism that aim to account for such a strong reconfiguration must necessarily affect the identity of the primary pyramidal cells, recruited by the OB.

As we saw in Chapter 3, there are fundamentally two different forms of plasticity mechanisms whereby neural networks can adapt their activity: synaptic plasticity, which affects the strength of the connections between neurons; and non-synaptic or intrinsic plasticity, which can alter the excitability of the neurons —typically through structural changes affecting voltage-dependent membrane conductances in the axon, dendrites, or soma. In view of the empirical evidence pointing at the fundamental role of synaptic modifications in different parts of the olfactory cortex (D. A. Wilson et al. 2004; Ito et al. 2008; T.-F. Ma et al. 2012; Y. Cohen et al. 2015; Jacobson et al. 2018; Kumar et al. 2021), we will focus on synaptic plasticity mechanisms as the most plausible source of RD. More specifically, in the remaining of this chapter we will see that a simple, although biologically realistic, spike-time dependent plasticity (STDP) rule (Fig. 5.3) can reproduce to a great extent the empirically observed measures of RD.

## 5.3 Experimental setup and methods

### 5.3.1 Modeling the olfactory cortex

In this section we present a fairly realistic model for the olfactory bulb (OB) and piriform cortex (PCx) based on the original work of Stern et al. 2018, which we will be using across the remainder of the chapter (see Fig. 5.2a).

Unlike other sensory stimuli such as sounds and images, which can be easily characterized computationally, simulating real molecules and their microscopic effects over biologically-realistic odorant receptors is a daunting task. Therefore, following the approach in Stern et al. 2018, here we model odorant identity directly in terms of a latency encoding in the OB. Unless otherwise stated, the OB will consist of 2000 mitral/tufted cells (MTCs)[1], each connected to one of the existing 100 glomeruli, so that on average each glomerulus projects into $\sim 20$ MTCs. To model the responses of MTCs to different odorants, each glomerulus was assigned an onset latency time with respect to the time of odor presentation, independent of the odor concentration, but specific to the odor identity. This onset latency is then inherited by all the MTCs projecting into the given glomerulus. In this way, the activity of cell $i$ in response to odor $o$ will be modeled as an inhomogeneous Poisson process whose rate is given by:

$$r_i^o(\Delta t^*) = \begin{cases} f_{bl}, & \text{if } \Delta t^* < \tilde{\tau}_i^o \text{ or } \Delta t^* > \tau_{inh} \\ f_{exc}, & \text{otherwise} \end{cases} \tag{5.1}$$

where $\Delta t^* = t - t_0$ is the time elapsed from the beginning of a respiration cycle, at $t_0$, and $\tilde{\tau}_i^o$ is the onset latency time of neuron $i$ under odor $o$. Each respiration cycle consists of an inspiration period of length $\tau_{ins} = 150$ms, followed by an expiration period lasting $\tau_{exp} = 350$ms (see Fig. 5.2c). Following the work in Stern et al. 2018, we chose $f_{bl} = 1.5$Hz and $f_{exc} = 100$Hz for the baseline and excited rates of the MTCs. For each odor, latency times with respect to the time of odor presentation, $\tilde{\tau}_i^o$, were randomly drawn from a uniform distribution such that, at reference odor concentration, only an average 10% of all MTCs will become responsive to a given odor within the inhalation period (i.e., will have a latency time $\tilde{\tau}_i^o < \tau_{inh}$). Thus, in this model, the *identity* of an odor $o$ becomes fully characterized by the set $\{\tilde{\tau}_i^o\}$ of all glomeruli latency times. An example of the emergent pattern of mitral activity during three consecutive respiration cycles under a particular odor is given in Fig. 5.2b (upper row).

---

[1]For simplicity, we will not distinguish between mitral and tufted cells, although it is known that the later present shorter response latencies.

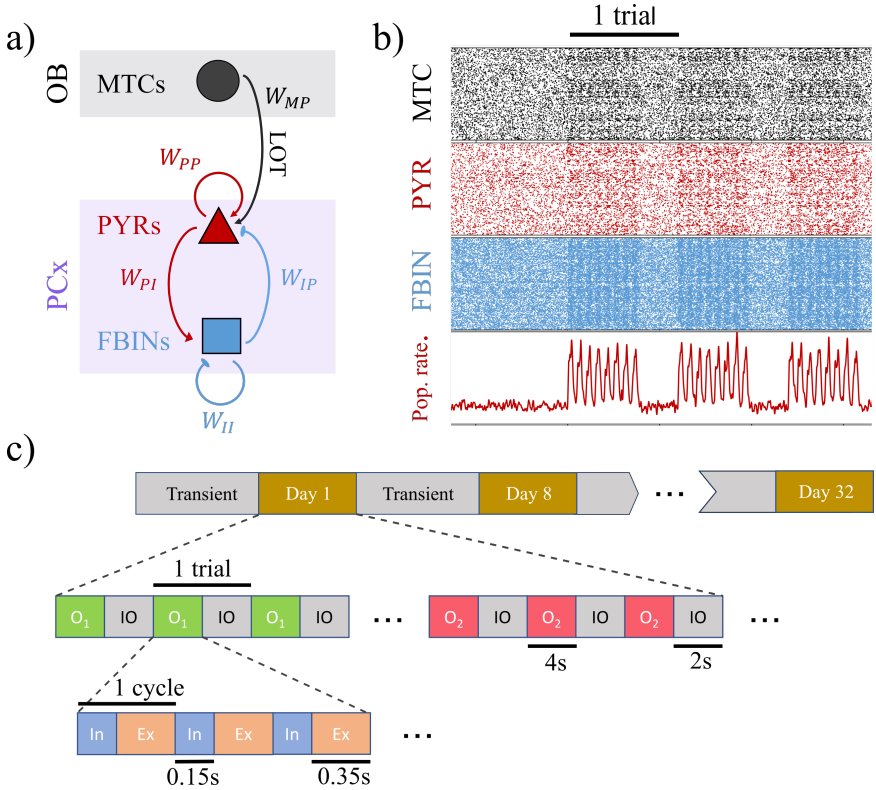

Figure 5.2: **Model and experimental setup.** **(a)** Diagram of the modeled olfactory cortex, depicting the different types of neural populations considered and their interactions (see Fig. 5.1 for a direct comparison with the biological architecture). **(b)** Raster plots for the activity of MTCs, pyramidal neurons and FBINs, together with a trace of the average pyramidal population activity across time. **(c)** Protocol for testing of PCx responses to odorant presentation, following the experimental setup in Schoonover et al. 2021.

To characterize the PCx and its inputs, the model proposed by Stern *et al.* simplifies the original architecture by disregarding the effects of semilunar cells, thus assuming that pyramidal neurons can only receive excitatory input directly from MTCs through LOT connections, or from recurrent connections with other pyramidal cells. In this work, we will further reduce the model by neglecting the effect of feedforward inhibitory neurons (FFINs), which receive excitatory inputs from MTCs and then inhibit pyramidal neurons in the PCx. These neurons have been shown to simply modulate the amplitude but not the shape of the pyramidal responses to an odor Stern et al. 2018. Thus, inhibition of pyramidal cell activity is limited to feedback inhibitory neurons (FBINs) (see Fig. 5.2a), which have been shown to be responsible of suppressing the effect of later MTCs inputs (i.e., having longer onset latencies), halting the explosive growth of activity in pyramidal neurons (Stern et al. 2018; Bolding et al. 2020; Bolding et al. 2018).

The below-threshold voltage dynamics of pyramidal and FBINs take the general form of a LIF equation:

$$\tau_m \frac{dV_i}{dt} = -(V_i(t) - V_{rest}) + \frac{I_i^{tot}(t)}{g_m} \ , \qquad (5.2)$$

where $I_i^{tot}(t) = I_i^{exc}(t) + I_i^{inh}(t)$ is the sum of all incoming excitatory and inhibitory currents to neuron $i$, $\tau_m$ is the membrane characteristic time scale, $g_m$ is the membrane conductance and $V_{rest}$ is the resting potential. Defining $\Lambda_j = \sum_f \delta(t - t_j^f)$, one can write for pyramidal neurons:

$$\tau_{exc} \frac{dI_i^{exc}}{dt} = -I_i^{exc} + \sum_{j=1}^{N_{MTC}} w_{ij}^{mp} \Lambda_j + \sum_{j=1}^{N_{PYR}} w_{ij}^{pp} \Lambda_j \ , \qquad (5.3)$$

$$\tau_{inh} \frac{dI_i^{inh}}{dt} = -I_i^{inh} + \sum_{j=1}^{N_{FBIN}} w_{ij}^{fp} \Lambda_j \ , \qquad (5.4)$$

whereas for FBINs:

$$\tau_{exc}\frac{dI_i^{exc}}{dt} = -I_i^{exc} + \sum_{j=1}^{N_{PYR}} w_{ij}^{pf}\Lambda_j \ , \qquad (5.5)$$

$$\tau_{inh}\frac{dI_i^{inh}}{dt} = -I_i^{inh} + \sum_{j=1}^{N_{FBIN}} w_{ij}^{ff}\Lambda_j \ . \qquad (5.6)$$

In the above equations, $\tau_{exc}$ and $\tau_{inh}$ represent the characteristic decay time for the excitatory and inhibitory input currents, while $w_{ij}^{mp}$, $w_{ij}^{pp}$, $w_{ij}^{fp}$, $w_{ij}^{pf}$, $w_{ij}^{ff}$, denote the synaptic efficacies between presynaptic neuron $j$ and postsynaptic neuron $i$ for the mitral-to-pyramidal, pyramidal-to-pyramidal, FBIN-to-pyramidal, pyramidal-to-FBIN and FBIN-to-FBIN connections, respectively. We remark that, although in the original work by Stern et al all synaptic weights connecting the same type of neurons were set to a common value, in our case we decided to model the weights as continuous values drawn from a lognormal probability distribution with mean $\mu = \langle w \rangle$ and standard deviation $\sigma = \mu/2$, different for each type of connection (see Appendix E.I for tables containing all the parameters).

In order to account for the baseline resting-state activity observed in the PCx in the absence of odor inputs, we further extended the original model to include random Poissonian spiking of all pyramidal neurons at a fairly slow rate, $f_{spont} = 0.05$Hz. For simplicity, we reabsorbed the membrane conductance $g_m$ into the definition of input current, so that all currents and synaptic efficacies are now in units of voltage.

Once a neuron reaches its firing threshold, $V_{th}$, at a given time, $t^f$, its membrane potential is reset and clamped to a value $V_{reset} = -65$mV for a refractory period, $\tau_{ref}$ , before it can evolve again according to Eq. (5.2). Likewise, we did not allow the membrane potential to decrease below a minimum voltage, $V_{min} = -75$mV.

Default values for all the parameters used in the simulations —including density and synaptic efficacies for all connectivity matrices, characteristic time-scales for excitation and inhibition currents, etc.— can be found in Appendix E.I.

### 5.3.2 Modeling the source of drift

It has been experimentally observed that long-term potentiation (LTP) and long-term depression (LTD) of synaptic weights depend on the exact timing of the pre- and postsynaptic spikes (Markram et al. 1997; Bi et al. 2001; Buchanan et al. 2010). LTP is typically induced when the presynaptic spike precedes the postsynaptic one on an interval of 10 to 20ms, whereas LTD occurs if the order of spikes is reversed (see Fig. 5.3a). This mechanism of STDP has been largely studied from a theoretical point of view (Rossum et al. 2000; Gilson et al. 2011a; Gilson et al. 2011b; Carlson et al. 2013; Effenberger et al. 2015) and many models have been proposed to investigate its functional implications (see Caporale et al. 2008 and Morrison et al. 2008 for reviews on the topic). Mathematically speaking, the change in the synaptic weight induced by pre- and postsynaptic spikes at time $t_{pre}$ and $t_{post}$, respectively, can be written as:

$$\Delta w = \eta \Gamma(w; t_{post} - t_{pre}) \tag{5.7}$$

where $\eta$ is the learning rate and $\Gamma(w; t_{post} - t_{pre})$ is the plasticity window, which can lead to potentiation (LTP) or depression (LTD) depending on the relative timing of the spikes (see Fig. 5.3a):

$$\Gamma(w; t_{post} - t_{pre}) = \begin{cases} f_+(w) \exp\left(-\dfrac{|t_{pre} - t_{post}|}{\tau_+}\right), & \text{if } t_{pre} < t_{post} \\ -f_-(w) \exp\left(-\dfrac{|t_{pre} - t_{post}|}{\tau_-}\right), & \text{if } t_{pre} > t_{post} . \end{cases} \tag{5.8}$$

Within the above expression different choices of the scaling functions for potentiation, $f_+(w)$, and depression, $f_-(w)$, can give rise to different models of STDP (Morrison et al. 2008). In this chapter in particular we will be using a *multiplicative* STDP, as originally proposed in Rossum et al. 2000 —on the basis of experimental observations in (Bi et al. 1998)—, for which the LTP and LTD scaling functions read:

$$f_+(w) = a_+ , \tag{5.9}$$

$$f_-(w) = a_- w , \tag{5.10}$$

for some constant values $a_+$ and $a_-$. Remarkably, one can use the Fokker-Plank formalism to derive the stationary probability distribution for the weights (see Fig. 5.3b) under a general model of STDPs described by Eqs. 5.7 and 5.8, assuming uncorrelated inputs to the network (Gilson et al. 2011a):

$$\mathcal{P}_{\sqcap}(w) = \frac{\mathcal{N}}{B(w)} \exp\left[\int_0^w \frac{2A(w')}{B(w')} dw'\right] , \tag{5.11}$$

where:

$$A(w) = \eta\{\tau_+ f_+(w) - \tau_- f_-(w)\} , \tag{5.12}$$

$$B(w) = \eta^2\{\frac{\tau_+}{2}[f_+(w)]^2 + \frac{\tau_-}{2}[f_-(w)]^2\} , \tag{5.13}$$

and, in particular, the average equilibrium weight for uncorrelated inputs under the multiplicative version of the rule is then given by $\langle w_{st} \rangle = (a_+ \tau_+)/(a_- \tau_-)$.

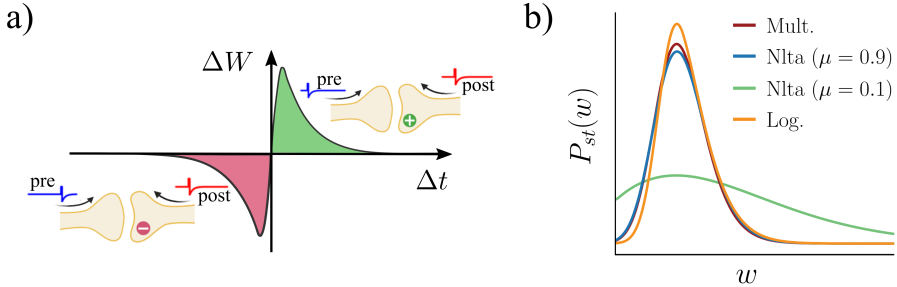

Figure 5.3: **Spike-time dependent plasticity as the source of drift (a)** In spike-time dependent plasticity (STDP), a given connection from a presynaptic to a postsynaptic neuron is potentiated when the postsynaptic neurons fires within a small time-window after a spike of the presynaptic neuron. Conversely, depression of the synaptic efficacy occurs if a postsynaptic spike is immediately followed by firing of the presynaptic neuron. **(b)** Stationary probability distributions for the synaptic weights under uncorrelated random inputs for a multiplicative STDP (red; Rossum et al. 2000), a non-linear temporally asymmetric STDP for two different values of the exponent $\mu$ (blue and green; J. Rubin et al. 2001), and a logarithmic STDP (yellow; Gilson et al. 2011a).

We remark that, although all the analysis and simulations presented in this chapter were derived using the multiplicative STDP proposed in Rossum et al. 2000, a similar phenomenology can also be obtained with other more complex STDP models such as non-linear temporally assymetric STDP, with relatively large values of the exponent $\mu$ (J. Rubin et al. 2001); and logarithmic STDP (Gilson et al. 2011a) (see Fig. 5.3b). On the other hand, application of the original additive STDP model as formulated in S. Song et al. 2000 eventually lead to run-away activity in the PCx, as a result of the emerging bi-modal stationary distribution of weights.

### 5.3.3 Experimental setup

In order to compare the results of our simulations with the existing empirical evidence, we will aim at reproducing the experimental setup and analyses proposed in Schoonover et al. 2021, introducing some modifications only when reducing the computational cost of the simulations becomes imperative.

Following the aforementioned study, the total simulated experiment consists of 32 intervals that, for simplicity, we will refer to as "days", although their duration is much shorter than an actual day of simulated time. Each of these days is divided consists of a transient period of 160s, where only background pyramidal activity is present. At 8-day intervals, a test period is also included during which 8 different odors are presented. We present each odor 7 times (trials) during the test period, spanning 8 respiration cycles (4s) in each trial, followed by a 2s inter-trial transient (see Fig. 5.2c).

For each test day $d$, odorant $o$ and trial $m$, population firing rates $\mathbf{x}_{d,o,m}$ are computed by averaging neural responses in time across four 2s-windows after odor onset, then concatenating the corresponding vectors so that $\mathbf{x}_{d,o,m} \in \mathbb{R}^{4N_{pyr}}$, and finally subtracting the average baseline spontaneous rate during odorant presentation. Unless otherwise stated, all measures are performed on such spontaneous, baseline-subtracted population vectors, which we may refer to simply as *representations*.

To quantify the amount of drift over time, we resort to several magnitudes as proposed in Schoonover et al. 2021. On the one hand, we define the correlation between same-odor representations at days $p$ and $q$, with $p \neq q$, as the average over all odors of the Pearson correlation coefficient between trial-averaged population vectors at the corresponding days:

$$c_{p,q} = \frac{1}{n_{odors}} \sum_{o=1}^{n_{odors}} \frac{\langle (\mathbf{x}_{p,o} - \overline{x}_{p,o})(\mathbf{x}_{q,o} - \overline{x}_{q,o}) \rangle}{\sigma_{\mathbf{x}_{p,o}} \sigma_{\mathbf{x}_{q,o}}} \tag{5.14}$$

where $\mathbf{x}_{p,o} = M^{-1} \sum_{m=1}^{M} \mathbf{x}_{p,o,m}$ is the trial-averaged population response to odor $o$ on day $p$, and $\sigma_{\mathbf{x}_{q,o}}$ its standard deviation. Thus, the average correlation between same-odor responses separated by a time interval of $\Delta$-days in their measurements, can be defined as:

$$\overline{c}_{\Delta} = \frac{1}{n_{\Delta}} \sum_{p,q \,:\, |p-q|=\Delta} c_{p,q} \tag{5.15}$$

where $n_{\Delta}$ is the number of pairs of test days separated by a time interval $\Delta$.

On the other hand, the average angle between a pair of population vectors representing the same odor at days $p$ and $q$ can be written as:

$$\theta_{p,q} = \frac{1}{n_{\Delta}} \sum_{o=1}^{n_{odors}} \theta_{p,q}^{o} = \frac{1}{n_{odors}} \sum_{o=1}^{n_{odors}} \cos^{-1}\left( \frac{\mathbf{x}_{p,o} \cdot \mathbf{x}_{q,o}}{\|\mathbf{x}_{p,o}\| \|\mathbf{x}_{q,o}\|} \right) , \tag{5.16}$$

and the average corrected angle between any two representations measured on tests separated by $\Delta$-days:

$$\overline{\theta}_{\Delta} = n_{\Delta}^{-1} \sum_{p,q \,:\, |p-q|=\Delta} \theta_{p,q} - \overline{\theta} , \tag{5.17}$$

where $\overline{\theta}$ is the average within-day angle between same-odor population responses, when compared across even and odd trials (see Appendix E.II, Eq. (E.5)).

We can now give a measure for the rate of drift (in angles per day and corrected for within-day fluctuations) in terms of this last quantity:

$$\overline{r} = \left\langle \frac{\overline{\theta}_\Delta}{\Delta} \right\rangle_\Delta \; . \tag{5.18}$$

where the average is taken across all possible time intervals, $\Delta$, between any pair of test days.

## 5.4 Understanding representational drift in the piriform cortex

For the results presented in the following section, we used networks of 1000 pyramidal neurons, 2250 mitral neurons and 125 FBINs. All results are derived considering STDP-induced changes at the lateral olfactory track (LOT) connections, following the arguments laid out in Section 5.2.2. Moreover, to respect the sparsity in the connectivity between MTCs and pyramidal cells, we only allowed weight changes in already existing connections. Parameters used for modeling the network structure (sparsity of connections, weight values...), their dynamics (baseline rates, threshold voltages, timescales for excitation and inhibition...) and STDP updates (characteristic time scales of LTP and LTD, maximum weight efficacy...) were kept at biologically plausible values when possible (see Appendix E.I for details of all the parameters values used in the simulations).

We can now proceed and, following the experimental protocol described in Section 5.3.3 (see also Fig. 5.2c), show different "odorants" —i.e., different patterns of MTC activity as input to the PCx— at 8-days intervals, recording for day $d$, odor $o$ and trial $m$ the population responses of the pyramidal neurons. We recall that, between these odorant presentations, which take place every 8 days, the PCx undergoes a sustained resting-state activity that arises from both, random incoming inputs from baseline activity at the MTCs, and low-rate random Poissionian spiking of pyramidal neurons. We thus hypothesize that a slow but constant drift in the representation of odorants would inevitably emerge from a drift of the weights in the LOT connections, due to spontaneous STDP-mediated changes under such noisy background of activity.

Fig. 5.4a shows the change in sensitivity for three odorant-unit pairs (i.e., the response of a particular pyramidal neuron to a given odorant) in test days across the 32 days of the experiment. See, for instance how neuron #758 loses its responsiveness to odorant one after the first 8 days, while responses of neuron #643 remains stable across all days. In Fig. 5.4b, we z-scored the activity of all odor-unit pairs in our simulations (substracting,

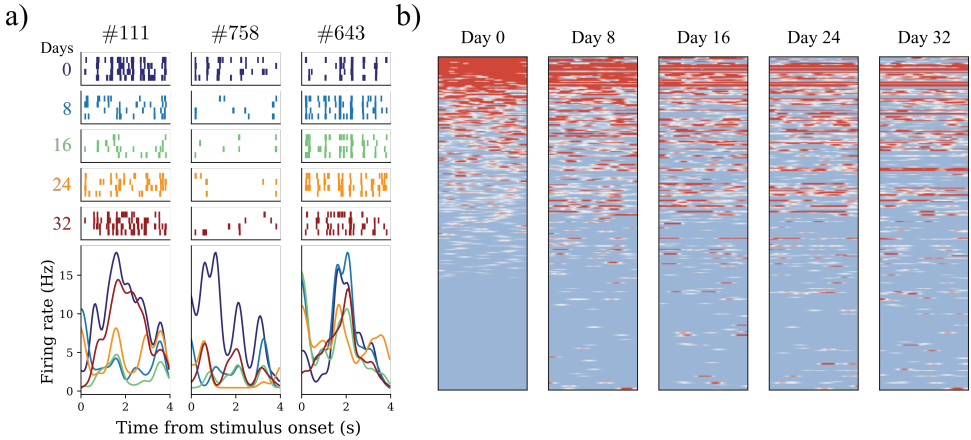

Figure 5.4: **Piriform cortex responses changes over 32 days. (a)** Raster plots and peristimulus time-histograms for three pyramidal neurons responses to a given odorant, at each of the test days. In each row in the raster plot represents a single trial of odorant presentation. **(b)** Z-scored activity of a subsample of 200 units, ordered in all cases by their value on day 0 from strong (red) to weak (blue) responses.

for each unit, the average spontaneous activity when no odor is presented, then dividing by its standard deviation) and ordered the responses from higher (red) to lower (blue) activity on the first day. As days go by in our simulations, we can see how the selectivity of units change, with new responsive odor-unit pairs emerging, while other —initially responsive— pyramidal neurons have lost their sensitivity by the end of the experiment.

Fig. 5.5 reproduces the main results observed in Schoonover et al. 2021 regarding the emergence of RD at an ensemble-encoding level. First, we notice how the firing rate of individual units changes across days, with neurons that both gain and loose sensitivity to particular odorants across the experiment (Fig. 5.5a). On the other hand, correlations between all trial-averaged population responses to a common odorant were computed across all possible pairs of days (Fig. 5.5b, left), and their average value plotted against the time interval between same-odorant representations (Fig. 5.5b, right). A similar measure, but for the average corrected angle (Eq. (5.17)) between any two representations of the same odorant on different days is depicted in Fig. 5.5d (right), together with the cumulative distribution functions for the angles across days, within-day, and across-odorants Fig. 5.5d (left). Drifting in the representations is thus reflected as a decrease of the correlations (and an increase of the angle) between representations corresponding to the same stimuli across days. In our simulations, we measured, using Eq. (5.18), an average angle drift rate $\bar{r} = 1.44° \pm 0.56°$ (Fig. 5.5c), in excellent agreement with the empirical observations, for which $\bar{r}_{exp} = 1.3° \pm 1.2°$.

In Schoonover et al. 2021 it was observed how, despite the individual drifting responses of neurons, the overall population statistics remained stable across days, suggesting that the general properties of odorant encoding changed only marginally over time. In fact, it was shown that the performance of a linear classifier trained to decode the odors from the PCx activity on earlier days, but then tested on later days, deteriorated as a function of the time interval between training and testing. On the contrary, within-day performance (i.e., decoding of the stimuli using a linear classifier trained on even trials and then tested in odd trials within the same day) remained fairly stable across the 32 days of the experiment, meaning that the new odor representations emerging through drift hold the same encoding capabilities as the original ones. To test whether this invariant population properties where also present in our model, we measured the fraction of responsive neurons, population sparseness (Eq. E.2), lifetime sparseness (Eq. E.3) and within-day correlations (Eq. E.4), showing how all these properties remained invariant across days (see Fig. 5.5f). Similarly, we show high and stable within-day performance in the classification of odors using a linear classifier, but a deterioration in the decoding ability when the classifier was trained in earlier days and then tested several days later (Fig. 5.5c). Definitions for all the above quantities and further details on the classification algorithm can be found in Appendix E.II.

Given the presented results —which are in excellent agreement with the empirical observations of Schoonover *et al.*— and assuming that the same phenomenology would most likely apply to human brains, how can a consistent perception of an objective reality (i.e., a given odorant in this case) emerge from the readout of constantly changing internal representations?

One possibility, for instance, considers that downstream regions performing such a readout could adapt to the drift provided there was some invariant geometry of the representation manifold (for example, if all induced changes could be mapped to a rotation or other type of transformation of an invariant manifold in a high-dimensional space; Qin et al. 2021).

Notably, the authors in Schoonover et al. 2021 also tested whether the geometry of odor responses was conserved despite drift. If one considers the mean response of all the neurons over the duration of the odorant presentation as a point in an $N$-dimensional space, then all points corresponding to different odorants will span a certain response manifold, as we already discussed in Chapter 4. One can then measure changes in the geometry of such a manifold by comparing the relative angle between pairs of triples of connected vertices on different days, where each vertex is the PCx representation of a given odorant (see Fig. 5.6a). Thus, within each individual day, $p$, an odor similarity matrix was defined as a matrix $A^p \in \mathbb{R}^{M \times n_{odors}}$ whose columns, $\mathbf{a}_{\bullet i}$, contain the $M = n_{odors}\binom{n_{odors}-1}{2}$ possible angles between any two edges connecting the representation (vertex) for odor $i$. Fig. 5.6b shows the values of these similarity matrices on different days of the experiment, reflecting the non-invariant geometry of the representation.

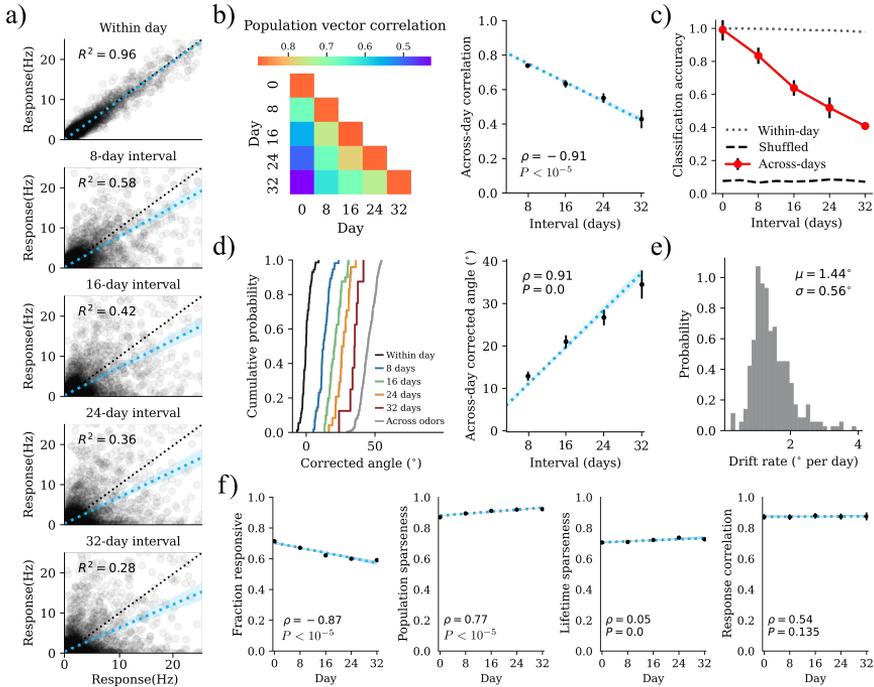

Figure 5.5: **Odorant representations drift despite invariant population statistics.**
**(a)** Firing rate responses of odorant-unit pairs compared across different intervals. **(b)** Average population vector correlations for same-odorant representations across days (left) and decay of the correlations against the time interval between representations (right). **(c)** Classification accuracy for a Support Vector Machine (SVM) trained to classify the odorants using the activity of 41 units on the first day, then tested on the activity of these same units on later days. Within-day classification of odorants (dotted line) and a surrogate example with shuffled labels (dashed line) are included as control cases. **(d)** Cumulative probability distribution for the across-days corrected angle between representations (left), and increase of the corrected angle with the time interval between same odor representations (right). **(e)** Histogram for the drift rate in degree angles per day. **(f)** Population statistics, including, on each test day and from left to right: the fraction of responsive neurons; the average population sparseness; the average lifetime sparseness and the average within-day correlations (see Appendix E.II for a detailed description of the above quantities).

One can further quantify this drift in the "geometry" of the representaton manifold through a measure of *matrix dissimilarity*, as given by the Frobenius norm between odor similarity matrices in days $p$ and $q$ :

$$\|A^{p,q}\|_F := \|A^p - A^q\|_F = \sqrt{\sum_{k=1}^{M} \sum_{i=1}^{n_{odors}} |a_{k,i}^p - a_{k,i}^q|^2} \ , \qquad (5.19)$$

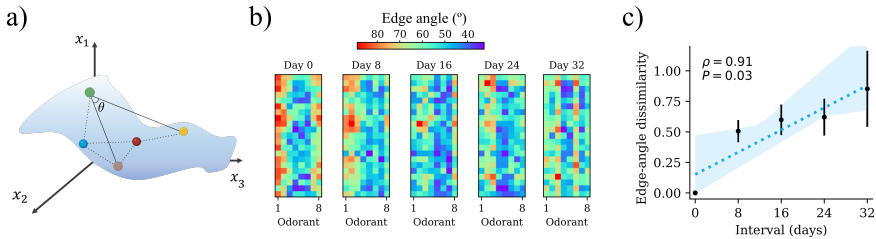

Figure 5.6: **Geometry of odor representation manifolds also drift in time.** (a) Schematic depiction of the representation manifold geometry, marking one of the angles defined by the population vector responses to three different odors. (b) Edge-angle similarity matrices at test days. (c) Edge-angle corrected dissimilarity measure as a function of the time interval between test days ($\rho$ and $P$ denote the correlation coefficient and $p$-value for the linear regression).

where $\|A^{p,q}\|_F = 0$ for identical matrices, whereas higher values of the Frobenius norm indicate an increasing difference between the odor similarity matrices. Following the methods in Schoonover et al. 2021 we computed a normalized angle matrix dissimilarity that corrected for within-day fluctuations of the manifold geometry (see Appendix E.II, Eq. E.1). Fig. 5.6c shows the value of this corrected matrix dissimilarity against the length of the interval (in days) between the considered similarity matrices. As we can see, changes in edge-angles between encoded odor responses accumulate in time, evidencing the inconsistency of the representation geometry.

## 5.5   The role of learning

We saw how the empirically observed effects of RD over the encoding of odors in the PCx could be successfully explained in terms of STDP-induced weight changes under the presence of a spontaneous noisy background of uncorrelated spiking activity. We remark that, for this first part of the experiments, the presented odorants had never been shown to the mice (nor to our simulated olfactory cortex) before the first test day (day 0).

Nevertheless, a yet more intriguing phenomenon was observed empirically when measuring the drift in the representations to odorants previously known to the mice. More specifically, a cohort A of 3 mice were presented with a panel of four odorants daily across 16 days prior to the beginning of the experiment. Beginning on day 0, the same set of already "familiar" odorants was still presented on a daily basis, but mice were also subject to a set of four "unfamiliar" odorants at 8-day intervals (see Fig. 5.7a, top). Interestingly, a slower drift rate for the representations of "familiar" odorants (i.e., those presented daily on the 16 days prior to the experiment) was observed when compared to the drift for the unfamiliar ones.

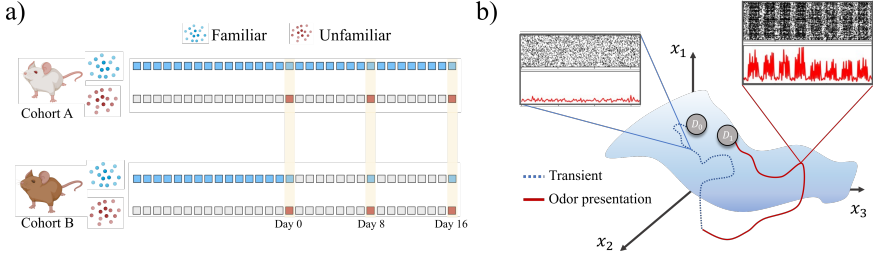

Figure 5.7: **A model for learning-dependent drift.** **(a)** Experimental setup to assess the dependence of the drift on the frequency of stimulus presentation. **(b)** Schematic depiction of the drifting representations during across-test-days transients and odorant presentation. Insets show raster plots (top) and average population responses (bottom) for the activity of pyramidal neurons in the PCx.

Notably, a second cohort of mice (B) in which the 4 familiar odorants were not presented daily after day 0, but at 8-days intervals instead (Fig. 5.7a, bottom), showed no statistically significant changes in the drift rate for familiar and unfamiliar odorants. Put it together, these results suggests that "learned" representations of familiar stimuli would naturally drift as fast as representations of new inputs, unless the familiar stimuli are presented with a relatively high frequency.

To understand this phenomenon, let us now go back to our representation manifold picture and consider that, through continued exposure to a particular odorant, LTP and LTD mechanisms at the MTCs-to-pyramidal connections lead to a stable, learnt representation, $\mathcal{D}_0$, of the given odorant in the PCx, characterized by a given stationary probability distribution of the LOT synaptic weights, $P_0^{st}(w_{mp})$. As we saw in the previous section, during the transient days in which the odorant is not being presented, STDP-mediated changes in the LOT synapses due to random spontaneous activity in both, the OB and PCx, will cause the original representation, $\mathcal{D}_0$, to drift in the activity space towards a new representation, $\mathcal{D}_t$, after a certain time $T$ has elapsed (see Fig. 5.7b).

We now hypothesize that, provided the accumulated changes in the weights did not drive the representation too far from the original point in the manifold, presenting back the odorant can induce new changes in a time scale $\Delta t \ll T$ towards the originally-learned stationary distribution of weights, $P_0^{st}(w)$, effectively "drifting back" the representations into a new point in the neural space, $\mathcal{D}_{T+\Delta t}$, such that $\|\mathcal{D}_0 - \mathcal{D}_{T+\Delta t}\| < \|\mathcal{D}_0 - \mathcal{D}_T\|$ (i.e., moving the representation "closer" to the originally learned one).

To assess the validity of this hypothesis, we mimicked once again the experimental setup in Schoonover et al. 2021, letting now changes in the LOT synapses through STDP learning during both, across-days transients of background, Poisson-like spikes, as well as during the presentation of odorants in test days. There is, however, a strong computational limitation regarding the relative time-scale of the drift with respect to the odor-induced changes. In the original experiments, odorants were presented 7 times per day, for a period of 4s each time, so that STDP changes induced by PCx responses to odors took place on a span of $\sim$ 30s in each test day. On the contrary, drifting effects during spontaneous activity occurred over a time span of days between consecutive measures, which translates into an effective rate around 3 to 4 orders of magnitude slower than the odor-induced changes.

Since running a fairly realistic model of thousands of integrate-and-fire neurons for a *simulated* time of actual days proves computationally unfeasible, one could think of increasing the learning rate of the STDP rule several orders of magnitude during the between-test intervals, to compensate for their much shorter duration with respect to the experiment. However, one needs to be specially careful when using this trick, as the width of the stationary probability distribution of the weights will increase with the value of the learning rate, and the emergence of very large weights can destabilize the network (see Eqs. 5.7 to 5.13 and Gilson et al. 2011a). Moreover, this means that even if the network was subject to the exact same spiking statistics during odorants presentation as in the transients between test days, STDP would be driving the network towards two different stationary probability weight distributions in each case. To minimize this undesired effect, we used a moderately increased learning rate for the transients between test days, while also increasing the spontaneous firing rate of PCx neurons (see Appendix E.I for the exact values of all parameters).

Moreover, to minimize the computational cost, we reduced by an order of magnitude the original size of the model to 100 pyramidal neurons, 250 MTCs and 25 FBINs. For such a small network size, it is likely that changes induced by one odorant will strongly affect the representation of other odorants, mostly because of the high percentage of neurons (in relation to the network size) with a shared response across stimuli. To avoid these undesired cross-stimuli effects due to the network limited representational capacity, we conducted the simulations one odorant at a time, resetting to the exact same initial conditions for the network at the beginning of each experiment. Furthermore, because we are interested in the differences that emerge *exclusively* due to changes in the frequency of stimulus presentation, we also got rid of the possible variability steaming from the use of different odorants for the familiar and unfamiliar cases (i.e., different spatiotemporal patterns of MTCs activity), employing the exact same set of inputs in both cases.

In Fig. 5.8 we show analysis for the simulations of the two experimental frameworks above, with results for the familiar and unfamiliar stimuli aver-aged over the same set of 8 different odorants. The first experiment (Fig. 5.8, top row) mimics mice in cohort A, in which familiar odorants are presented every day for the 32-days of the simulations (16 days of "familiarization" + 16 days of experiment), while "unfamiliar" odorants are presented at 8-days interval after day 0. We notice how responses of pyramidal neurons to the unfamiliar odorants became more dissimilar across days as compared to the familiar odorants (Fig. 5.8a). Notably, it can be seen in Fig. 5.8b-c how the drift for the representation of the unfamiliar odorants was was almost twice as fast ($\bar{r} = 2.52 \pm 0.91 \,°/day$) for the representations of unfamiliar odorants as compared to the case in which the odorants were presented daily ($\bar{r} = 1.38 \pm 0.58 \,°/day$).

Results for a second experiment mimicking cohort B, in which familiar odorants are presented at 8-days interval after day 0, are shown in Fig. 5.7 (bottom row). Just as in the experiments of Schoonover *et al*, when the familiar odorants are not presented frequently after the beginning of the experiment, no statistically significant differences are found in the average drift rates with respect to the unfamiliar odorants case.

## 5.6 Conclusions and perspectives

In this chapter we developed a fairly realistic spiking model of the mouse olfactory cortex to study the phenomenon of representational drift (RD), trying to understand the fundamental mechanisms that can give rise to the empirically observed drifting pyramidal population responses during odor encoding

In particular, we showed that a relatively simple weight-dependent STDP rule at the LOT connections, combined with large periods of noisy, resting-state spiking activity at both, the OB and PCx levels, could reproduce not only qualitatively, but also quantitatively, the experimental findings in Schoonover et al. 2021. Within our model —based on the work of Stern et al. 2018, but with added modifications such as the addition of synaptic plasticity— we also showed that no invariant geometry seemed to be present in the across-days comparison of neural population responses, even when population statistics remained fairly invariant across the length of the experiment.

Moreover, we were able to characterize the observed dependence of the drift rate on the frequency of stimulus presentation in terms of the antagonistic effects of STDP-mediated changes during odor presentation and noisy, resting-state baseline activity. Within this picture, drifting representations in the activity space are a result of drifting synaptic weights. This means that long-term potentiation of the excitatory synapses carrying OB inputs to pyramidal neurons that were originally responsive for a given odorant, could potentially correct in some directions of the weight space the accumulated changes due to random drift. This hypothesis, based on the possibility of a learning mechanism involving synaptic plasticity in the LOT connections between the OB and the PCx, is also backed experimentally by recent findings showing that LOT synapses could undergo strong and robust long-term potentiation mediated by only a few local NMDA-spikes (Kumar et al. 2021). Although we are still conducting further analysis to reach a deeper insight into the weight dynamics under STDP rules in our model of olfactory cortex, from the empirical side new experiments that can track the evolution of synaptic efficacies over long periods of time are much needed to elucidate the relation between the observed drift in the neural responses and the hypothesized drift at the synaptic weight level.

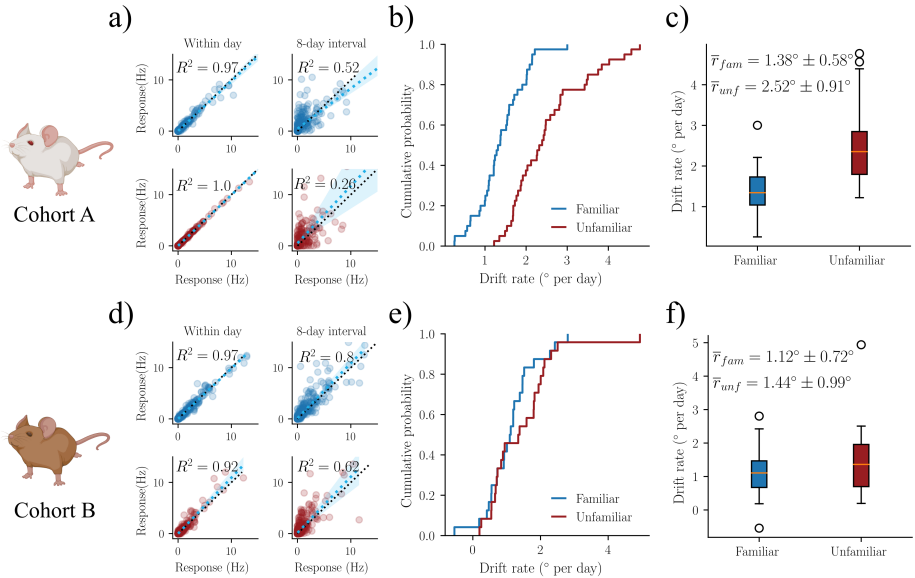

Figure 5.8: **Drift rate changes with the frequency of odor presentation.** For cohort A: **(a)** firing rates for odor-unit pairs within-day and across 8-days intervals; **(b)** cumulative probability distribution for the drift rate; **(c)** box-plot for the corrected rate drift (Eq. 5.18), measured in angles per day. The yellow line inside each box represents the sample median and the whiskers reach the non-outlier maximum and minimum values; the distance between the top (upper quartile) and bottom (lower quartile) edges of each box is the inter-quartile range (IQR). Open circles represent outliers, defined as values that are more than 1.5-IQRs away from the limits of the box. Plots **(d-f)** show the exact same quantities for simulations of cohort B. In all cases, values for familiar (unfamiliar) odorants are plotted in blue (red).

In any case, we remark that, even when we have focused here on synaptic plasticity as the potential mechanism behind RD, preliminary results suggest that forms of structural plasticity such as synaptic turnover —which in biological terms involves the birth and depletion of synaptic spines at the dendritic trees— could account as well for the observed RD in the mouse olfactory cortex. This type of mechanism, however, would not be sufficient on its own to explain the dependence of the drift rate with the frequency of stimulus presentation.

Although, in more general terms, several works have pointed to the possibility of stable memories arising from dynamic synapses (Pantic et al. 2002; Susman et al. 2019), the question also remains of how this changing responses at the piriform cortex can then be readout to generate an stable perception of the encoded odorants. Moreover, do these representations show similar properties of continuity and differentiability of the manifolds, as discussed in the previous chapter? If so, how does RD affect the topology of the original neural manifolds?

Despite our findings, the phenomenon of representational drift can still be considered a very fertile ground which, in my opinion, will benefit strongly in the years to come from further efforts on the theoretical and computational side.

# Epilogue: the artificial *vs* biological brain

It is beyond doubt that we are facing the beginnings of a new revolution in technology, led by incredibly fast-paced developments in artificial intelligence (AI) and machine learning (ML). In fact, from the advent of learning machines —as discussed in our first chapter— to the present day, AI has swiftly permeated its way from the realms of academic research to our everyday life.

As a matter of fact, during fall 2023, just a few weeks before the writing of this epilogue, lexicographers from the *Collins English Dictionary* chose the term *AI* to be the word of the year, as defined by:

> The modelling of human mental functions by computer programs.

Meanwhile, for the *Cambridge Dictionary*, it was not the appearance of a new word, but rather the adaptation of existing ones and their connotations to this new AI-driven era, what made them choose the term *hallucinate* as word of the year 2023:

> When an artificial intelligence (= a computer system that has some of the qualities that the human brain has, such as the ability to produce language in a way that seems human) hallucinates, it produces false information.

Notably, what both these words have in common is the idea that AI seeks to reproduce "the qualities that the human brain has". But how far can we go in emulating the brain? Will AIs ever be able to, not only process and understand information in an intelligent way, but also hold consciousness and self-perception?

In his book, *Shadows of the mind*, Nobel Prize in Physics Roger Penrose distinguished four possible attitudes towards the answer to the above question (Penrose 1994), which can be summarized as:

$\mathcal{A}$. All thinking is computation, and so consciousness can emerge once we reach a certain level of computational complexity.

$\mathcal{B}$. Consciousness can emerge from the brain's physical processes, but even if these process could be modeled computationally, a simulation by itself could not evoke conscious awareness.

$\mathcal{C}$. The physical processes that allow for the emergence of consciousness in human brain cannot be understood in terms of reproducible computations.

$\mathcal{D}$. Consciousness cannot be explained in scientific terms, be physical or computational.

According to $\mathcal{A}$ —commonly known as the *strong*-AI viewpoint—, if an AI can simulate all the external manifestations of a conscious brain, so that even under extensive questioning it behaves as though it possesses consciousness, then one should conclude that such an AI is indeed conscious. On the other hand, supporters of $\mathcal{B}$ (or the *soft*-AI viewpoint) argue that an AI could eventually reproduce the behavior of a conscious individual without ever possessing the emergent quality of consciousness that we observe in human brains. In the case of argument $\mathcal{C}$, as endorsed by Penrose in Penrose 1994, such a perfect simulation of human conscious behavior could just never be achieved from a purely computational approach. As a scientist, I cannot help but follow the author in rejecting the mystical viewpoint sustained by $\mathcal{D}$.

At the basis of arguments $\mathcal{A}$, $\mathcal{B}$ and $\mathcal{C}$ lies the famous Turing test, proposed by the British mathematician Alan Turing in 1950 (Turing 1950). In this thought-experiment, known as the imitation game, human judges would have to hold text-based conversations with a hidden computer and an unseen person. If the judges could not reliably detect which conversation belonged to a human and which to a machine, then the machine should be thought as capable of "thinking".

Remarkably, the latest version of the now widely known generative AI-model created by OpenAI, ChatGPT-4.0 —although still very far from any type of consciousness, self-awareness or genuine understading— has been claimed to have already passed the Turing test (Biever 2023). Such an achievement leaves obsolete the arguments in favor of $\mathcal{C}$, unless one redefines what it means to simulate "human behavior" in any terms different from a Turing test.

Nevertheless, even if one chose a *hard*-AI viewpoint into the problem, the question remains of whether proper consciousness can emerge from machine learning algorithms and architectures that do not seek to reproduce the hu-man brain, but just its functional capabilities, as with the case of Generative Pretrained Transformers (GPTs) and, more generally, Large Language Models (LLMs). And if this is not the case, what makes human brains so unique in that respect?

Along this thesis, we have employed tools from statistical mechanics and the physics of complex systems to reach theoretical explanations for the emergence of scale-invariant properties in the brain, working upon the hypothesis of an existing close-to-critical dynamics in biological neural networks. We have also observed how this type of dynamical regime could be further linked to the existence of optimal neural encodings of external stimuli, in terms of topological properties of the representation manifolds. Interestingly, several works have also identified consciousness as an emergent property in brains with a dynamics that is close to a phase transition (Toker et al. 2022; Walter et al. 2022), going as far as proposing the framework of self-organized criticality (SOC) as a neurodynamical parameter for consciousness (Walter et al. 2022). Thus, a close-to-critical dynamics could be a necessary (although definitely not sufficient) condition for the emergence of consciousness in the human brain.

Even while emerging AI models such as ChatGPT-4.0 are now crossing the trillion parameters, I would argue (perhaps in alignment with some *soft*-AI perpectives) that new insights into the unique properties of the brain will not arise from even larger artificial models, but rather from the introduction of new biologically-inspired algorithms. One of such examples are plasticity-based learning rules which, as we have seen in this thesis, not only can explain empirically-observed phenomena such as the existence of representational drift, but can also improve the performance in certain types of machine learning architectures.

It might be the case that, this time, "more is not different" when it comes to developing AI that can shadow the computational capabilities of human brains. Will we need to uncover what makes our brains so unique before we are capable of building truly intelligent machines? Maybe the time has arrived to finally understand why "different is more".

# APPENDICES

# Making sense of correlations

## I Linking correlations and structure

In this first appendix we will derive a simple proof for Eq. (2.15), which relates the long time window covariance matrix of the linear-rate model (LRM) described by Eq. (2.12), to the connectivity matrix describing the interactions among the units.

We start by rewriting the linear rate model in vector form as:

$$\dot{\mathbf{x}}(t) = (-\mathbb{I} + J)\mathbf{x}(t) + \boldsymbol{\xi}(t) \ , \tag{A.1}$$

where $J$ defines the connectivity matrix, $\mathbb{I}$ is the identity, and $\boldsymbol{\xi}(t)$ is an uncorrelated Gaussian noise with variance $\sigma^2$, such that $\boldsymbol{\xi}(t)\boldsymbol{\xi}^T(s) = \sigma^2\delta(t-s)$. Let us define $A \equiv -\mathbb{I} + J$. The solution for the above stochastic differential equation, which is just a multivariate Ornstein-Uhlenbeck process, can then be written as:

$$\mathbf{x}(t) = e^{-A(t-t_0)}\mathbf{x}(t_0) + e^{-A(t-t_0)}\int_{t_0}^{t} e^{-As}\boldsymbol{\xi}(s) \ . \tag{A.2}$$

Taking $\mathbf{x}(t_0) = 0$ (as in Hu et al. 2020), and assuming the stationary regime has been reached ($t_0 \to \infty$) the solution is just:

$$\mathbf{x}_{st}(t) = e^{-At}\int_{-\infty}^{t} e^{-As}\boldsymbol{\xi}(s) \ . \tag{A.3}$$

**Definition I.1 (Covariance matrix in the stationary regime).** For the Ornstein-Uhlenbeck process defined by Eq. (A.1), the covariance in the stationary regime is the following matrix of scalar products:

$$\hat{\sigma} = \langle \mathbf{r}_{st}(t)\mathbf{r}_{st}^T(t)\rangle \ , \tag{A.4}$$

which is time-independent in the stationary regime, as we will see later.

We now set out to prove that the above covariance matrix follows a Lyapunov equation.

**Lemma I.2.** The covariance matrix for the Ornstein-Uhlenbeck process in the stationary regime, $\hat{\sigma}$, verifies the following **Lyapunov equation**:

$$A\hat{\sigma} + \hat{\sigma}A^T = \sigma^2\mathbb{I} \ . \tag{A.5}$$

*Proof.* This is just a simple calculation:

$$
\begin{aligned}
A\hat{\sigma} + \hat{\sigma}A^T &= \int_{-\infty}^{t} Ae^{-A(t-s)}\int_{-\infty}^{t} \overbrace{\langle\boldsymbol{\xi}(s)\boldsymbol{\xi}^T(t)\rangle}^{\sigma^2\delta(s-z)}e^{-A^T(t-z)}dz \\
&\quad + \int_{-\infty}^{t} e^{-A(t-s)}\sigma^2 e^{-A^T(t-s)}A^T ds \\
&= \int_{-\infty}^{t} \frac{d}{ds}[e^{-A(t-s)}e^{-A^T(t-s)}]ds \\
&= \sigma^2[e^{-A(t-s)}e^{-A^T(t-s)}]_{-\infty}^{t} = \sigma^2
\end{aligned}
\tag{A.6}
$$

$\square$

In particular, from the last lemma and the fact that $A$ is time-independent, one can already derive that $\hat{\sigma}$ must be also time-independent.

To provide a measure of the distance to criticality, in both the works of Dahmen et al. 2019 and Hu et al. 2020, the authors resorted to the computation of long time window covariances. Let us give now a more formal definition of this matrix.

**Definition I.3** (**Long time window covariance**). We define the deviation of the activity of unit $i$ with respect to its mean value in the time window $[t, t + \Delta t]$ as:

$$\Delta s_i(t) = \int_t^{t+\Delta t} x_i(t') - \langle x_i(t') \rangle dt' \ , \tag{A.7}$$

and the long time window (or noise) covariance as:

$$C_{i,j} = \lim_{\Delta t \to \infty} \frac{1}{\Delta t} \langle \Delta s_i(t) \Delta s_j(t) \rangle \ . \tag{A.8}$$

Let us show now how the above definition can be easily treated to find a recipe that allows us to calculate long time window covariances from empirical data. Going back to our cherished LRM, the long time window covariance can now be written as:

$$C_{i,j} = \lim_{\Delta t \to \infty} \frac{1}{\Delta t} \left\langle \int_t^{t+\Delta t} (x_i(s) - \langle x_i(s) \rangle) \, ds \int_t^{t+\Delta t} (x_j(z) - \langle x_j(z) \rangle) \, dz \right\rangle$$

$$= \lim_{\Delta t \to \infty} \frac{1}{\Delta t} \left( \left\langle \int_t x_i(s) \, ds \int_t x_j(z) \, dz \right\rangle - \left\langle \int_t x_i(s) \, ds \right\rangle \int_t \langle x_j(z) \rangle \, dz \right.$$

$$\left. - \left\langle \int_t x_j(z) \, dz \right\rangle \int_t \langle x_i(s) \rangle \, ds + \int_t \langle x_i(s) \rangle \, ds \int_t \langle x_j(z) \rangle \, dz \right) , \tag{A.9}$$

where we took into account that noise averages can only affect variables that have not been averaged over time yet. We can define now the vectors:

$$I_i := \int_t^{t+\Delta t} x_i(s) ds, \quad \chi_i := \int_t^{t+\Delta t} \langle x_i(s) \rangle ds \ , \tag{A.10}$$

where, once again, $\langle \cdot \rangle$ refer to averages over realizations of the noise (i.e., samples). Now, since the averages commute with the integrals, we can finally rewrite the expression for the long time window covariance as:

$$C_{i,j} = \lim_{\Delta t \to \infty} \frac{1}{\Delta t} \left( \langle I_i I_j \rangle - (\chi_i \langle I_j \rangle + \chi_j \langle I_i \rangle) + \chi_i \chi_j \right)$$

$$= \lim_{\Delta t \to \infty} \frac{1}{\Delta t} \left( \langle I_i I_j \rangle - \chi_i \chi_j \right) \ . \tag{A.11}$$

Now, if one calculates the $\chi_i$ terms for the stationary solution we found in Eq. (A.3), it is straightforward to see that they all vanish (in real measurements, this terms are of order $1/\sqrt{M}$, where $M$ is the number of measurements), meaning that only the integrals $I_i$ need to be computed. The above result serves as a recipe to calculate the long time window covariance matrix using the time series data obtained from different realizations of the firing rates, provided the time window is sufficiently big compared to the typical time scale of the series.

At this point, we just need one more result on correlations before setting out to prove Eq. (2.15).

**Proposition I.4 (Correlations at different times).** Consider the firing rates $\mathbf{x}(t)$ defined by Eq. (A.1). We then have:

- **If s>t:** Then $\langle \mathbf{x_{st}}(t)\mathbf{x_{st}^T}(s)\rangle = \hat{\sigma}e^{-A^T(s-t)}$ .

- **If t>s:** Then $\langle \mathbf{x_{st}}(t)\mathbf{x_{st}^T}(s)\rangle = e^{-A^T(s-t)}\hat{\sigma}$ .

*Proof.* We prove the case $s > t$ (the second case is analogous).

$$\langle \mathbf{x_{st}}(t)\mathbf{x_{st}}^T(t)\rangle = \int_{-\infty}^{t} e^{A(z-t)}\boldsymbol{\xi}(z)dz \int_{-\infty}^{s} e^{A^T(u-s)}\xi^T(u)du$$

$$= \int_{-\infty}^{t}\int_{-\infty}^{s} e^{A(z-t)}e^{A^T(u-s)}\sigma^2\delta(z-u)dzdu \qquad \text{(A.12)}$$

$$= \left(\int_{-\infty}^{t} e^{A(z-t)}\sigma^2 e^{A^T(z-t)}\right)e^{-A^T(s-t)} .$$

Now, using the fact that the covariance matrix fulfils the Lyapunov equation, $A\hat{\sigma} + \hat{\sigma}A^T = \sigma^2$, we obtain:

$$\langle \mathbf{x_{st}}(t)\mathbf{x_{st}}^T(t)\rangle = A\Bigg(\int_{-\infty}^{t} e^{-A(t-z)}\hat{\sigma}e^{-A^T(t-z)}dz$$

$$+ \left[\int_{-\infty}^{t} e^{-A(t-z)}\hat{\sigma}e^{-A^T(t-z)}dz\right]A^T\Bigg)e^{-A^T(s-t)} . \qquad \text{(A.13)}$$

And, finally, since:

$$\frac{d}{dz}\left[e^{-A(t-z)}\hat{\sigma}e^{-A^T(t-z)}\right] = Ae^{-A(t-z)}\hat{\sigma}e^{-A^T(t-z)}$$

$$+ e^{-A(t-z)}\hat{\sigma}e^{-A^T(t-z)}A^T , \qquad \text{(A.14)}$$

one can write:

$$\langle \mathbf{x_{st}}(t)\mathbf{x_{st}}^T(t)\rangle = \left\{\int_{-\infty}^{t} \frac{d}{dz}\left[e^{-A(t-z)}\hat{\sigma}e^{-A^T(t-z)}\right]dz\right\}e^{-A^T(s-t)}$$

$$= \hat{\sigma}e^{-A^T(s-t)} . \qquad \text{(A.15)}$$

$\square$

With all the above steps we can proceed and give a formal proof of the relation between covariances and connectivity in the linear rate model framework.

**Theorem I.5** (**Relation between connectivity and long time window covariance**). For a linear-rate model driven by external white noise, as described by Eq. (A.1), the long time window covariance is related to the connectivity matrix $J$ as:

$$C(J) = \sigma^2 (\mathbb{I} - J)^{-1} (\mathbb{I} - J)^{-T} . \tag{A.16}$$

*Proof.* This is now a simple calculation:

$$
\begin{aligned}
C &= \lim_{\Delta t \to \infty} \frac{1}{\Delta t} \int_0^{\Delta t} \int_0^{\Delta t} \langle \mathbf{x_{st}}(t) \mathbf{x_{st}}^T(s) \rangle ds dt \\
&= \lim_{\Delta t \to \infty} \frac{1}{\Delta t} \overbrace{\int_0^{\Delta t} ds \int_0^s dt \hat{\sigma} e^{-A^T(s-t)}}^{t<s(\text{We use previous result})} + \overbrace{\int_0^{\Delta t} dt \int_0^t ds \hat{\sigma} e^{-A^T(s-t)}}^{t>s} \\
&= \lim_{\Delta t \to \infty} \frac{1}{\Delta t} \Delta t \left( \hat{\sigma} A^{-T} + A^{-1} \hat{\sigma} \right) \Rightarrow ACA^T = A\hat{\sigma} + \hat{\sigma} A^T
\end{aligned}
\tag{A.17}
$$

Meaning that:

$$C = \sigma^2 A^{-1} A^{-T} = \sigma^2 (\mathbb{I} - J)^{-1} (\mathbb{I} - J)^{-T} \tag{A.18}$$

$\square$

Although derived here on the basis of linear response theory, we remark that this relation has been proven to be very general, holding as a very good approximation for many different (nonlinear) models, including spiking neural networks (see, for instance, Pernice et al. 2011; Trousdale et al. 2012; Dahmen et al. 2019 and Ocker et al. 2017 for a recent review).

# II   Rank-ordered eigenvalues vs probability densities

In Section 2.2 we argued that if correlations are self-similar inside a cluster of size $K$, then the $n$-th largest eigenvalue, $\lambda_n$, decays as a power-law of its rank, $n$, as:

$$\lambda_n = A \left( \frac{n}{K} \right)^{-\mu} . \tag{A.19}$$

On the other hand, Section 2.4.4 reported on the probability density of the covariance matrix eigenvalues (Eq. (2.17)), maintaining that the tail of the distribution converges to a power-law provided the system is close to the edge of instability:

$$p(\lambda) \sim B\lambda^{-\nu}. \tag{A.20}$$

Here we make explicit the relationship between these two power-laws. We first notice that the spectrum of a matrix of size $K$ is composed by $K$ discrete eigenvalues:

$$p(\lambda) = \frac{1}{K} \sum_{n=1}^{K} \delta\left(\lambda - \lambda_n\right) \ . \tag{A.21}$$

In the limit of $K \to \infty$ the above discrete distribution can be approximated by a continuous one:

$$p_{RC}^{(c)}(\lambda) \approx \frac{1}{K} \int_1^\infty \delta\left(\lambda - \lambda_n\right) dn = \frac{1}{K} \int_1^\infty \delta\left(\lambda - A\left(\frac{n}{K}\right)^{-\mu}\right) dn \ , \tag{A.22}$$

where we used Eq. (A.19) for the rank-ordering of the spectrum . The function $g(n) \equiv \lambda - A\left(\frac{n}{K}\right)^{-\mu}$ has roots $n_0 = K\left(\frac{\lambda}{A}\right)^{-\frac{1}{\mu}}$. Using the following property of the Dirac-delta distribution:

$$\delta\left(g(n)\right) = \frac{\delta\left(n - n_0\right)}{|g'\left(n_0\right)|} = \frac{\delta\left(n - n_0\right)}{K^{-1}A\mu\left(\frac{\lambda}{A}\right)^{\frac{\mu+1}{\mu}}} \ , \tag{A.23}$$

the spectral density reads:

$$p_{RC}^{(c)}(\lambda) \approx \frac{1}{K\mu}A^{\frac{1}{\mu}}K\lambda^{-1-\frac{1}{\mu}} \int_1^\infty \delta\left(n - n_0\right) dn = \frac{1}{\mu}A^{\frac{1}{\mu}}\lambda^{-1-\frac{1}{\mu}} \ . \tag{A.24}$$

Thus, by comparing Eq. (A.20) with Eq. (A.24), one readily finds a relationship between the power law exponent $\mu$ of the rank-ordered eigenvalues and the exponent of the spectral distribution $\nu$:

$$\nu = 1 + \frac{1}{\mu} \ . \tag{A.25}$$

For a more detailed discussion of this scaling relation see, for instance, W. Li 2002 and De Marzo et al. 2021.

## III   Cross-validated PCA

Let us consider two observation matrices, $X_{(1)}, X_{(2)} \in \mathbb{R}^{N \times T}$, corresponding to two identical realizations or trials of an experiment. For simplicity, we further assume that the mean activity of each neuron across images has been subtracted, so that both matrices have zero-mean rows. The

Singular Value Decomposition (SVD) theorem states that any observation matrix $X$ can be decomposed as:

$$X = USV^T , \qquad (A.26)$$

so that

$$XV = US \implies X\mathbf{v}_i = \sigma_i \mathbf{u}_i , \qquad (A.27)$$

where $U \in \mathbb{R}^{N \times r}$ contains the $\mathbf{u}_i$ eigenvectors by columns, $V \in \mathbb{R}^{T \times r}$ contains the $\mathbf{v}_i$ eigenvectors by columns, and $S \in \mathbb{R}^{r \times r}$ is a diagonal matrix containing the singular values $s_{ii} = \sigma_i = \sqrt{\lambda_i}$.

Let us assume one has performed SVD on the first trial observation matrix $X_{(1)}$, obtaining the singular vectors $U_{(1)}$ and $V_{(1)}$. Defining the projection matrix $P = U_{(1)}^T \in \mathbb{R}^{r \times N}$, which diagonalizes $C_u^{(1)} = \frac{1}{T-1} X_{(1)} X_{(1)}^T$, one can then compute:

$$Y_{(1)} = PX_{(1)} \implies \mathbf{y}_i^{(1)} = (\mathbf{u}_i^{(1)})^T X_{(1)} \in \mathbb{R}^{1 \times T} , \qquad (A.28)$$

where $Y_{(1)} \in \mathbb{R}^{r \times T}$ contains the projection of the first trial activity over its $r$ principal components. Now, the idea of cv-PCA is to project on the same subspace the observations coming from the second trial (Stringer et al. 2019a):

$$Y_{(2)} = PX_{(2)} \implies \mathbf{y}_i^{(2)} = (\mathbf{u}_i^{(1)})^T X_{(2)} \in \mathbb{R}^{1 \times T} , \qquad (A.29)$$

and ask to form a covariance matrix with the product of both projections:

$$C_\psi = \frac{1}{T-1} Y_{(1)} Y_{(2)}^T = \frac{1}{T-1} \begin{pmatrix} \mathbf{y}_1^{(1)} \cdot \mathbf{y}_1^{(2)} & \mathbf{y}_1^{(1)} \cdot \mathbf{y}_2^{(2)} & ... \\ \mathbf{y}_2^{(1)} \cdot \mathbf{y}_1^{(2)} & \mathbf{y}_2^{(1)} \cdot \mathbf{y}_2^{(2)} & ... \\ ... & ... & \mathbf{y}_r^{(1)} \cdot \mathbf{y}_r^{(2)} \end{pmatrix} . \qquad (A.30)$$

One can now hypothesize that the activity at time $t$ of any neuron $i$ during trial $k$ can be linearly decomposed as:

$$\mathbf{x}_i^{(k)}(t) = \psi_i(t) + \epsilon_i^{(k)}(t) , \qquad (A.31)$$

where $\psi_i$ denotes the input-related activity, which should be independent of the trial for experiments carried in identical conditions; and $\epsilon_i^{(k)}$ is defined as background or trial-to-trial variable activity, spanning a subspace that is orthogonal to the input-related one. Let us now show that the i-th diagonal element of $C_\psi$ is a non-biased estimator of the input-related variance $\omega_i$:

$$\omega_i = \frac{1}{T-1} \mathbf{y}_i^{(1)} \cdot \mathbf{y}_i^{(2)} = \frac{1}{T-1} (\mathbf{u}_i^{(1)})^T X_{(1)} X_{(2)}^T \mathbf{u}_i^{(1)} . \qquad (A.32)$$

Rewriting the observation matrices as $X_{(k)} = \Psi + \Sigma_{(k)}$, where $\Psi$ contains vectors $\psi_i$ in rows and so does $\Sigma_{(k)}$ with vectors $\epsilon_i^{(k)}$:

$$
\begin{aligned}
\omega_i &= \frac{1}{T-1} \mathbf{u}_i^T \left( \Psi + \Sigma_{(1)} \right) \left( \Psi + \Sigma_{(2)} \right)^T \mathbf{u}_i \\
&= \frac{1}{T-1} \mathbf{u}_i^T \Psi \Psi^T \mathbf{u}_i + \frac{1}{T-1} \mathbf{u}_i^T \Sigma_{(1)} \Psi^T \mathbf{u}_i \ ,
\end{aligned}
\tag{A.33}
$$

where all the terms containing $\Sigma_{(2)}$ can be dismissed due to the statistical independence of the realizations. Under reasonable approximations shown in Stringer et al. 2019a, one can prove that if the singular vectors $\mathbf{u}_i$ approach the singular vectors of $\Psi$, then the first term converges to the actual variance of the input-related activity along the principal direction $\mathbf{u}_i$, while the second term converges to zero. We have thus seen how the cvPCA method proposed by Stringer *et al.* allows one to estimate the input-related covariances.

Is it possible to take a step further and find a proxy for the input-related activity observation matrix $\Psi \in \mathbb{R}^{N \times T}$, such that $\tilde{C}_\psi = \frac{1}{T-1} \Psi \Psi^T$ has the same eigenvalue spectrum as $C_\psi$? To do so, we first define a new basis of projected vectors given by:

$$
\mathbf{z}_i = \sqrt{\mathbf{y}_i^{(1)} \cdot \mathbf{y}_i^{(2)}} \ \frac{\mathbf{y}_i^{(2)}}{\left\| \mathbf{y}_i^{(2)} \right\|} \in \mathbb{R}^{1 \times T} \ .
\tag{A.34}
$$

which clearly fulfills $\mathbb{E} \left[ \mathbf{z}_i \cdot \mathbf{z}_i^T \right] = \omega_i$ under the same conditions as above. If $Z \in \mathbb{R}^{r \times T}$ is the matrix composed by the $r$ vectors $\mathbf{z}_i$ in rows, then:

$$
\Psi = P^T Z \ ,
\tag{A.35}
$$

is the input-related activity of the neurons. From there, it is straightforward using Eq. (A.31) that one can estimate the background activity just by subtracting the input-related activity from the raw data:

$$
\Sigma_{(k)} = X_{(k)} - \Psi \ .
\tag{A.36}
$$

# Supplementary Information to Chapter 2

## I   Experimental protocols

### ∗   Steinmetz *et al.*'s dataset

In the study by Steinmetz *et al.*, initial surgery was performed under anesthesia to implant a steel headplate and a 3D-printed recording chamber on mice, which facilitated the stability in the recording conditions. Recordings of neural activity were then performed by inserting Neuropixels electrode arrays into the left hemisphere of the brain. The use of several Neuropixels probes, each coated with DiI for later track localization and presenting 384 selectable recording sites, allowed for a number of brain regions being recorded simultaneously in each session (Steinmetz et al. 2019). Neural data was then processed with Kilosort (Pachitariu et al. 2016) and curated manually, with recording sites localized to brain regions through histological analysis, complemented by alignment with the Allen Institute Common Coordinate Framework.

The study was conducted in the context of a two-alternative unforced choice task, although for the analysis presented in Chapter 2 we filtered the data leaving only the intervals in the recordings corresponding to spontaneous, resting-state activity. Adding all sessions and mice together, a total of 42 brain regions were recorded, capturing the activity of 29,134 neurons. In our analysis, we limited the study to those sessions and regions in each mice containing more than 128 simultaneously recorded neurons.

### †   Soriano's group dataset

The dataset analyzed in Section 2.6 is based on the unpublished master thesis work of M. Olives Verger, from J. Soriano's lab. In this section we provide details of the corresponding protocols employed, which are in any case similar those used in previous works of J. Soriano's group (see, for instance, Montalà-Flaquer et al. 2022).

For each type of topographical pattern, masks were created from graphic designs and used in a lithography process to transfer the design onto a silicon wafer. polydimethylsiloxane (PDMS) was then mixed, poured onto the patterned wafer, and cured to form a negative topographic mold with the desired structure, whose surface was treated with oxygen plasma and coated with poly-D-lysine to enhance neuron adhesion.

Neurons from rat cerebral cortex were then dissociated and seeded onto the PDMS surfaces. Recording of neural activity using fluorescence calcium imaging techniques did not begin until day 5 after the culture seeding (to allow for reliable fluorescence signals), and lasted for 13 more days, after which neurons began to degrade or detach from the substrate. During the recording process, 732 region of interests (ROIs) were identified, and for each of these regions the mean fluorescence intensity was measured over a 15-minute period each day.

## ‡   The OMEGA dataset

The full OMEGA dataset contains about 900 resting-state magnetoencephalography (MEG) recording sessions, for a total of over 75 hours of data (Niso et al. 2016). The access to the dataset was granted for the present study for 12 months,and reviewed and approved by the internal research ethics board.

The recordings were collected for 294 volunteering participants, out of which 161 subjects are healthy controls, 127 subjects were diagnosed Parkinson's disease, and 7 participants were diagnosed with chronic pain. The recordings were taken with a 275-channel 2005 series CTF MEG system at the McConnell Brain Imaging Center, at a time resolution of 2400Hz. The data is structured according to the standard BIDS 1.7.0 and was preprocessed with the open-source software *Brainstorm* (Tadel et al. 2011).

Following the recommended protocol in Tadel et al. 2011, we applied a Notch filter at frequencies of 60, 120, 180, 240 and 300Hz to remove the noise due to the AC power line frequency in Canada, and a high-pass filter at 0.3Hz, 60dB. Moreover, since the dataset contains simultaneous bipolar Electrocardiogram (ECG) and vertical and horizontal bipolar Electrooculogram (EOG) recordings, we cleaned the MEG data from the artifacts due to heartbeats and eye blinks. The artifact cleaning procedure consists in detecting reproducible stereotyped and localized topographies, that correlate with the signals of the EOG and ECG, through a Signal-Space Projection (SSP).

# II   Choosing the right time bin

In order to determine pairwise correlations from empirical neural activity data, it is necessary to discretize the time into bins of certain length $\Delta t$. In Chapter 2, we obtained a spike train vector from a list of spike times $t_j^*$ using a common value of $\Delta t$ for all units belonging to the same brain region. The number of bins for a neuron can be therefore chosen as $T_i = \frac{max(t_j^*)}{\Delta t}$ and, more generally, we set $T = max(T_i)$ so that all neurons have the same amount of bins in their spike trains.

Choosing the time bin that best transforms the spike times into discrete trains of spikes is an open problem in neuroscience for which several solutions have been proposed: from the use of random bins (Tamura et al. 2012), to methods that find the best bin size by minimizing a certain cost function (Omi et al. 2011; Cubero et al. 2020; Schölkopf et al. 2007; Ghazizadeh et al. 2020), and bin-less approaches (Victor 2002; Paiva et al. 2010). The problem becomes even more prominent when working with the simultaneously-recorded activity of a large number of cells as, typically, one finds neurons operating at broadly different time scales within the same area (see Fig. B.1), as well as a hierarchy of timescales across regions (Kiebel et al. 2008; Spitmaan et al. 2020).

Since we are interested in a measure of the "typical" time scale at which neurons in a population operate, and given also the large neuron-to-neuron variability (see single-cell histograms in Fig. B.1), we define the optimal time bin $\Delta t$ as the geometric mean of all inter-spike interval (ISI) distribution from neurons in the population which (black dashed line in Fig. B.1). This geometric mean value is computed for each region and subsequently used in the corresponding RG analysis of Chapter 2 (see Table B.1 for a summary of all regions with their corresponding selected time bin).

On the other hand, the computation of each region's distance to criticality relies on the equivalence between time-lagged covariances and spike-count covariances, a relation that holds exactly only in the limit of very large observation times $\Delta t \to \infty$, but can be applied in practice provided the autocorrelation functions for the neurons decay within the chosen time bin. Thus, we seek to find the minimum bin size, $\Delta t$, that grants stationary spiking statistics while maximizing the number of samples. For this, spike-trains were first smoothed using a Gaussian kernel with $\sigma = 50ms$ to obtain a firing rate time series. Then, for each neuron, the autocorrelation function of the activity was computed. For each region, single-cell curves were then averaged across all neurons to obtain an estimate of the autocorrelation for the population activity (see inset in Fig. 2.4). The resulting decaying function of the lag was fitted to an exponential, extracting a characteristic time scale, $\tau_{Corr}$, for each experiment within a region. Finally, these values were averaged across experiments to obtain the mean autocorrelation decay time in each region (see Table B.1). At the light of these results, we chose an optimal bin width $\Delta t = 1s$, which we then employed across all distance-to-criticality estimations when computing spike-counts covariances.

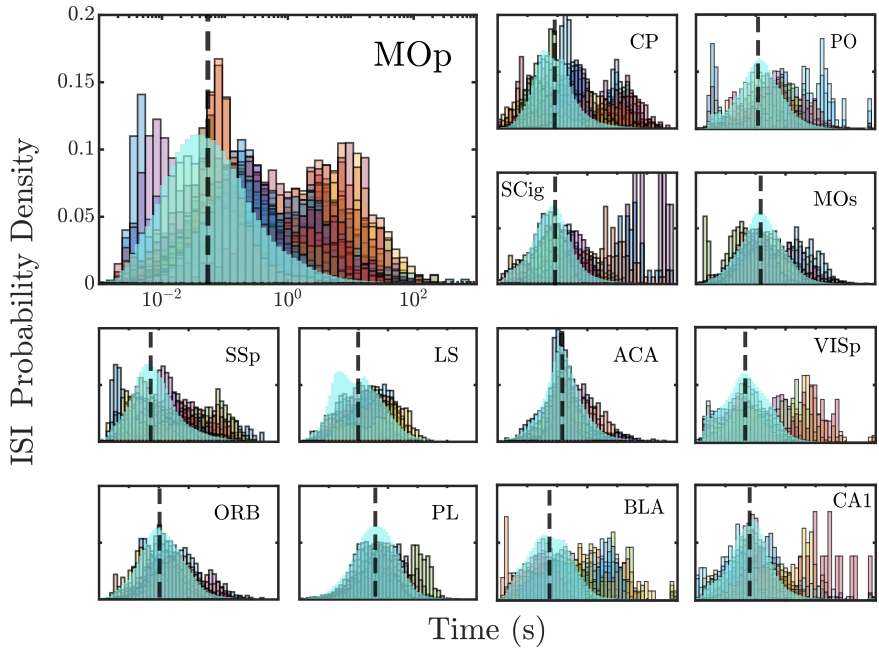

Figure B.1: Histograms estimating the probability of finding a certain value of the inter-spike interval (ISI) in the resting-state activity of individual neurons belonging to a particular area (results shown for 13 out of 16 regions). Each colored histogram in the background represents the distribution for a single neuron (only a few neurons are shown for the sake of clarity). The distribution for the ISIs across all neurons is plotted in light blue, with a black dashed line marking the geometric mean of the ISIs distribution in each area. The later will define our measure for the "optimal" time bin.

To assess whether this choice of $\Delta t$ is indeed sufficient to guarantee the stationarity of the recordings in each region, we performed an augmented Dickey-Fuller (ADF) test (Dickey et al. 1979). This type of statistical test evaluates the null hypothesis that a unit root is present in a time series (i.e., it has some time-dependent structure and it is therefore non-stationary). The alternative hypothesis, should the null hypothesis be rejected, is that the time series is stationary. The *adfuller* function from the *statsmodel* Python library was used to perform the analysis.

Thus, for each neuron in a particular region and experiment we tested the null hypothesis at a 5% significance level, then counted the fraction of neurons in the experiment for which the null hypothesis could not be rejected (i.e., showed non-stationary activity). This fraction was then averaged over all experiments to estimate the percentage of neurons with non-stationary activity for a given region (see Table B.1), so that only those experiments with less than 10% of the neurons showing non-stationary spiking statistics were considered in subsequent analyses.

# III   Extended results on scale invariance

## ∗   Scaling of autocorrelations

Fig. B.2 shows the auto-correlation function at different steps $k$ of the PRG, computed as:

$$C^{(k)}(t) = \frac{1}{N_k} \sum_{i=1}^{N_k} \frac{\langle x_i^{(k)}(t_0) x_i^{(k)}(t_0+t) \rangle - \langle x_i^{(k)} \rangle^2}{\langle \left( x_i^{(k)} \right)^2 \rangle - \langle x_i^{(k)} \rangle^2} , \qquad (B.1)$$

and normalized to lie within the unit interval. In each subplot, corresponding to a region of the mouse brain, the above quantity is plotted against the re-scaled time, $t/\tau_c$, where $\tau_c$ is the characteristic time as obtained from a fit of the autocorrelations to an exponential decay. For ease of visualization, the errors —computed as the standard deviation over random (non-shuffled) quarters of the data— are only shown in the last step of coarse-graining procedure.

In Fig. B.3, the characteristic correlation time, $\tau_c$, is plotted for all the considered regions at different steps of the the PRG analysis as a function of the number of neurons inside the coarse-grained variables, $K$, together with a best fit of $\tau_c(K)$ to Eq. (2.9) in the main text.

## †   Scaling of eigenvalue spectra

In Fig. B.4 we show the observed scaling in the spatial correlations inside the clusters, as measured by the evolution of the average correlation eigenspectrum across PRG steps. Observe not only the decay of the rank-ordered eigenvalues as a power law of the rank, but also the excellent collapse of the cut-offs obtained after rescaling the eigenvalue rank by the total size $K$.

# IV   Control analysis on quasi-universal exponents

## ∗   A test for powerlaw distributions

In the limit of $N \to \infty$ there exists a direct relationship between the exponent of the rank-ordered eigenvalues and their probability density (see

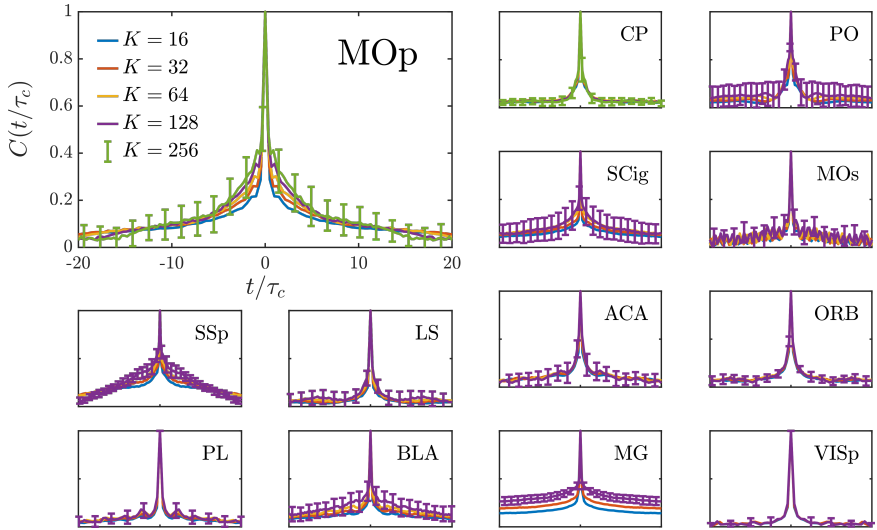

Figure B.2: Mean normalized correlation function of coarse-grained variables during the RG flow, measured in 13 different areas of the mouse brain. Time is re-scaled for each curve by the characteristic time scale $\tau_c(K)$, computed as the $1/e$ point of the decay. Errors shown in the last step of coarse-graining are computed as the standard deviation over random quarters of the data

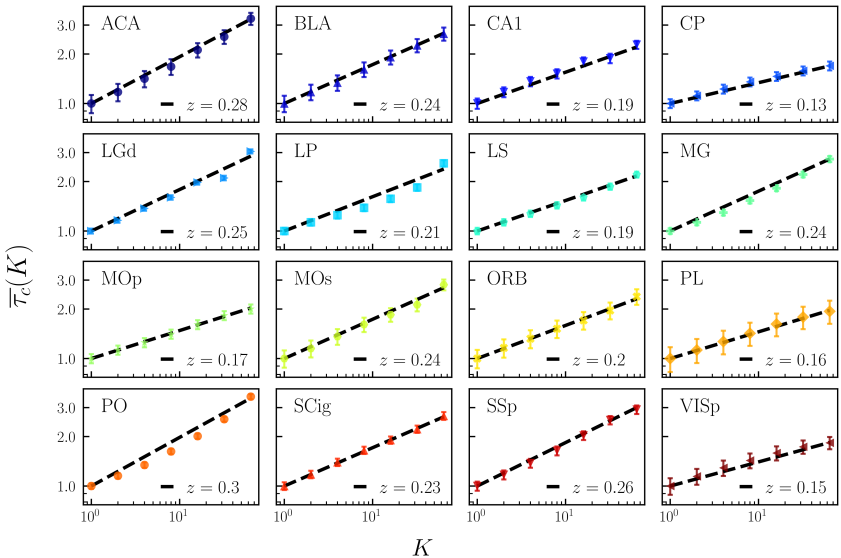

Figure B.3: Scaling of the characteristic correlation time as a function of $K$ in double logarithmic scale for the different regions of the mouse brain. To facilitate the comparison between regions, correlation time as been normalized as $\overline{\tau}_c(K) = \tau_c(K)/\tau_c(K = 2)$. Errorbars are computed as the standard deviation over random split-quarters of the data.

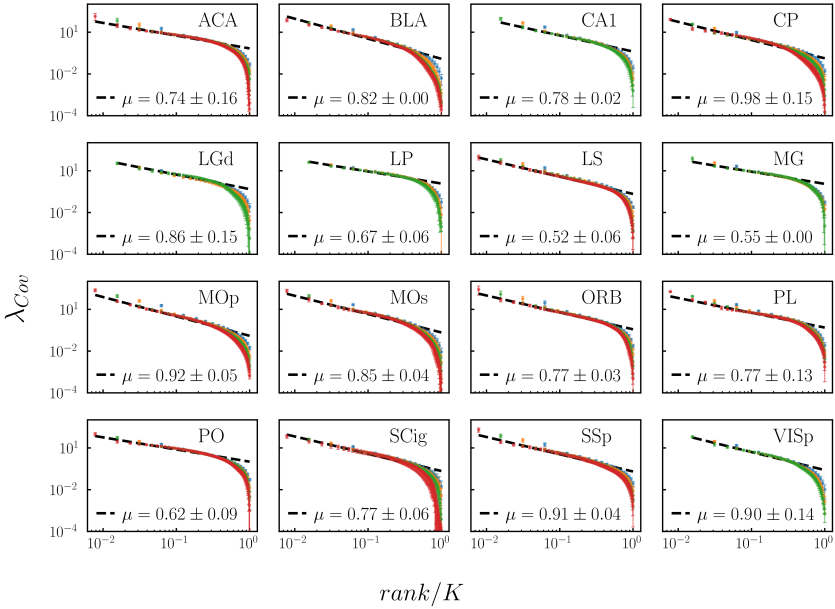

Figure B.4: Scaling of the covariance matrix spectrum for the resting-state activity in clusters of size $K = 16, 32, 64, 128$ (blue, yellow, green and red markers, respectively) in 16 different brain regions. For each region, we measured the average powerlaw exponent and its standard deviation across experiments (see Table B.2), while errorbars (often smaller than the marker size) are computed as the standard deviation across split-quarters of data.

Appendix A.II). Therefore, we leveraged the approach in Clauset et al. 2009 to test whether a power-law distribution is indeed the best description of the density of eigenvalues for the different regions in the data set. Following Clauset et al. 2009 we computed the loglikelihood ratio (LR) between the estimated power-law and equivalent exponential and lognormal candidate distributions, together with the the $p$-value for the significance of the test. In 14 out of the 16 regions, the exponential distribution provides a worse fit —tested at a 5% significance level— than the corresponding power-law (LR > 0 ; $p \leq 0.05$). On the other hand, lognormal distributions always provide an equally good description of the eigenvalue density ($p > 0.05$), and cannot be ruled out as the actual underlying distribution. These values, together with the $R^2$ value for the best powerlaw and exponential fits of the rank-ordered eigenvalues in each region, are listed in Table B.2, whereas the corresponding fitted curves are plotted in Fig. B.5.

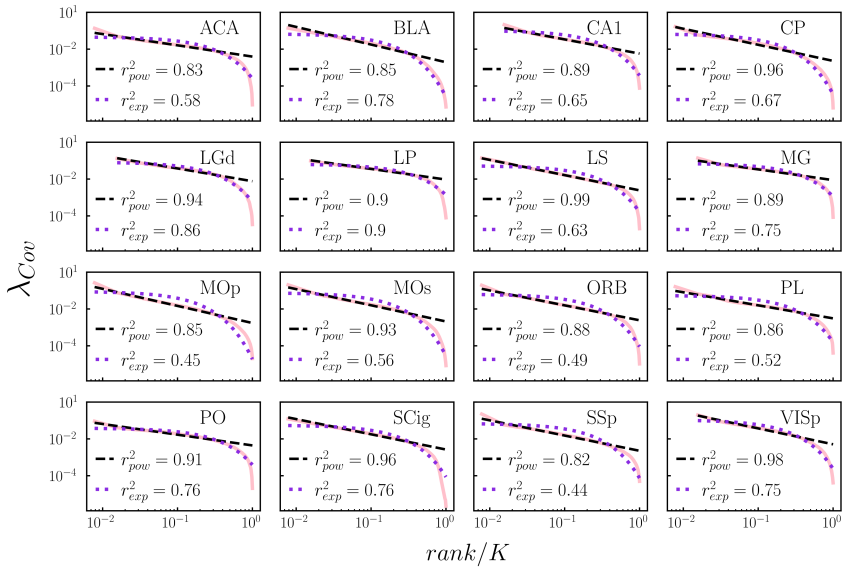

Figure B.5: Best power-law (black dashed line) and exponential (purple dotted line) fits for the covariance matrix eigenvalues at the last step of RG (pink line), together with their corresponding R-Squared values. For each region, the experiment with the greatest number of recorded neurons was considered.

## † Sensibility to time bin choice

Activity of neurons, even within the same brain region, can span many time scales, with some neurons typically firing in the range of milliseconds and with a characteristic time scale of the order of seconds (see Fig. B.1). Thus, in selecting a common time bin to convert the spike times of all neurons within a region into discrete trains of spikes, we irretrievably lose some information.

To assess whether the exponent values reported in the main text are robust against our choice of $\Delta t$, we repeated the same RG analyses using different time-windows $\Delta t \in [0.01s, 4s]$ to bin the resting-state activity of each region.

As illustrated in Fig. B.6, the values of the exponents $\alpha$, $\beta$ and $z$ turn out to be fairly robust within a broad range of biologically plausible time scales. As one could trivially expect, broader time bins increase the probability of finding neurons spiking simultaneously (i.e., appearing as more strongly correlated). This, in turn, has the effect of shifting the values of the exponents $\alpha$ and $\beta$ for the variance and free-energy scalings slightly towards those expected for a fully correlated scenario. On the other hand, the exponent $\mu$ remains fairly unchanged for binning times up to $\sim 100ms$, but then increases on longer time scales beyond the typical geometric mean

ISI of the regions.

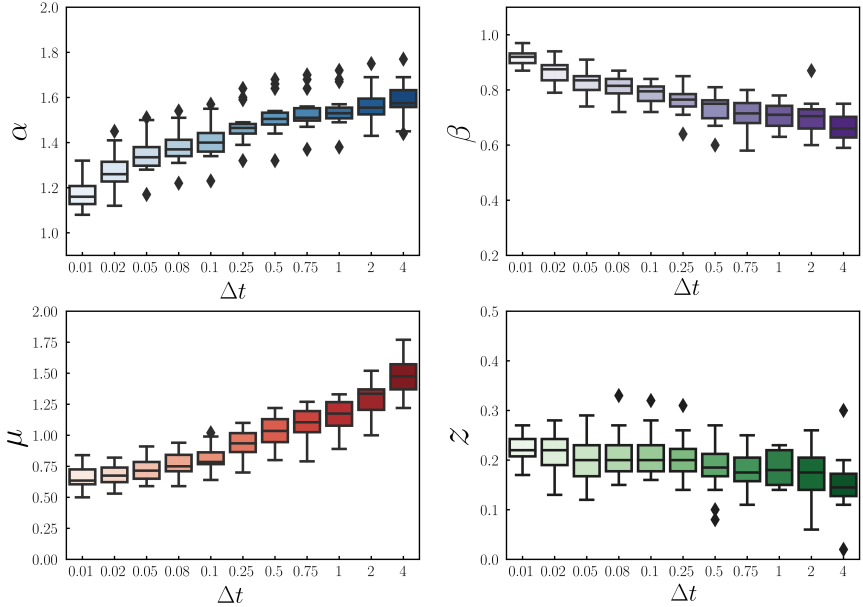

Figure B.6: Change of average scaling exponents as a function of the employed time bin within the range $\Delta t \in [0.01s, 4s]$ for each region. The line inside of each box represents the sample median, the whiskers reach the non-outlier maximum and minimum values; the distance between the top (upper quartile) and bottom (lower quartile) edges of each box is the inter-quartile range (IQR). Black diamonds represent outliers, defined as values that are more than 1.5 times the IQR away from the limits of the box.

## ‡ Sensibility to samples-to-neurons ratio

Fig. B.7 shows the dependence of the scaling exponent $\mu$ for the rank-ordered eigenvalues of the covariance matrix with the samples-to-neurons ratio, $a_0 = T/N$. Estimates for the exponent converge to an asymptotic value only for sufficiently large ratios $a_0 > 20$, while $\mu$ typically increases as one moves towards the sub-sampled regime.

In the inset of Fig. B.7 we compare the fitted exponents for the rank-ordered plot of eigenvalues in the example region MOp when i) taking the activity recorded from each of the 4 intervals of spontaneous activity in the recording session, ii) merging the 4 intervals together and iii) averaging over quarter splits of the data set where the intervals were merged together. For individual spontaneous intervals, the requirement $a_0 > 20$ is not always fulfilled, thus causing variability in the fitted exponents. By merging together all recordings of spontaneous activity belonging to the same experiment we can typically ensure that $a_0 > 20$, even when analyses are performed over quarter-splits of data. This confirms the consistency and robustness of the estimated exponents.

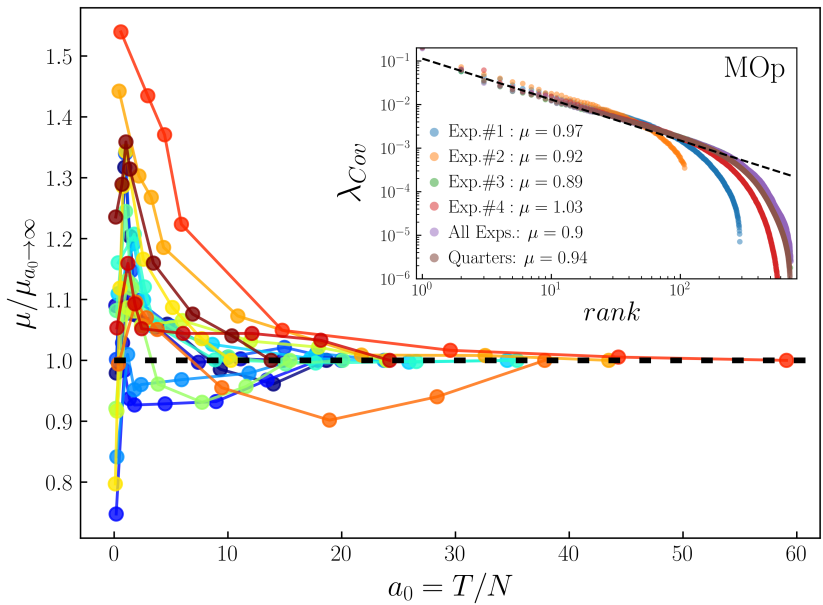

Figure B.7: For each region, the scaling exponent for the covariance matrix eigenvalues at the last step of RG is plotted against the ratio samples-to-neurons in sub-sampled experiments, then re-scaled by its expected value when $a_0 \to \infty$ (estimated by using the full length of the time series). For each region, the experiment with the greatest number of recorded neurons was considered. Inset: Rank-ordered plot of eigenvalues in the MOp region using all the neurons while taking (i) each of the four recordings belonging to a experiment separately; (ii) merging them together; and (iii) averaging over quarter-splits of the data. Line fit is over quarters of the data.

## § Resting-state vs task-induced activity

We plotted in Fig. B.8 the four measured scaling exponents in task-induced vs resting-state type of activity, with results that clearly support the idea that the observed exponents are almost invariant with the mouse behavioral state, since only small deviations are observed in a few regions.

## ¶ RG analyses on surrogated data

As a sanity check, we performed the RG analysis over surrogated data in three different ways:

- Randomly shuffling all the spikes for each neuron.

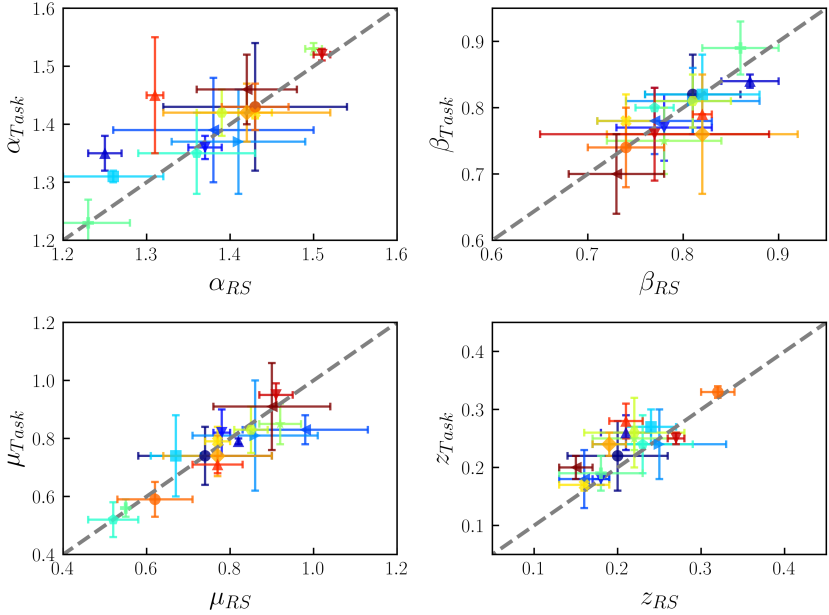

Figure B.8: Scaling exponents for task-induced vs resting-state type of activity in 16 different regions of the mouse brain. Markers and colors for each area have been chosen as defined by Fig. 2.2 in Chapter 2. Error bars are defined as the standard deviation of the corresponding quantity across all experiments belonging to the same region.

- Shifting all the spikes of each neuron by a random time interval, so that the structure within the spike train is preserved.

- Randomizing the spikes across neurons but not time, so that the mean and variance of the population firing rate is preserved.

For these analysis, we chose the MOp region for having the greatest number of recorded neurons. We observe that shuffling the spikes in each neuron (Fig. B.9b); shifting them (Fig. B.9c); or randomizing their label across neurons (Fig. B.9d) destroys the non-trivial scaling properties observed in the unsurrogated case (Fig. B.9a), leading to exponents $\alpha$, $\beta$ and $z$ akin to those expected for a model of uncorrelated neurons. The power-law spectrum of the clusters covariance matrix still seems to decay with a smaller, non-trivial exponent, in the shuffling and shifting scenarios —likely due to the firing rate heterogeneity in the population activity—, but closer inspection reveals that a power-law dependence is no longer suitable when compared to the unsurrogated case.

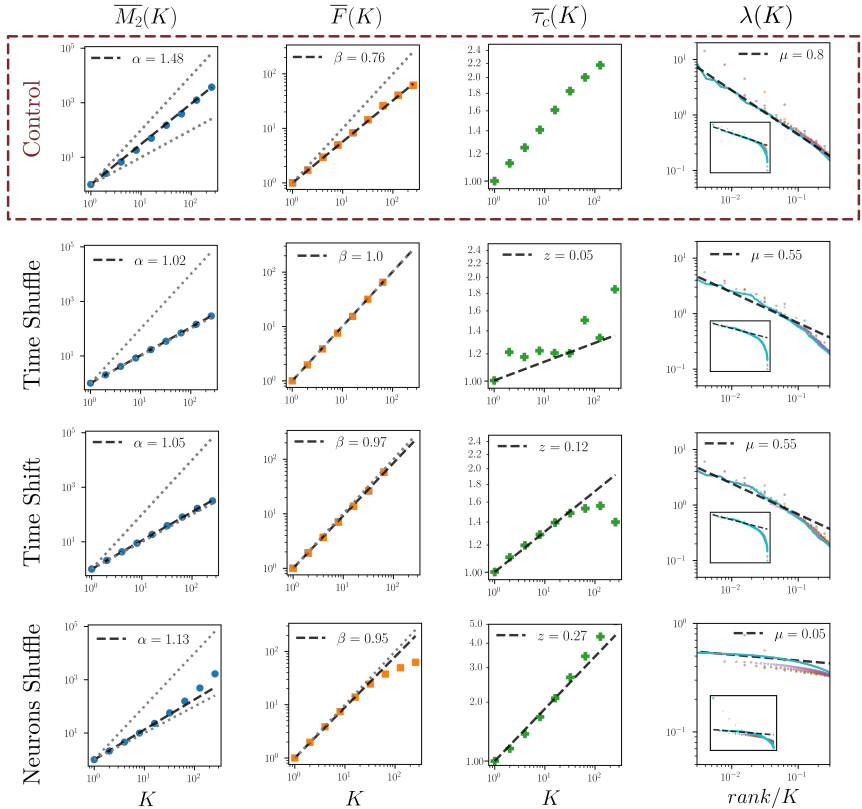

Figure B.9: RG analysis over **a** original, control data, and surrogate data created by **(a)** randomly shuffling all the spikes within each neuron; **(b)** shifting all the spikes series of each neuron by random time lag, so that the structure within the spike train is preserved; and **(c)** randomizing the label of the spikes across neurons but not time. In columns, from left to right, we plot the scaling of the normalized variance for the coarse-grained activity, $\overline{M_2}(K)$ (Eq. (2.4)); the scaling of the normalized "free-energy" with the cluster size (Eq. (2.5)); the scaling of the normalized characteristic autocorrelation time, $\overline{\tau_c}(K)$ (Eq. (2.9)); and the covariance matrix spectrum for different steps of the RG flow, with he expected power law fit against the normalized rank (Eq. (2.11), inset: full spectrum, main: close-up).

‖ **Tables of analysis and results**

Table B.1: **Distance to criticality analysis.** For each region we show: **(i)** number, $M$, of available experiments; **(ii)** average number of neurons, $N$; **(iii)** average duration, $t_{max}$, of the experimental recordings; **(iv)** characteristic time scale of the autocorrelations; **(v)** percentage of non-stationary neurons; **(vi)** distance to criticality, using the method in Dahmen et al. 2019; **(vii)** distance to criticality, using the method in Hu et al. 2020; **(viii)** distance between the sampled eigenvalue distribution and the theoretical expression proposed in Hu et al. 2020, as given by the Cramer-von Mises statistic; **(ix)** same as in (viii), but comparing the empirical distribution to a best-fitting MP distribution; **(x)** distance to criticality extrapolated to a common system size of $N = 10^4$ neurons. All errors are given as the standard deviation across available experiments. For a given region each experiment corresponds to a different mouse, and only mice with more than $N = 128$ recorded neurons were selected.

| Full name | Abbrev. | $M$ | $N (\times10^2)$ | $t_{max} (\times10^2 s)$ | $\tau_{Corr}(ms)$ | % N.S. | $\hat{g}_d$ | $\hat{g}_s$ | $L^2_{HC}$ | $L^2_{MP}$ | $\hat{g}^{N=1e4}$ |
|---|---|---|---|---|---|---|---|---|---|---|---|
| Anterior cingulate area | ACA | 3 | 1.8 ± 0.8 | 7.5 ± 3.2 | 0.10 ± 0.02 | 1.2 ± 0.5 | 0.70 ± 0.09 | 0.74 ± 0.09 | 0.07 ± 0.02 | 0.11 ± 0.04 | 0.965 ± 0.013 |
| Basolateral amygdalar nucleus | BLA | 2 | 2.7 ± 0.0 | 8.1 ± 0.5 | 0.09 ± 0.01 | 1.3 ± 0.6 | 0.81 ± 0.00 | 0.86 ± 0.04 | 0.07 ± 0.02 | 0.17 ± 0.02 | 0.969 ± 0.000 |
| Cornu ammonis | CA1 | 2 | 1.9 ± 0.2 | 8.8 ± 2.9 | 0.07 ± 0.01 | 7.7 ± 0.3 | 0.85 ± 0.04 | 0.83 ± 0.05 | 0.06 ± 0.01 | 0.15 ± 0.02 | 0.980 ± 0.004 |
| Caudoputamen | CP | 4 | 4.1 ± 1.0 | 11.0 ± 1.2 | 0.09 ± 0.01 | 2.6 ± 1.4 | 0.91 ± 0.02 | 0.93 ± 0.01 | 0.09 ± 0.02 | 0.23 ± 0.01 | 0.983 ± 0.002 |
| Lateral septal nucleus | LS | 3 | 2.6 ± 0.5 | 6.4 ± 2.9 | 0.05 ± 0.03 | 2.1 ± 0.5 | 0.90 ± 0.02 | 0.83 ± 0.02 | 0.05 ± 0.00 | 0.15 ± 0.00 | 0.972 ± 0.007 |
| Dorsal part of the lateral geniculate complex | LGd | 5 | 1.6 ± 0.1 | 8.3 ± 1.0 | 0.08 ± 0.01 | 1.6 ± 1.6 | 0.77 ± 0.04 | 0.82 ± 0.06 | 0.09 ± 0.02 | 0.12 ± 0.03 | 0.962 ± 0.009 |
| Lateral posterior nucleus of the thalamus | LP | 2 | 1.8 ± 0.1 | 9.4 ± 0.7 | 0.08 ± 0.01 | 0.3 ± 0.3 | 0.72 ± 0.06 | 0.66 ± 0.06 | 0.07 ± 0.04 | 0.06 ± 0.00 | 0.983 ± 0.002 |
| Medial geniculate complex of the thalamus | MG | 2 | 2.2 ± 0.2 | 10.0 ± 2.4 | 0.09 ± 0.01 | 3.4 ± 3.4 | 0.74 ± 0.07 | 0.73 ± 0.00 | 0.09 ± 0.02 | 0.08 ± 0.01 | 0.962 ± 0.009 |
| Primary motor area | MOp | 3 | 4.2 ± 2.2 | 11.5 ± 9.9 | 0.12 ± 0.02 | 1.3 ± 0.5 | 0.93 ± 0.01 | 0.91 ± 0.02 | 0.09 ± 0.01 | 0.27 ± 0.02 | 0.987 ± 0.002 |
| Secondary motor area | MOs | 5 | 2.5 ± 0.7 | 8.5 ± 3.2 | 0.10 ± 0.01 | 4.3 ± 1.8 | 0.84 ± 0.02 | 0.84 ± 0.03 | 0.06 ± 0.01 | 0.16 ± 0.02 | 0.975 ± 0.000 |
| Orbital area | ORB | 3 | 2.7 ± 0.3 | 9.5 ± 0.8 | 0.09 ± 0.02 | 1.8 ± 0.7 | 0.87 ± 0.02 | 0.80 ± 0.02 | 0.05 ± 0.01 | 0.13 ± 0.01 | 0.980 ± 0.003 |
| Prelimbic area | PL | 4 | 2.4 ± 0.5 | 8.7 ± 1.6 | 0.12 ± 0.04 | 3.9 ± 2.3 | 0.83 ± 0.05 | 0.80 ± 0.08 | 0.04 ± 0.01 | 0.14 ± 0.05 | 0.974 ± 0.010 |
| Posterior complex of the thalamus | PO | 3 | 2.8 ± 1.1 | 10.1 ± 1.0 | 0.10 ± 0.03 | 0.9 ± 0.6 | 0.89 ± 0.02 | 0.78 ± 0.05 | 0.09 ± 0.02 | 0.12 ± 0.03 | 0.983 ± 0.005 |
| Superior colliculus, intermediate gray layer | SCig | 2 | 2.8 ± 0.9 | 8.4 ± 0.4 | 0.11 ± 0.00 | 3.6 ± 2.0 | 0.91 ± 0.01 | 0.86 ± 0.05 | 0.08 ± 0.02 | 0.19 ± 0.04 | 0.986 ± 0.001 |
| Primary somatosensory area | SSp | 2 | 3.3 ± 0.0 | 11.6 ± 1.0 | 0.12 ± 0.01 | 1.7 ± 0.7 | 0.93 ± 0.01 | 0.91 ± 0.02 | 0.08 ± 0.01 | 0.20 ± 0.02 | 0.988 ± 0.002 |
| Primary visual area | VISp | 4 | 1.9 ± 0.3 | 9.5 ± 2.1 | 0.08 ± 0.01 | 4.0 ± 2.5 | 0.84 ± 0.03 | 0.85 ± 0.02 | 0.07 ± 0.00 | 0.16 ± 0.02 | 0.978 ± 0.004 |

Table B.2: **Analysis of scale invariance.** For each region we show the average time bin, $\Delta t$, used in the RG analysis (error as standard deviation across experiments). For each scaling exponent we collect: **(i)** average value across experiments; **(ii)** MAE, computed as the average across experiments of the experiment-specific errors measured over split-quarters of data; **(iii)** standard deviation across experiments; **(iv)** $R^2$ value of best powerlaw fit for the experiment with a greater number of recorded neurons. For the exponent $\mu$, we also provide the $R^2$ value of the best exponential fit, as well as the LR between the estimated powerlaw fits and candidate exponential and lognormal distributions, with positive ratios indicating that the data is best fitted by a powerlaw distribution (statistically significant $p$-values highlighted in boldface).

| Abbrev. | $\Delta t(s)$ | $\langle\alpha\rangle$ | MAE | $\sigma$ | $R^2$ | $\langle\beta\rangle$ | MAE | $\sigma$ | $R^2$ | $\langle z\rangle$ | MAE | $\sigma$ | $R^2$ | $\langle\mu\rangle$ | MAE | $\sigma$ | $R^2_{pow}$ | $R^2_{exp}$ | Exponential LR | $p$ | Lognormal LR | $p$ |
|---|---|---|---|---|---|---|---|---|---|---|---|---|---|---|---|---|---|---|---|---|---|---|
| ACA | 0.16 ± 0.07 | 1.43 | 0.02 | 0.11 | 1.00 | 0.81 | 0.03 | 0.05 | 1.00 | 0.2 | 0.02 | 0.06 | 0.98 | 0.74 | 0.05 | 0.16 | 0.83 | 0.58 | 90.5 | **0.00** | -0.1 | 0.22 |
| BLA | 0.13 ± 0.01 | 1.25 | 0.03 | 0.02 | 1.00 | 0.87 | 0.04 | 0.03 | 1.00 | 0.21 | 0.03 | 0.03 | 1.00 | 0.82 | 0.03 | 0.00 | 0.85 | 0.78 | 9.4 | 0.12 | -2.9 | 0.09 |
| CA1 | 0.13 ± 0.01 | 1.37 | 0.03 | 0.02 | 1.00 | 0.78 | 0.04 | 0.05 | 1.00 | 0.18 | 0.03 | 0.01 | 0.99 | 0.78 | 0.08 | 0.02 | 0.89 | 0.65 | 55.4 | **0.00** | -0.3 | 0.59 |
| CP | 0.08 ± 0.02 | 1.38 | 0.02 | 0.12 | 1.00 | 0.77 | 0.05 | 0.06 | 0.98 | 0.16 | 0.02 | 0.03 | 0.99 | 0.98 | 0.1 | 0.15 | 0.96 | 0.67 | 43.3 | **0.00** | -2.5 | 0.17 |
| LS | 0.17 ± 0.13 | 1.41 | 0.05 | 0.08 | 1.00 | 0.81 | 0.04 | 0.07 | 1.00 | 0.25 | 0.05 | 0.08 | 0.99 | 0.86 | 0.08 | 0.15 | 0.94 | 0.86 | 8.4 | **0.05** | -0.9 | 0.32 |
| LGd | 0.07 ± 0.03 | 1.26 | 0.03 | 0.06 | 1.00 | 0.82 | 0.04 | 0.06 | 1.00 | 0.24 | 0.06 | 0.03 | 1.00 | 0.67 | 0.03 | 0.06 | 0.9 | 0.9 | 0.5 | 0.86 | -1.5 | 0.21 |
| LP | 0.08 ± 0.02 | 1.36 | 0.03 | 0.07 | 1.00 | 0.77 | 0.02 | 0.02 | 1.00 | 0.23 | 0.05 | 0.06 | 0.99 | 0.52 | 0.06 | 0.06 | 0.99 | 0.63 | 152.3 | **0.00** | -3.5 | 0.08 |
| MG | 0.06 ± 0.01 | 1.23 | 0.02 | 0.05 | 1.00 | 0.86 | 0.02 | 0.04 | 1.00 | 0.18 | 0.04 | 0.05 | 0.99 | 0.55 | 0.05 | 0.00 | 0.89 | 0.75 | 24.3 | **0.00** | -0.0 | 0.90 |
| MOp | 0.06 ± 0.01 | 1.5 | 0.03 | 0.01 | 1.00 | 0.78 | 0.03 | 0.06 | 0.99 | 0.22 | 0.02 | 0.05 | 1.00 | 0.92 | 0.05 | 0.05 | 0.85 | 0.45 | 138.8 | **0.00** | -0.2 | 0.66 |
| MOs | 0.16 ± 0.03 | 1.39 | 0.04 | 0.03 | 1.00 | 0.81 | 0.04 | 0.04 | 1.00 | 0.22 | 0.03 | 0.06 | 0.99 | 0.85 | 0.07 | 0.04 | 0.93 | 0.56 | 85.1 | **0.00** | -0.0 | 0.87 |
| ORB | 0.12 ± 0.02 | 1.43 | 0.02 | 0.02 | 1.00 | 0.74 | 0.02 | 0.03 | 0.99 | 0.16 | 0.02 | 0.03 | 0.99 | 0.77 | 0.03 | 0.03 | 0.88 | 0.49 | 146.5 | **0.00** | -0.6 | 0.50 |
| PL | 0.13 ± 0.08 | 1.42 | 0.04 | 0.10 | 1.00 | 0.82 | 0.04 | 0.1 | 0.99 | 0.19 | 0.03 | 0.02 | 0.99 | 0.77 | 0.05 | 0.13 | 0.86 | 0.52 | 180.9 | **0.00** | -0.6 | 0.48 |
| PO | 0.07 ± 0.02 | 1.43 | 0.02 | 0.04 | 1.00 | 0.74 | 0.03 | 0.04 | 1.00 | 0.32 | 0.01 | 0.02 | 0.99 | 0.62 | 0.03 | 0.09 | 0.91 | 0.76 | 22.8 | **0.00** | 0.0 | 0.85 |
| SCig | 0.07 ± 0.02 | 1.31 | 0.04 | 0.01 | 1.00 | 0.82 | 0.04 | 0.01 | 0.98 | 0.21 | 0.04 | 0.02 | 1.00 | 0.77 | 0.08 | 0.06 | 0.96 | 0.76 | 26.6 | **0.00** | -2.8 | 0.11 |
| SSp | 0.06 ± 0.01 | 1.51 | 0.04 | 0.01 | 1.00 | 0.77 | 0.02 | 0.12 | 0.99 | 0.27 | 0.02 | 0.01 | 1.00 | 0.91 | 0.03 | 0.04 | 0.82 | 0.44 | 121.4 | **0.00** | -0.5 | 0.54 |
| VISp | 0.12 ± 0.04 | 1.42 | 0.02 | 0.06 | 1.00 | 0.73 | 0.02 | 0.05 | 1.00 | 0.15 | 0.02 | 0.02 | 0.98 | 0.90 | 0.04 | 0.14 | 0.98 | 0.75 | 22.2 | **0.00** | -1.2 | 0.26 |

# Supplementary Information to Chapter 3

## I  The Mackey-Glass series

The Mackey-Glass chaotic time series used in Chapter 3 was generated from the following delay differential equation:

$$\frac{dx}{dt} = \left[ \frac{\alpha x(t-\tau)}{1 + x(t-\tau)^{\beta}} - \gamma x(t) \right], \tag{C.1}$$

where $\tau$ represents the delay and the parameters are set to $\alpha = 0.2$, $\beta = 10$ and $\gamma = 0.1$, a common choice for this type of prediction tasks (Yusoff et al. 2016; Ortín et al. 2015).

To construct the temporal series, the above equation was solved using $Matlab$ dde23 delay differential equation solver, generating 10000 points with an initial washout period of 1000 points. Integration was performed using a time step of $\Delta t = 0.1$, with an absolute error tolerance of $\varepsilon = 10^{-16}$, and the resulting time series was sampled every 10 points, following the methodology in Jaeger 2001b. Before feeding them to the network, all input series were re-scaled by their maximum value to lay in the range $[0, 1]$.

Forecasting performance of the ESN was evaluated for the MG-17 time series (with a delay $\tau = 17$ and mildly chaotic behavior) using $T = 4000$ points as training set, then testing over $T_{test} = 400$ points.

## II  Derivation of Oja's rule

Originally, Oja proposed the plasticity rule that takes his name as a way of deriving a local rule that implements Hebbian learning while normalizing the synaptic weights at each step of the training (Oja 1982). With this goal in mind, he considered that the total weight connecting pre-synaptic neuron $j$ to post-synaptic neuron $k$ is normalized at each step by the sum of all incoming weights to the post-synaptic neuron:

$$w_{kj}(t+1) = \frac{w_{kj}(t) + \eta y_k(t)x_j(t)}{\sqrt{\sum_j \left(w_{kj}(t) + \eta y_k(t)x_j(t)\right)^2}} \ , \tag{C.2}$$

where $y_k \equiv x_k(t+1)$ in the RC notation. From a biological point of view, this type of normalization could arise, for instance, from a limited number of synaptic resources that must be shared among all synapses. Writing now $w_{kj}(t+1) = g(\eta)$ as a Taylor series expansion around $\eta = 0$:

$$w_{kj}(t+1) = g(0) + g'(0)\eta + \frac{g''(0)}{2}\eta^2 + \cdots \ , \tag{C.3}$$

where:

$$g'(0) \equiv \frac{\partial g}{\partial \eta}\Big|_{\eta=0} = \frac{y_k(t)x_j(t)}{\sqrt{\sum_j \left(w_{kj}(t)\right)^2}} - \frac{w_{kj}(t)\sum_j w_{kj}(t)y_k(t)x_j(t)}{\left(\sum_j w_{kj}^2(t)\right)^{3/2}} \ . \tag{C.4}$$

Using now the condition of normalized incoming weights to the post-synaptic neuron (i.e., $\sqrt{\sum_j \left(w_{kj}(t)\right)^2} = 1$):

$$\frac{\partial g}{\partial \eta}\Big|_{\eta=0} = y_k(t)x_j(t) - w_{kj}(t)y_k(t)\sum_j w_{kj}(t)x_j(t) \tag{C.5}$$

pluggin the above expression into Eq. (C.3) and disregarding al terms of order $\mathcal{O}(\eta^2)$:

$$w_{kj}(t+1) \approx w_{kj}(t) + \eta \left( y_k(t)x_j(t) - w_{kj}(t)y_k(t)\sum_j w_{kj}(t)x_j(t) \right) \ . \tag{C.6}$$

Finally, assuming a linear activation function and no external inputs, so that $y_k(t) = \sum_j w_{kj}(t)x_j(t)$, we obtain the local update equation known as Oja's rule:

$$w_{kj}(t+1) \approx w_{kj}(t) + \eta \left( y_k(t)x_j(t) - w_{kj}(t)y_k^2(t) \right) \ . \tag{C.7}$$

# Supplementary Information to Chapter 4

## I    The representation of a chaotic attractor: beyond the third dimension

In the following plots we show two-dimensional projections of reservoir units activities when subject to the Lorenz's series as input, over all pairwise combinations of the first five principal axes.

Each figure corresponds to a reservoir initialization in one of the four points highlighted in Fig. 4.7, corresponding to a stable dynamical regime (point A, Fig. D.1); a close-to-critical regime (point B, Fig. D.2); an unstable regime, but not too far from the edge of instability (point C, Fig. D.3); and a dynamical state far into the unstable regime (point D, Fig. D.4). The same axes limits have been chosen for all figures and for all subplots within each figure. We can see how, when the dynamics is very stable (point A), most of the variability in the activity of the units can be captured with barely two components, as the reservoir representation inherits the dimensionality of the original input. Closer to the critical point, the trajectories still move within a continuous and differentiable manifold, but of much higher dimensionality as evidenced by the non-trivial dynamics taking place along lower-rank components.

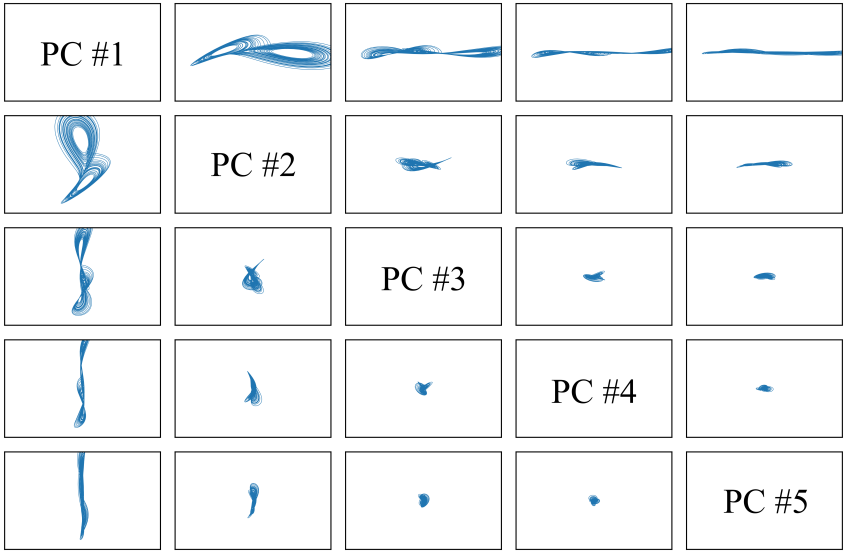

Figure D.1: Projections of reservoir units trajectories onto different pairwise combinations of the first 5 principal components of the activity, during presentation of $T = 6000$ points of the Lorenz's attractor. The ESN was initialized with parameters according to point A in Fig. 4.7, with a measured MLE $= -0.798$.

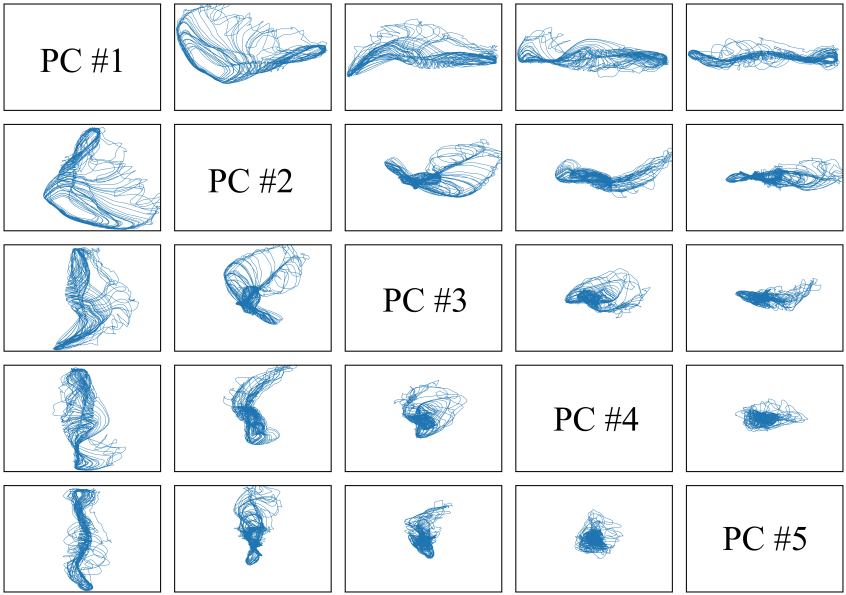

Figure D.2: Projections of reservoir units trajectories onto different pairwise combinations of the first 5 principal components of the activity, during presentation of $T = 6000$ points of the Lorenz's attractor. The ESN was initialized with parameters according to point B in Fig. 4.7, with a measured MLE $= -0.0007$.

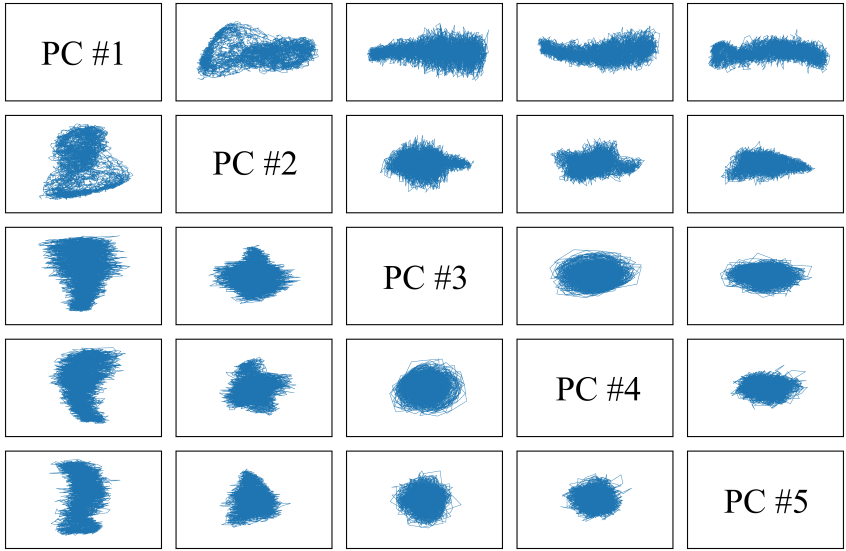

Figure D.3: Projections of reservoir units trajectories onto different pairwise combinations of the first 5 principal components of the activity, during presentation of $T = 6000$ points of the Lorenz's attractor. The ESN was initialized with parameters according to point C in Fig. 4.7, with a measured MLE = 0.028.

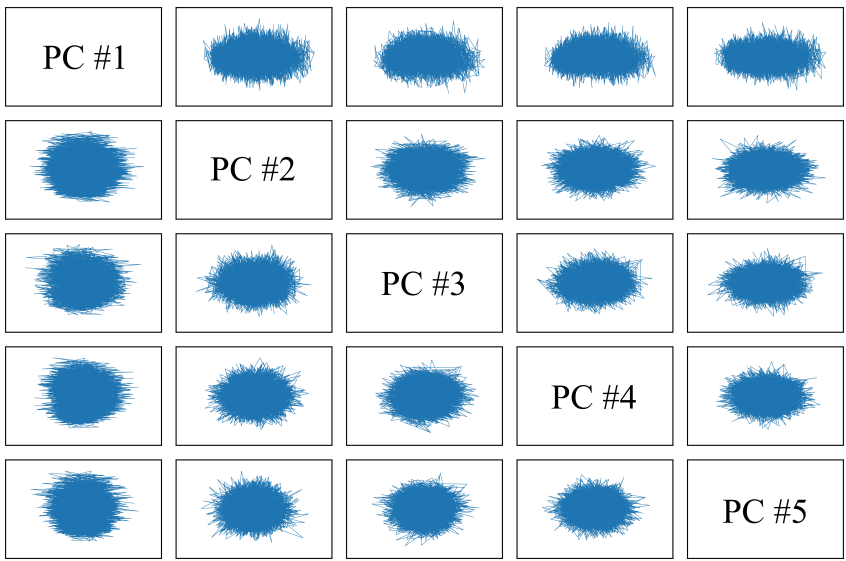

Figure D.4: Projections of reservoir units trajectories onto different pairwise combinations of the first 5 principal components of the activity, during presentation of $T = 6000$ points of the Lorenz's attractor. The ESN was initialized with parameters according to point D in Fig. 4.7, with a measured MLE = 0.122.

# Supplementary Information to Chapter 5

## I  Model parameters

In the following tables, all time parameters have units of seconds, rates are in herzs and potentials in millivolts.

| Experimental Setup | | | | | | | | |
|---|---|---|---|---|---|---|---|---|
| $n_{days}$ | $n_{odors}$ | $n_{trials}$ | $n_{cycles}$ | $\tau_{ins}$ | $\tau_{exp}$ | $\Delta t_{IO}$ | $\Delta t_{ID}$ | dt |
| 32 | 8 | 7 | 8 | 0.150 | 0.350 | 2 | 100 | 0.0005 |

| Weights | | | | | Sparsity | | | | |
|---|---|---|---|---|---|---|---|---|---|
| $\langle w_{mp} \rangle$ | $\langle w_{pp} \rangle$ | $\langle w_{pf} \rangle$ | $\langle w_{fp} \rangle$ | $\langle w_{ff} \rangle$ | $p_{mp}$ | $p_{pp}$ | $p_{pf}$ | $p_{fp}$ | $p_{ff}$ |
| 5 | 1.25 | 10 | 10 | 10 | 0.1 | 0.022 | 0.1 | 0.1 | 0.065 |

| STDP changes | | | | | | | Background noise | | |
|---|---|---|---|---|---|---|---|---|---|
| $\eta_{trans}$ | $\eta_{odor}$ | $\tau_+$ | $\tau_-$ | $A_+$ | $A_-$ | $w_{mp}^{max}$ | $f_{spont}$ | $f_{bl}$ | $f_{exc}$ |
| 0.2 | 1 | 0.020 | 0.050 | 0.001 | 0.083 | 2 | 0.05 | 1.5 | 100 |

## II  Extended methods

### *  Classifier analysis

Following the methods in Schoonover et al. 2021, we trained a Support Vector Machine (SVM) with linear kernel and L2-regularization to classify the odorants from the population responses of pyramidal neurons in the PCx. For within-day classification, we used leave-one-out cross-validation, training on all but one of the 56 trials on a given test day (8 odorants, 7 trials) and then testing on the trial that was left out. This procedure was repeated until all trials on a given day had been tested in this way. For across-day classification, the model was trained using the 56 trials on one day, and then tested on all 56 trials on another day. As control case, we measured performance on shuffled data by randomly permuting the odorant stimulus labels on the test set.

Notably, in the experimental results of Schoonover *et al*, training and testing of SVMs was limited to the lowest number of stable single units for any across-day comparison and mouse (41 units). To produce a sensible comparison, we randomly selected the same number of units in our simulations to train and test the SVM, mimicking the heavily subsampled regime in which the experiments take place.

## † Matrix dissimilarity

The corrected matrix dissimilarity between days $p$ and $q$ is defined as:

$$\|\hat{A}^{p,q}\|_F := \frac{\|A^{p,q}\|_F - \overline{\|A^w\|}_F}{\|A^s\|_F - \overline{\|A^w\|}_F} \ , \tag{E.1}$$

where $\overline{\|A^w\|}_F := (1/n_{days}) \sum_{p=1}^{n_{days}} \|A_{odd}^p - A_{even}^p\|$ is the mean across all days of the within-day Frobenius norm between similarity matrices computed in odd and even trials, and $\|A^s\|_F$ is the Frobenius norm between similarity matrices measured on the first and last day of the experiment after odors shuffling. In this way, $\|\hat{A}^{p,q}\|_F$ is (on average) bounded between zero, for angle drifts on the order of intra-day fluctuations, and one, for the shuffled case.

## ‡ Population statistics

To identify responsive odour–unit pairs, a Wilcoxon rank-sum test (Haynes 2013) was performed between the spike count during the 2-s epoch before stimulus onset for all trials on a given day and the spike count on the seven trials during the odorant stimulus, using a significance level of $\alpha = 0.001$.

Given spontaneous baseline-substracted responses, $r_{j,o}$, for each unit, $j$, to a given odorant, $o$, , average population sparseness was defined as:

$$S_p = \frac{N-1}{n_{odors}N} \sum_{o=1}^{n_{odors}} \left( 1 - \frac{\left( N^{-1} \sum_{j=1}^{N} r_{j,o} \right)^2}{N^{-1} \sum_{j=1}^{N} r_{j,o}^2} \right) \ , \tag{E.2}$$

whereas the average lifetime sparseness across all units was given by:

$$S_{lt} = \frac{n_{odors}-1}{n_{odors}N} \sum_{j=1}^{N} \left( 1 - \frac{\left( n_{odors}^{-1} \sum_{o=1}^{n_{odors}} r_{j,o} \right)^2}{n_{odors}^{-1} \sum_{o=1}^{n_{odors}} r_{j,o}^2} \right) \ . \tag{E.3}$$

To compute within-day correlations between odor responses, population vectors were averaged across even and odd trials separately. Thus, at each day, we defined:

$$c_p = \frac{1}{n_{odors}} \sum_{o=1}^{n_{odors}} \frac{\langle (\mathbf{x}_{p,o}^{even} - \overline{x}_{p,o}^{even})(\mathbf{x}_{p,o}^{odd} - \overline{x}_{p,o}^{odd}) \rangle}{\sigma_{\mathbf{x}_{p,o}^{even}} \sigma_{\mathbf{x}_{p,o}^{odd}}} \ , \tag{E.4}$$

where $\mathbf{x}_{p,o}^{even}$ ($\mathbf{x}_{p,o}^{even}$) is the population response to odor $o$ on day $p$ averaged over all even (odd) trials.

Similarly, the average within-day angle at each day $p$ was defined as:

$$\theta_p = \frac{1}{n_{odors}} \sum_{o=1}^{n_{odors}} \cos^{-1} \frac{\mathbf{x}_{p,o}^{even} \cdot \mathbf{x}_{p,o}^{odd}}{\|\mathbf{x}_{p,o}^{even}\| \|\mathbf{x}_{p,o}^{odd}\|} \ . \tag{E.5}$$

# Acronyms

**ADF** augmented Dickey-Fuller 136

**AI** artificial intelligence 9, 10, 119–121

**BPTT** backpropagation-through-time 11

**cvPCA** cross-validated Principal Component Analysis 77, 78, 81, 83, 84, 91, 132

**EPSP** excitatory post-synaptic potential 4

**ESN** Echo State Network 11, 50–54, 58–62, 64–66, 68, 74, 78, 80, 81, 83–86, 88–91, 93, 147, 150, 151

**FBIN** feedback inhibitory neuron 99, 100, 102–104, 108, 114

**FFIN** feedforward inhibitory neuron 99, 100, 103

**FFNN** feed-forward Neural Network 9–11

**FPP** furthest predicted point 57, 61, 62, 88, 90

**H&H** Hodgkin-Huxley 4, 5, 7

**I&F** integrate-and-fire 7

**IP** intrinsic plasticity 51, 60, 61, 64–69

**IPSP** inhibitory post-synaptic potential 4

**IQR** inter-quartile range 116, 141

**ISI** inter-spike interval 23, 26, 41, 135, 136

**LIF** leaky integrate-and-fire 7, 8, 34, 97, 103

**LOT** lateral olfactory track 96, 99, 100, 103, 108, 113, 115, 116

**LR** loglikelihood ratio 139, 145

**LRM** linear-rate model 29–31, 35–39, 45, 125, 127, 129

**LSM** Liquid State Machine 11, 50

**LTD** long-term depression 104, 105, 108, 113

**LTP** long-term potentiation 104, 105, 108, 113

**MAE** mean absolute error 145

**MC** memory capacity 65

**MEG** magnetoencephalography 43, 45, 48, 134

**MG** Mackey-Glass 60

**ML** machine learning x, 8–11, 50–52, 61, 75, 80, 84, 88, 119

**MLE** maximum Lyapunov exponent x, 63–65, 68, 81, 82, 84–86, 89, 90

**MLP** multilayer perceptron 10

**MP** Marchenko-Pastur 35, 145

**MTC** mitral/tufted cell 99–103, 108, 114

**OB** olfactory bulb 99–101, 113, 115, 116

**OSN** olfactory sensory neuron 98, 99

**PCA** Principal Component Analysis 44, 76, 77, 81, 83, 91

**PCx** piriform cortex 99–104, 106, 108, 110, 112–116, 153

**PDMS** polydimethylsiloxane 39, 134

**PRG** Phenomenological Renormalization Group 21, 22, 25, 28, 37–39, 41, 42, 48, 137

**PSP** post-synaptic potential 4

**RC** Reservoir Computing x, 8, 11, 50, 51, 58, 75, 148

**RD** representational drift 96–98, 100, 101, 109, 112, 115–117

**RG** Renormalization Group 13–15, 20, 21, 28, 50

**RMSE** root mean-square error 57, 61, 62, 88, 90

**RNN** recurrent neural network 10, 11, 50

**ROI** region of interest 40–42, 48, 134

**STDP** spike-time dependent plasticity 96, 101, 105, 106, 113, 114, 116

**SVD** Singular Value Decomposition 131

**SVM** Support Vector Machine 79, 111, 153, 154

# Guillermo B. Morales

## The Dynamics of Neural Codes

How can our external reality be translated into a language of millions of firing neurons? Is such a translation even stable in time? How does an optimal encoding of information depend on the dynamical regime of the brain? Could the same theory that explains magnetization in spin systems unravel new insights into the dynamical properties of neural networks?

From the historical foundations of neuroscience to the implementation of neural plasticity in AI, in this book the reader will come across fractal networks of neurons, quasi-universal scaling relations in the brain, chaotic attractors, and drifting neural representations of the external world. The results presented combine analytical approaches based on the theory of statistical mechanics with data-driven analyses from in-vivo and in-vitro experiments. Additionally, different forms of artificial neural networks are considered to simulate the observed phenomenology in the brain.

Physics, neuroscience and machine learning come together in this book to shed some light on one of the most fundamental yet complex problems of modern science: understanding the language of the brain.

**"***A truly beautiful book.***"**
—The author's grandmother

Cover image and illustrations generated with the assistance of a stable diffusion AI.